\documentclass[]{article}
\usepackage[margin=1 in]{geometry}

\usepackage{lettrine}

\usepackage{etex}
\usepackage{fleqn}
\usepackage{cite}
\usepackage{amsmath,amssymb,amsfonts}
\usepackage{algorithmic}
\usepackage{graphics}
\usepackage{textcomp}
\usepackage{tikz}
\usepackage{color}
\usepackage{comment}
\usepackage{xcolor}
\usepackage{filecontents}
\usepackage{float}
\usetikzlibrary{arrows.meta,patterns}
\usepackage{graphics}
\usepackage{pict2e,color}
\usepackage{subcaption}
\usepackage{nccmath}
\usetikzlibrary{fit}
\usepackage{nccmath}
\usepackage{pgfplots}
\usepackage{lipsum}
\usepackage{stackengine}
\usepackage{cleveref}
\usepackage[ruled,vlined]{algorithm2e}
\usepackage{amsthm}
\usepackage{balance}
\usepackage{color,soul}
\newtheorem{theorem}{Theorem}
\newtheorem{lemma}{Lemma}
\newtheorem{Corollary}{Corollary}
\newtheorem{assumption}{Assumption}
\newtheorem{definition}{Definition}
\newtheorem{proposition}{Proposition}

\setul{0.5ex}{0.3ex}
\setulcolor{blue}

\tikzset{%
  highlight/.style={rectangle,rounded corners,fill=white!15,draw,
    fill opacity=0.2,inner sep=0pt}
}
\newcommand{\mathleft}{\@fleqntrue\@mathmargin0pt}

\newcommand{\tikzmark}[2]{\tikz[overlay,remember picture,
  baseline=(#1.base)] \node (#1) {#2};}

\def\BibTeX{{\rm B\kern-.05em{\sc i\kern-.025em b}\kern-.08em
    T\kern-.1667em\lower.7ex\hbox{E}\kern-.125emX}}

\def\T{\mathrm{T}}
\def\E{\mathrm{E}}

\def\e{\mathrm{e}}

\def\D{\mathbf{D}}
\DeclareMathOperator*{\argmax}{arg\,max}
\usepackage[colorinlistoftodos,prependcaption,textsize=footnotesize]{todonotes}
\usepackage{regexpatch}
\makeatletter
\xpatchcmd{\@todo}{\setkeys{todonotes}{#1}}{\setkeys{todonotes}{inline,#1}}{}{}
\makeatother

\begin{document}

\date{}

\title{On the Impact of Network Delays on Time-to-Live Caching
%On the Impact of Network Delays on Time-to-Live Caching
}

%\author{\IEEEauthorblockN{Karim Elsayed and Amr Rizk}\\
%\IEEEauthorblockA{University of Duisburg-Essen,
%Germany}\\
%{\emph{\small first.last@uni-due.de}}
%}
\author{Karim Elsayed and Amr Rizk\\
{University of Duisburg-Essen} \\
{\emph{\small first.last@uni-due.de}}
}
\maketitle

\begin{abstract}
We consider Time-to-Live (TTL) caches that tag every object in cache with a specific (and possibly renewable) expiration time.
State-of-the-art models for TTL caches assume zero object fetch delay, i.e., the time required to fetch a requested object \textit{that is not in cache} from a different cache or the origin server.
%This object request is usually forwarded according to some routing strategy to a different cache or to the source origin.
Particularly, in cache hierarchies this delay has a significant impact on performance metrics such as the object hit probability.
Recent work suggests that the impact of the object fetch delay on the cache performance will continue to increase due to the scaling mismatch between
shrinking
%ever shrinking possible
inter-request times (due to higher data center link rates) in contrast to processing and memory access times.

In this paper, we analyze tree-based cache hierarchies with random object fetch delays and provide an exact analysis of the corresponding object hit probability.
Our analysis allows understanding the impact of random delays and TTLs on cache metrics for a wide class of request stream models characterized through Markov arrival processes.
This is expressed through a metric that we denote \textit{delay impairment} of the hit probability.
In addition, we analyze and extend state-of-the-art approximations of the hit probability to take the delay into account.
We provide numerical and trace-based simulation-based evaluation results showing that larger TTLs do not efficiently compensate the detrimental effect of object fetch delays.
Our evaluations also show that unlike our exact model the state-of-the-art approximations do not capture the impact of the object fetch delay well especially for cache hierarchies.
Surprisingly, we show that the impact of this delay on the hit probability is not monotonic but depends on the request stream properties, as well as, the TTL.
% \textcolor{red}{Moreover, we provide trace based results using repositories from a network file system at Microsoft.
% Furthermore, we use state lumpability to potentially reduce the computational complexity of the MAP modelling a tree-based cache hierarchy.
% We show that the computational complexity significantly decreases when the tree has symmetric caches.}

% of the TTL and the fetch delay distributions for different canonical caching hierarchies.
% We note that the shown results are not limited to this timer-driven caching strategy as extensive research have shown remarkable relations (in expectation) of TTL caching and classical capacity driven caching strategies such as LRU and FIFO.
% Notes so far on the efficient approximation of caching inter-miss time taking into account the time delay to fetch the object. In addition, the application of such approximation to line caching networks to recursively calculate the closed form inter-miss time at any cache.
\end{abstract}

%\begin{IEEEkeywords}
%Caching under network delays, TTL caching, Markov arrival processes, State space Lumpability
%\end{IEEEkeywords}

\vspace{-10pt}
\section{Introduction}

%\begin{itemize}
%  \item general context of the paper, why is the topic or context important?
%  \item what is the impact of solving that paper?
%  \item contributions of this paper
%\end{itemize}

% Why is caching important?
Content orientation is the raison d'être for a wide scale of caching infrastructures in the Internet~\cite{AndreevMMS03,CDN,Bhat:SABR} in addition to being a foundation for new Internet architectures\cite{ICNsurvey}.
Recently, a renewed interest in a class of caching systems, namely Time-to-Live (TTL) caches, triggered a line of analytical work on the performance evaluation of such systems~\cite{fofack:DNS,Fofack,Daniel,Martina:2014,Fricker,Mostafa}.
%\todo[inline]{@Karim: please add appropriate references}
A remarkable property of TTL caches is that some of their performance measures are equivalent to their counterparts for classical caching algorithms such as Least Recently Used (LRU) under mild conditions\cite{che:02:approx,Fricker,Martina:2014}.

The analysis of large-scale, interconnected TTL caching systems is a difficult task that does not simply allow for exact results.
%due to the problem complexity.
This difficulty stems from the calculation of the forwarded object request stream of each cache, i.e., the cache miss process, and the aggregation of multiple of such processes as input to parent caches. To this end, related work derives approximations that may deviate significantly from the caching system behavior \cite{Daniel}.
A significant contribution to the analysis of TTL cache is provided by \cite{Daniel} through an exact formulations for hierarchies under the assumption of  \textit{zero object fetch delay}.
%We consider TTL caches, which can be accurately modeled  using the method in~\cite{Daniel}.
We note that in general TTL-based caches are often analyzed under idealized communication assumptions such as negligible latency, no packet loss, no network dynamics, and no capacity constraints~\cite{Schomp2020,Fofack,Daniel,Fricker,Martina:2014,rosensweig2013steady,che:02:approx,Rizk:caching_hier_2017}.

Observations show, however, that particularly in cache hierarchies the object fetch delay may have a significant adverse effect on the cache performance especially in the tail.
The importance of modelling the object fetch delay was pointed at in a recent work~\cite{AtreSWB20} that has shown that there is a growing mismatch between data center link rates, thus the ever shrinking request inter-arrival times, and the processing/memory access times in caching systems~\cite{LiRJ18}.
This is reflected in form of empirically observed request waiting behavior and service batching~\cite{AtreSWB20}.
This suggests that the impact of the delay on the cache performance will continue to increase due to this scaling mismatch.

In this paper, we analyze tree-based TTL cache hierarchies under random delays and provide an exact model extending the previous approaches to a joint consideration of caching and communication.
Our analysis provides the object hit probability under link-based delays allowing to understand the impact of random network delays and TTLs on cache metrics for a wide class of request stream models.
% To this end, we provide an algorithm to construct a Markov arrival process description of the output of a given caching hierarchy, i.e., its miss process for random link delays.
To incorporate the object fetch delays in our proposed model we provide a recursive algorithm to construct a Markov Arrival Process (MAP) description of the output of a given caching hierarchy, i.e., its miss process.
To relate the impact of the delay to the object hit probability we coin the metric \textit{delay impairment}.

In addition to the exact analysis of TTL cache hierarchies under random link delays we analyze state-of-the-art  approximations of the object hit probability that aim to encompass these delays.
Our numerical as well as trace-based evaluation results shows that unlike our exact model the state-of-the-art approximations do not capture the object fetch delays well.
Moreover, we show that larger TTLs are not efficient in compensating for the object fetch delay especially in  caching hierarchies.
Interestingly, for request processes that approach a periodic behavior it can be shown that the hit probability is not monotonically decreasing in the delay.
To reduce the computational complexity of the calculation of performance measures for TTL caching trees we provide an analytical approach to state lumpability that takes advantage of symmetric sub-trees of caches within a hierarchy.

Our contributions can be summarized as follows:
\begin{itemize}
    \item We provide an exact analysis of TTL cache hierarchies under non-zero object fetch delay. In addition, we generalize the existing models to Phase-type distributed inter-request times,  TTLs, and delays.
    \item We analyze the computation complexity of the exact performance model and provide a rigorous computational speedup method for calculating exact cache performance metrics based on a MAP lumpability argument taking into account cache tree symmetry and recursion.
    \item We provide numerical and trace-driven simulation results that show the significant detrimental impact of network delays on cache hierarchy performance, as well as, the accuracy of our model. Further results also show the non-trivial approximation error due to state-of-the art approximation methods.
\end{itemize}

The remainder of the paper is structured as follows: We first provide an overview of related work on modeling and analyzing TTL caches in Sect.~\ref{sec:background} before describing the system model and the problem statement in Sect.~\ref{Section:MAP_approach}. In Sect.~\ref{sec:exact_models} we analyze cache hierarchies under object fetch delays.  In Sect.~\ref{sec:approx_hierarchies} we derive performance metric approximations for single caches and for cache hierarchies and in Sect.~\ref{sec:lumpability} we provide a model-based approach to speed up the numerical calculation of cache hierarchy performance metrics. Finally, in Sect.~\ref{sec:evaluations} we show numerical and trace-based simulation results before concluding the paper in Sect.\ref{sec:conclusions}.

\vspace{-10pt}
\section{Background and Related work}
\label{sec:background}
%Unlike classical caching algorithms such as LRU, LFU, RANDOM, ARC, etc., objects in
Time-to-Live caches decouple the management of objects in the cache.
The inclusion and removal of an object from the cache is based solely on the request process \textit{for that object} and its TTL, i.e., the validity timestamp assigned to the object.
An object request generates either a \textit{hit} if the object is present in the cache's local memory, or a \textit{miss} if not.
In the case of a hit, the object is delivered to the request source immediately.
In the case of a miss, the object is requested from another cache in the cache  hierarchy (or finally the origin server) according to a specified routing algorithm.
Upon receiving the object the cache that generated the miss in the first place forwards the object to the request source and adds it to local storage with a  validity timestamp, the TTL.
%\footnote{Established applications of TTL caching include DNS \cite{fofack:DNS} and ICN~\cite{ICNsurvey}.}
Objects are removed from local storage upon TTL expiry.
TTL can be \textit{regenerative} or \textit{fixed}, i.e., renewing upon an object hit or not, respectively.
TTLs can also set deterministically or randomly with a given distribution.
Different methods exist for selecting and optimizing TTLs~\cite{Daniel,FerragutRP18} to optimize the cache performance.
The work in \cite{Fricker} shows that the number of TTL cached objects in one cache follows a concentrated distribution.
%Note that a simple argument on the average occupancy can be used to relate the TTL model to capacity driven caching as discussed, e.g. in \cite{Daniel}.
Further work shows how parameterization rules for TTLs lead to statistically equivalent cache performance when compared to some classical caching algorithms such as LRU.
An established argument for this relates the expected cache occupancy to cache capacity
%in order to link the TTL model to capacity-driven caching
\cite{Fofack,JiangNT18,Daniel,Gelenbe73a,Che}.

\vspace{-10pt}
\subsection{Single Cache Analysis}
\label{subsec:single_cache_analysis}

State-of-the-art analytical models of TTL caches usually assume idealized communication such as negligible object fetch latency, no losses, no network dynamics, and no link capacity constraints~\cite{Schomp2020,Fofack,Daniel,Fricker,Martina:2014,rosensweig2013steady,che:02:approx}.
Further assumptions on caching networks may even suspend the dynamics of cache contents over certain time windows~\cite{MullerAS017,MaghsudiS20,Ji16}.

\begin{figure}[t]
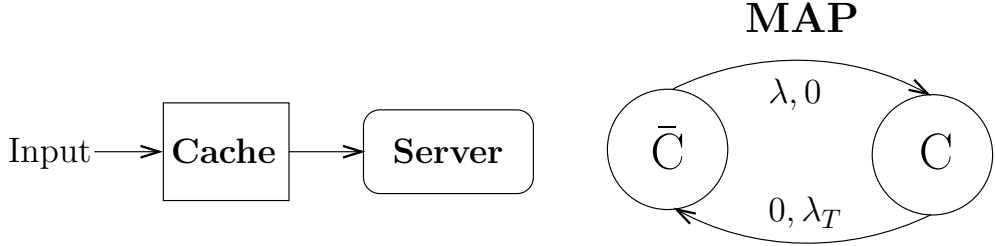

\centering
\resizebox{0.8\linewidth}{!}{\ifx\XFigwidth\undefined\dimen1=0pt\else\dimen1\XFigwidth\fi
\divide\dimen1 by 6404
\ifx\XFigheight\undefined\dimen3=0pt\else\dimen3\XFigheight\fi
\divide\dimen3 by 1603
\ifdim\dimen1=0pt\ifdim\dimen3=0pt\dimen1=4143sp\dimen3\dimen1
  \else\dimen1\dimen3\fi\else\ifdim\dimen3=0pt\dimen3\dimen1\fi\fi
\tikzpicture[x=+\dimen1, y=+\dimen3]
{\ifx\XFigu\undefined\catcode`\@11
\def\temp{\alloc@1\dimen\dimendef\insc@unt}\temp\XFigu\catcode`\@12\fi}
\XFigu4143sp
% Uncomment to scale line thicknesses with the same
% factor as width of the drawing.
%\pgfextractx\XFigu{\pgfqpointxy{1}{1}}
\ifdim\XFigu<0pt\XFigu-\XFigu\fi
\pgfdeclarearrow{
  name = xfiga0,
  parameters = {
    \the\pgfarrowlinewidth \the\pgfarrowlength \the\pgfarrowwidth},
  defaults = {
	  line width=+7.5\XFigu, length=+120\XFigu, width=+60\XFigu},
  setup code = {
    % miter protrusion = thk * sqrt(wd^2 + (tipmv*len)^2) / (2 * wd)
    \dimen7 2.15\pgfarrowlength\pgfmathveclen{\the\dimen7}{\the\pgfarrowwidth}
    \dimen7 2\pgfarrowwidth\pgfmathdivide{\pgfmathresult}{\the\dimen7}
    \dimen7 \pgfmathresult\pgfarrowlinewidth
    \pgfarrowssettipend{+\dimen7}
    \pgfarrowssetbackend{+-\pgfarrowlength}
    \dimen9 -0.5\pgfarrowlinewidth
    \pgfarrowssetvisualbackend{+\dimen9}
    \pgfarrowssetlineend{+-0.5\pgfarrowlinewidth}
    \pgfarrowshullpoint{+\dimen7}{+0pt}
    \pgfarrowsupperhullpoint{+-\pgfarrowlength}{+0.5\pgfarrowwidth}
    \pgfarrowssavethe\pgfarrowlinewidth
    \pgfarrowssavethe\pgfarrowlength
    \pgfarrowssavethe\pgfarrowwidth
  },
  drawing code = {\pgfsetdash{}{+0pt}
    \ifdim\pgfarrowlinewidth=\pgflinewidth\else\pgfsetlinewidth{+\pgfarrowlinewidth}\fi
    \pgfpathmoveto{\pgfqpoint{-\pgfarrowlength}{0.5\pgfarrowwidth}}
    \pgfpathlineto{\pgfqpoint{0pt}{0pt}}
    \pgfpathlineto{\pgfqpoint{-\pgfarrowlength}{-0.5\pgfarrowwidth}}
    \pgfusepathqstroke
  }
}
\clip(749,-4107) rectangle (7153,-2504);
\tikzset{inner sep=+0pt, outer sep=+0pt}
\pgfsetarrows{[line width=15\XFigu, width=75\XFigu]}
\pgfsetarrowsend{xfiga0}
\pgfsetlinewidth{+7.5\XFigu}
\draw (5085,-3105) arc[start angle=+116.15, end angle=+60.75, radius=+1791.6];
\draw (6750,-3915) arc[start angle=+-63.85, end angle=+-119.25, radius=+1791.6];
\draw  (5050,-3495) circle [radius=+388];
\draw  (6757,-3530) circle [radius=+388];
\draw (1800,-3195) rectangle (2610,-3825);
\draw (2610,-3510)--(3105,-3510);
\draw (4185,-3780) [rounded corners=+105\XFigu] rectangle (3090,-3240);
\draw (1355,-3510)--(1800,-3510);
\pgftext[base,left,at=\pgfqpointxy{3280}{-3585},rotate=+360] {\fontsize{14}{12.4}\textbf{Server}}
\pgftext[base,left,at=\pgfqpointxy{800}{-3575},rotate=+360] {\fontsize{15}{13.4}$\text{Input}$} 
\pgftext[base,left,at=\pgfqpointxy{1850}{-3585},rotate=+360] {\fontsize{14}{12.4}\textbf{Cache}}
\pgftext[base,left,at=\pgfqpointxy{6670}{-3620},rotate=+360] {\fontsize{20}{14.4}$\mathrm{C}$}
\pgftext[base,left,at=\pgfqpointxy{4940}{-3620},rotate=+360] {\fontsize{20}{14.4}${\bar{\mathrm{C}}}$}
\pgftext[base,left,at=\pgfqpointxy{5700}{-3960},rotate=+360] {\fontsize{15}{14.4}$0,\lambda_T$}
\pgftext[base,left,at=\pgfqpointxy{5700}{-3195},rotate=+360] {\fontsize{15}{14.4}$\lambda,0$}
\pgftext[base,left,at=\pgfqpointxy{5550}{-2740},rotate=+360]{\fontsize{18}{14.4}\textbf{MAP}}
\endtikzpicture%
}
\caption{MAP for a single cache with zero object fetch delay. C denotes object in cache, $\bar{C}$ denotes object out of cache. The parameters $\lambda,\lambda_T$ are related to the request process resp. TTL.}
\label{Single_MAP_no_D}
\vspace{10pt}
\end{figure}

\begin{figure}[t]
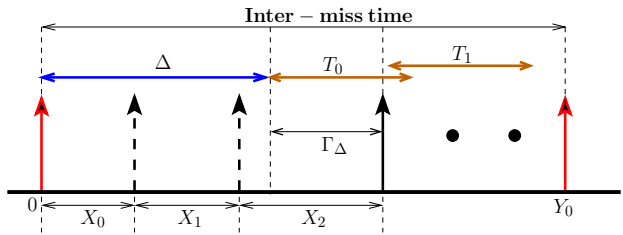

\centering
\resizebox{0.5\textwidth}{!}{\ifx\XFigwidth\undefined\dimen1=0pt\else\dimen1\XFigwidth\fi
\divide\dimen1 by 7202
\ifx\XFigheight\undefined\dimen3=0pt\else\dimen3\XFigheight\fi
\divide\dimen3 by 2637
\ifdim\dimen1=0pt\ifdim\dimen3=0pt\dimen1=3977sp\dimen3\dimen1
  \else\dimen1\dimen3\fi\else\ifdim\dimen3=0pt\dimen3\dimen1\fi\fi
\tikzpicture[x=+\dimen1, y=+\dimen3]
{\ifx\XFigu\undefined\catcode`\@11
\def\temp{\alloc@1\dimen\dimendef\insc@unt}\temp\XFigu\catcode`\@12\fi}
\XFigu3977sp
% Uncomment to scale line thicknesses with the same
% factor as width of the drawing.
%\pgfextractx\XFigu{\pgfqpointxy{1}{1}}
\ifdim\XFigu<0pt\XFigu-\XFigu\fi
\pgfdeclarearrow{
  name = xfiga0,
  parameters = {
    \the\pgfarrowlinewidth \the\pgfarrowlength \the\pgfarrowwidth},
  defaults = {
	  line width=+7.5\XFigu, length=+120\XFigu, width=+60\XFigu},
  setup code = {
    % miter protrusion = thk * sqrt(wd^2 + (tipmv*len)^2) / (2 * wd)
    \dimen7 2.15\pgfarrowlength\pgfmathveclen{\the\dimen7}{\the\pgfarrowwidth}
    \dimen7 2\pgfarrowwidth\pgfmathdivide{\pgfmathresult}{\the\dimen7}
    \dimen7 \pgfmathresult\pgfarrowlinewidth
    \pgfarrowssettipend{+\dimen7}
    \pgfarrowssetbackend{+-\pgfarrowlength}
    \dimen9 -0.5\pgfarrowlinewidth
    \pgfarrowssetvisualbackend{+\dimen9}
    \pgfarrowssetlineend{+-0.5\pgfarrowlinewidth}
    \pgfarrowshullpoint{+\dimen7}{+0pt}
    \pgfarrowsupperhullpoint{+-\pgfarrowlength}{+0.5\pgfarrowwidth}
    \pgfarrowssavethe\pgfarrowlinewidth
    \pgfarrowssavethe\pgfarrowlength
    \pgfarrowssavethe\pgfarrowwidth
  },
  drawing code = {\pgfsetdash{}{+0pt}
    \ifdim\pgfarrowlinewidth=\pgflinewidth\else\pgfsetlinewidth{+\pgfarrowlinewidth}\fi
    \pgfpathmoveto{\pgfqpoint{-\pgfarrowlength}{0.5\pgfarrowwidth}}
    \pgfpathlineto{\pgfqpoint{0pt}{0pt}}
    \pgfpathlineto{\pgfqpoint{-\pgfarrowlength}{-0.5\pgfarrowwidth}}
    \pgfusepathqstroke
  }
}
\pgfdeclarearrow{
  name = xfiga2,
  parameters = {
    \the\pgfarrowlinewidth \the\pgfarrowlength \the\pgfarrowwidth\ifpgfarrowopen o\fi},
  defaults = {
	  line width=+7.5\XFigu, length=+120\XFigu, width=+60\XFigu},
  setup code = {
    % miter protrusion = thk * sqrt(wd^2 + (tipmv*len)^2) / (2 * wd)
    \dimen7 2.6\pgfarrowlength\pgfmathveclen{\the\dimen7}{\the\pgfarrowwidth}
    \dimen7 2\pgfarrowwidth\pgfmathdivide{\pgfmathresult}{\the\dimen7}
    \dimen7 \pgfmathresult\pgfarrowlinewidth
    \pgfarrowssettipend{+\dimen7}
    \pgfarrowssetbackend{+-1.25\pgfarrowlength}
    \dimen9 -\pgfarrowlength\advance\dimen9 by-0.5\pgfarrowlinewidth
    \pgfarrowssetlineend{+\dimen9}
    \dimen9 -\pgfarrowlength\advance\dimen9 by-0.5\pgfarrowlinewidth
    \pgfarrowssetvisualbackend{+\dimen9}
    \pgfarrowshullpoint{+\dimen7}{+0pt}
    \pgfarrowsupperhullpoint{+-1.25\pgfarrowlength}{+0.5\pgfarrowwidth}
    \pgfarrowssavethe\pgfarrowlinewidth
    \pgfarrowssavethe\pgfarrowlength
    \pgfarrowssavethe\pgfarrowwidth
  },
  drawing code = {\pgfsetdash{}{+0pt}
    \ifdim\pgfarrowlinewidth=\pgflinewidth\else\pgfsetlinewidth{+\pgfarrowlinewidth}\fi
    \pgfpathmoveto{\pgfqpoint{-1.25\pgfarrowlength}{-0.5\pgfarrowwidth}}
    \pgfpathlineto{\pgfqpoint{0pt}{0pt}}
    \pgfpathlineto{\pgfqpoint{-1.25\pgfarrowlength}{0.5\pgfarrowwidth}}
    \pgfpathlineto{\pgfqpoint{-\pgfarrowlength}{0pt}}
    \pgfpathclose
    \ifpgfarrowopen\pgfusepathqstroke\else\pgfsetfillcolor{.}
	\ifdim\pgfarrowlinewidth>0pt\pgfusepathqfillstroke\else\pgfusepathqfill\fi\fi
  }
}
\definecolor{brown3}{rgb}{0.75,0.38,0}
\clip(2801,-6087) rectangle (10003,-3450);
\tikzset{inner sep=+0pt, outer sep=+0pt}
\pgfsetlinewidth{+7.5\XFigu}
\filldraw  (8010,-4950) circle [radius=+68];
\filldraw  (8730,-4950) circle [radius=+68];
\pgfsetdash{{+60\XFigu}{+60\XFigu}}{++0pt}
\draw (3241,-5661)--(3241,-6075);
\pgfsetlinewidth{+30\XFigu}
\pgfsetstrokecolor{red}
\pgfsetdash{}{+0pt}
\pgfsetarrows{[line width=30\XFigu, width=126\XFigu, length=144\XFigu]}
\pgfsetarrowsend{xfiga2}
\draw (3241,-5606)--(3241,-4472);
\pgfsetstrokecolor{.}
\draw (7200,-5589)--(7200,-4455);
\pgfsetdash{{+120\XFigu}{+120\XFigu}}{++0pt}
\pgfsetarrows{[width=126\XFigu, length=144\XFigu]}
\draw (4320,-5589)--(4320,-4455);
\draw (5535,-5589)--(5535,-4455);
\pgfsetstrokecolor{blue}
\pgfsetdash{}{+0pt}
\pgfsetarrows{[width=90\XFigu]}
\pgfsetarrows{xfiga0-xfiga0}
\draw (3196,-4275)--(5895,-4275);
\pgfsetstrokecolor{brown3}
\pgfsetarrows{[length=120\XFigu]}
\draw (5850,-4275)--(7560,-4275);
\draw (7245,-4140)--(8955,-4140);
\pgfsetlinewidth{+7.5\XFigu}
\pgfsetstrokecolor{.}
\pgfsetdash{{+60\XFigu}{+60\XFigu}}{++0pt}
\pgfsetarrows{-}
\draw (7200,-3690)--(7200,-4481);
\pgfsetlinewidth{+30\XFigu}
\pgfsetstrokecolor{red}
\pgfsetdash{}{+0pt}
\pgfsetarrows{[width=126\XFigu, length=144\XFigu]}
\pgfsetarrowsend{xfiga2}
\draw (9315,-5606)--(9315,-4472);
\pgfsetlinewidth{+7.5\XFigu}
\pgfsetstrokecolor{.}
\pgfsetdash{{+60\XFigu}{+60\XFigu}}{++0pt}
\pgfsetarrowsend{}
\draw (9315,-3690)--(9315,-4571);
\pgfsetlinewidth{+45\XFigu}
\pgfsetdash{}{+0pt}
\draw (2845,-5606)--(9959,-5606);
\pgfsetlinewidth{+7.5\XFigu}
\pgfsetdash{{+60\XFigu}{+60\XFigu}}{++0pt}
\draw (3241,-3690)--(3241,-4526);
\draw (5895,-3690)--(5895,-5634);
\pgfsetdash{}{+0pt}
\pgfsetarrows{[line width=7.5\XFigu, width=72\XFigu, length=144\XFigu]}
\pgfsetarrows{xfiga0-xfiga0[line width=1.5*\the\XFigu]}
\draw (3241,-3690)--(9315,-3690);
\pgfsetarrowsend{xfiga0}
\draw (3241,-5760)--(4320,-5760);
\draw (4321,-5760)--(5535,-5760);
\draw (5536,-5760)--(7200,-5760);
\pgfsetdash{{+60\XFigu}{+60\XFigu}}{++0pt}
\pgfsetarrows{-}
\draw (4320,-5661)--(4320,-6075);
\draw (5535,-5661)--(5535,-6075);
\draw (7200,-5661)--(7200,-6075);
\pgfsetdash{}{+0pt}
\pgfsetarrows{[line width=15\XFigu, width=75\XFigu, length=120\XFigu]}
\pgfsetarrows{xfiga0-xfiga0}
\draw (5895,-4905)--(7200,-4905);
\pgftext[base,left,at=\pgfqpointxy{3078}{-5820}] {\fontsize{15}{15.6}$0$}
\pgftext[base,left,at=\pgfqpointxy{4555}{-4175}] {\fontsize{15}{54.4}$\Delta$}
\pgftext[base,left,at=\pgfqpointxy{6505}{-4200}] {\fontsize{15}{14.4}$T_0$}
\pgftext[base,left,at=\pgfqpointxy{8005}{-4075}] {\fontsize{15}{14.4}$T_1$}
\pgftext[base,left,at=\pgfqpointxy{9180}{-5850}] {\fontsize{15}{15.6}$Y_0$}
\pgftext[base,left,at=\pgfqpointxy{5600}{-3630}] {\fontsize{15}{20.4}$\mathbf{Inter-miss \ time}$ }
\pgftext[base,left,at=\pgfqpointxy{3695}{-5980}] {\fontsize{15}{15.6}$X_0$}
\pgftext[base,left,at=\pgfqpointxy{4820}{-5980}] {\fontsize{15}{15.6}$X_1$}
\pgftext[base,left,at=\pgfqpointxy{6255}{-5980}] {\fontsize{15}{15.6}$X_2$}
\pgftext[base,left,at=\pgfqpointxy{6490}{-5125}] {\fontsize{15}{14.4}$\Gamma_\Delta$}
\endtikzpicture%
}
\caption{TTL cache model including the object fetch delay. For one object the inter-request times are given as $X_i$, TTLs as $T_i$ and inter-miss times $Y_i$. The delay is given by $\Delta$. We count requests arriving during $\Delta$ (dashed arrows) as misses.}
\label{TTL_D}
\vspace{10pt}
\end{figure}

Considering a single cache as depicted in Fig.~\ref{Single_MAP_no_D} (left), the standard analysis assumes a renewal request process for any given object as depicted in Fig.~\ref{TTL_D}.
The inter-request time is given by the independent and identically distributed (IID) random variables $X_i \sim X$.
The IID random variables $T_i \sim T$ denote the TTLs, whereas the object fetch delay is zero (i.e. $\Delta=0$ in Fig.~\ref{TTL_D}) and $Y$ represents the inter-miss time.
Given this model, a cache hit occurs when the inter-request time is less than the TTL and the object hit probability $P_h$ is~\cite{Fofack}
\vspace{-5pt}
\begin{equation*}
P_h = \mathbb{P}(X<T)= \int_{0}^{\infty} (1-F_T(t)) dF_X(t) \;,
\label{hit rate no delay}
\end{equation*}
where $F_T(t)$ and $F_X(t)$ are the CDFs of the TTL and the inter-request time, respectively.
% The inter-miss time CDF can be exactly described as in \cite{Fofack,Daniel} as
% \begin{equation}
% 	F_Y(t)=F_X(t)-L(t)+\int_0^t F_Y(t-x)dL(x) \;,
% \end{equation}
% where  $L(t)= \int_0^t (1-F_T(t)) dF_X(t)$.
% % \begin{equation}
% % 	L(t)= \int_0^t (1-F_T(t)) dF_X(t) \;.
% % 	\label{eq:L(t)}
% % \end{equation}
% In a more compact form,
Hence, the Laplace-Stieltjes Transform (LST) $F_Y^*(s)$ of the inter-miss time with CDF $F_Y(t)$ is given by \cite{Fofack}
\vspace{-5pt}
\begin{equation}
	F_Y^*(s)= \frac{F_X^*(s)-L^*(s)}{1-L^*(s)} \;,
	\label{Miss_process laplace}
\end{equation}
where $F_X^*(s)$ is the LST of the inter-request time and $L^*(s)$ is the LST of $L(t)= \int_0^t (1-F_T(t)) dF_X(t)$.

\vspace{-10pt}
\subsection{Analysis of TTL Cache Hierarchies}

At the core of the analysis of cache hierarchies lies the problem of describing the superposition of miss processes coming from different child caches as input to one parent cache.
The standard approximation for this superposition is given by the assumption that the miss process of any cache follows a Poisson process parameterized with the cache miss probability.
This approximation, however, deviates significantly from the exact miss process, especially when there are multiple levels of caching within the hierarchy~\cite{Daniel}.
The work in \cite{Fofack} provides an exact expression for the miss process of a single cache under zero fetch delay as a function of a renewal input process and the TTL, however, this \textit{exact} approach is not extended to hierarchies.
Relaxing the Poisson assumption of the request process to renewal processes allows to model a large class of request processes~\cite{Melazzi:2014,Martina:2014,Fofack,Rizk:caching_hier_2017}, but due to the superposition properties of renewal processes, assuming that the multiplexed miss process is renewal leads to significant approximation errors in the analysis of caching hierarchies.

The authors of \cite{Daniel} provide an exact analysis of caching hierarchies under zero fetch delay including miss process superposition using Markov arrival processes (MAPs).
A MAP is defined using two transition matrices ($\D_0$,$\D_1$) of same size that contain hidden and active transitions,  respectively. The matrix $\D_0$ controls a background Markov process $J(t)$ while $\D_1$ controls $J(t)$, as well as, a counting process $N(t)$ (for an introduction see~\cite{Asmussen}).
For example, for a given cache with Poisson request process with rate $\lambda$ and exponentially distributed TTL with parameter $\lambda_T$ the state for one object is shown in Fig.~\ref{Single_MAP_no_D} (right).
% $\left\{in,out\right\}$.
The elegant modelling using MAPs in \cite{Daniel} allows to exactly express the superposition of miss-processes using the Kronecker sum and Kronecker multiplication operations on the corresponding MAPs.
The object hit probability is given thus as \cite{Daniel}
\begin{equation}
\label{eq:object_hit_probability}
	P_h=1-\frac{\mathbf{\pi} \mathbf{D}_1 \mathbf{1}}{\mathbf{\pi}^{(I)} \mathbf{D}_1^{(I)} \mathbf{1}^{(I)}} \;,
\end{equation}
where $\mathbf{D}_1$ is the active matrix of the \textit{caching system MAP} and $\mathbf{\pi}$ is the vector of the corresponding steady state probabilities.
The active matrix $\mathbf{D}_1^{(I)}$ and the probability vector $\mathbf{\pi}^{(I)}$ are similarly defined for the MAP representing the input process of requests arriving to the caching system.
The vectors $\mathbf{1}$ and $\mathbf{1}^{(I)}$ are all ones.

%refer to All one vectors of size $M \times 1$ and $N \times 1$ respectively.
%\todo[inline]{@Karim please include the hit prob for MAPs here}

Taking the object fetch delay into account, the approach in \cite{Mostafa} showed first analytical results to incorporate the communication latency for a single TTL cache into the exact analysis.
However, to extend this analysis to hierarchies the work \cite{Mostafa} resorts to the inaccurate Poisson multiplexing approximation, i.e., assuming that the miss process of every cache is Poisson.
Recall that this assumption has been shown to produce significant errors (even in the zero object fetch delay regime)~\cite{Daniel}.
Further work that considers the download delay time for objects in single LRU caches includes \cite{PIT_CS,PIT_Cs_1}, where the authors propose variants of the LRU algorithm that work better under non-zero delay by jointly analyzing single LRU caching and pending interest table (PIT) that contain object requests that are not yet fetched.
The authors of \cite{RAM} consider the object download delay and propose an algorithm based on q-LRU to optimize the average latency based on the popularity of the object and its size.

On a different note, approaches that optimize TTL caching hierarchies, e.g., using utility functions can be found in~\cite{DehghanJSHSKTS17,DehghanMTMT19}.
Starting from a formulation as an optimization problem, TTL values that maximize the hit probability are sought in~\cite{DehghanMTMT19,FerragutRP18,FerragutRP16}. This problem is solved for heavy-tailed arrivals in~\cite{FerragutRP16}. In \cite{NegliaCM18}, a caching optimization problem with linear cost functions is proposed and solved. In \cite{WangTKC16}, a utility-based approach is proposed to model the caching problem as a Nash bargaining game, where in \cite{ChuDTZ16}, a utility-based approach to cache partitioning is proposed for cache resource allocation.
Other related caching algorithms, but not TTL specific~\cite{LiC15,DuJGZRQ19}, attempt to minimize the average cost of misses, where the cost of an object is given by, for example, the variability in latency or computation cost.
Other approaches that consider cache optimization such as ~\cite{BeckmannGHM20,ChengDSWDL16,WangBBS18,Koch:category_aware_caching,Koch:Mira}, attempt to optimize given cache utility offline. This is different from the approach of the paper at hand as we model caching under non-zero random object fetching delays.

\textit{How our contributions differ from \cite{Mostafa,Daniel,PIT_CS,PIT_Cs_1,RAM}:}  While \cite{Mostafa} analyzes cache hierarchies under non-zero object fetch delays using approximations, our work differs significantly as we \textit{exactly} analyze the caching hierarchies under non-zero delay. In Sect.\ref{sec:approx_hierarchies} and in the evaluations in Sect.~\ref{sec:evaluations} we discuss and show the significant approximation error for cache hierarchies. Note that the work in \cite{Mostafa} is exact in case of a single cache.
Further, our work is different from  \cite{PIT_CS,PIT_Cs_1,RAM} due to the same technical reasons, as well as, the fact that we consider a fundamentally different caching algorithm (TTL instead of LRU).
Our contributions also differ from the work in \cite{Daniel} which assumes zero object fetch delay and focuses on the derivation of cache performance metrics. Calculating the performance metrics as in the work \cite{Daniel} suffers from high computational complexity due to state explosion of the Markovian model.
In contrast we generalize the exact model here to non-zero object fetch delays and analyze the cache performance metrics under the effects of network delays.
To tackle the problem of the exploding computational complexity of the calculation of exact cache performance metrics we derive a model-based \textit{exact} speed-up method that is based on lumpable partitions and automorphisms in the corresponding MAP.
Finally, we include numerical and trace-based evaluation results to show the accuracy of our model as well as the impact of the network delays on the cache performance.

\vspace{-10pt}
\section{System Model and Problem Statement}
\label{Section:MAP_approach}
When cache hierarchies are formed on-top of (or within) communication networks, cache misses require some random latency until the sought objects are retrieved from other caches or from the origin server.
This latency results from processing and forwarding miss requests and  objects, along predefined routing paths between leaf and parent caches.
%and thus also the packet transmission of the objects along a predefined routing path.

Fig.~\ref{TTL_D} shows one cycle of the miss process where, without loss of generality, the first request comes at time $t=0$ and the object is initially not in cache.
The figure shows that the object is admitted after some random object fetch delay $\Delta$ and is evicted when the TTL expires. We also depict regenerative TTL that is refreshed upon each hit (black arrow). We count requests that arrive during the object fetch delay (dashed arrows) as misses as the object is not fetched yet.
The excess lifetime of the renewal request process at time $\Delta$ is $\Gamma_\Delta$.

% Note that the impact of the delay on the cache performance increases when the delays become significant in relation to the request rate and the expected TTL.
% It is this shift towards relatively large request latencies in comparison to inter-request times that essentially motivates the TTL cache model under non-zero object fetch delays.
% Note that the impact of the delay on the cache performance increases with more significant delays (in relation to the request rate and the expected TTL).

Essentially, the shift towards relatively large request latencies in comparison to inter-request times that has been reported recently in~\cite{AtreSWB20} implies that the impact of the delay on the cache performance is significant and increasing.
This shift is due to extremely fast data center communication compared to memory access, processing and forwarding of miss requests~\cite{LiRJ18}.
Recent empirical approaches~\cite{AtreSWB20} attempt to incorporate this behavior in a problem description based on a minimum-cost multi-commodity flow formulation.
This motivates our approach to an \textit{exact model of TTL cache hierarchies under non-negligible object fetch delays}.

The work in~\cite{Mostafa} calculates the object hit probability at \textit{a single cache} under non-zero delay as
\begin{equation}
	P_h= \frac{\E[N]}{1+\E[m(\Delta)]+\E[N]} \;,
	\label{hit rate delay}
\end{equation}
% refer to m(\Delta) from the  paper
where $N$ is a counting random variable representing the number of hits between two misses and $m(\Delta)$ is the expected number of requests within the delay $\Delta$.
%\todo{again better not use D }
Moreover, the work in~\cite{Mostafa} specifies the inter-miss distribution as
\vspace{-5pt}
\begin{align}
&F_Y(y) = \int_0^y \bigg[H_0^\delta(y) \mathbb{P}(N=0|\Delta=\delta) +
\nonumber \\
&\underset{n=1}{\overset{\infty}{\sum}}(H_1^\delta * L_2 * L_1^{(n-1)}\big)(y) \mathbb{P}(N=n|\Delta=\delta) \bigg] dF_\Delta (\delta) \;,
 \label{miss process delay}
\end{align}
%\vspace{-5pt}
with $F_\Delta(\delta)$ being the CDF of the delay, and $H_0^\Delta(.)$ is the distribution of the excess lifetime $\Gamma_\Delta$ given no hits in one cycle.
%(see. Fig~\ref{TTL_D} for $\Gamma_\Delta$).
%\todo{What is $\Gamma_\Delta$? Need to define it in the same sentence}
Similarly, $H_1^\Delta(.)$ is the distribution of the excess lifetime given at least one hit in the cycle.
Further, $L_1$ and $L_2$ denote the conditional distribution of the inter-request time $X$ given that it is less or it is larger than the TTL $T$, respectively.
% and $L_2$ is the conditional distribution of $X$ given that it is larger than the TTL $T$.
To obtain an exact analysis for cache hierarchies, a closed-form expression is needed for the superposition of request processes at a cache under random object fetch  delays.
These request processes may themselves be the miss processes of children caches.
While \cite{Mostafa} provides an expression for the miss process under random fetch delay at a \textit{single cache}, it is notoriously difficult to extend this expression towards the exact superposition of miss processes.
The analysis in \cite{Mostafa}
%provides an exact result for a single cache under delays, it
hence resorts to approximating the aggregation of miss processes as a renewal process to analyze cache hierarchies.
Similarly, in \cite{Mostafa_2} the cache hierarchy analysis under non-zero fetch delay is based on the assumption of Poisson miss processes.

\vspace{-5pt}
\section{Exact Cache model with Object Fetch Delays}
\label{sec:exact_models}
In this section, we use a technique based on Markov arrival processes to model TTL caching hierarchies when random object fetch delays are present. We first show how the object fetch delay is incorporated into the MAP representation of a single cache before expanding to tree-like caching hierarchies.
To introduce our approach, we first consider exponentially distributed inter-request times, TTLs and delays. In resemblance to Kendall's notation we call such a system M/M/M caching system (in the order \textit{inter-request times at the system input, TTLs, delays}). Consequently, we generalize the approach to caching systems under Phase-type (PH) distributions that describe the request process as well as the delays. We choose PH distributions as these are dense in the class of distributions describing requests and delays. At the end of this section we discuss the computational complexity of our approach.

\begin{figure}[t!]
\centering
\resizebox{0.6\linewidth}{!}{\input{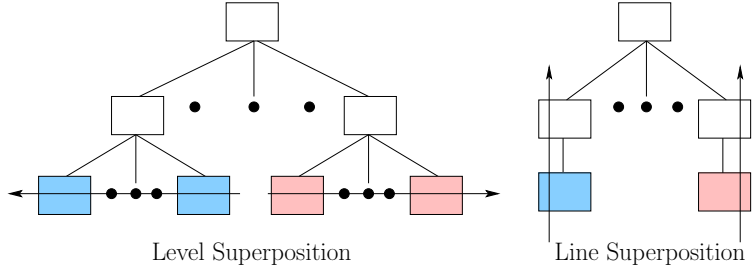}}
\caption{Construction of the hierarchy MAP through iteration of level and line superposition.}
\label{MAP formation}
\end{figure}

\vspace{-10pt}
\subsection{Single M/M/M Cache}

\begin{figure*}
     \centering
     \begin{subfigure}[b]{0.32\textwidth}
         \centering
	\resizebox{0.99\linewidth}{!}{\input{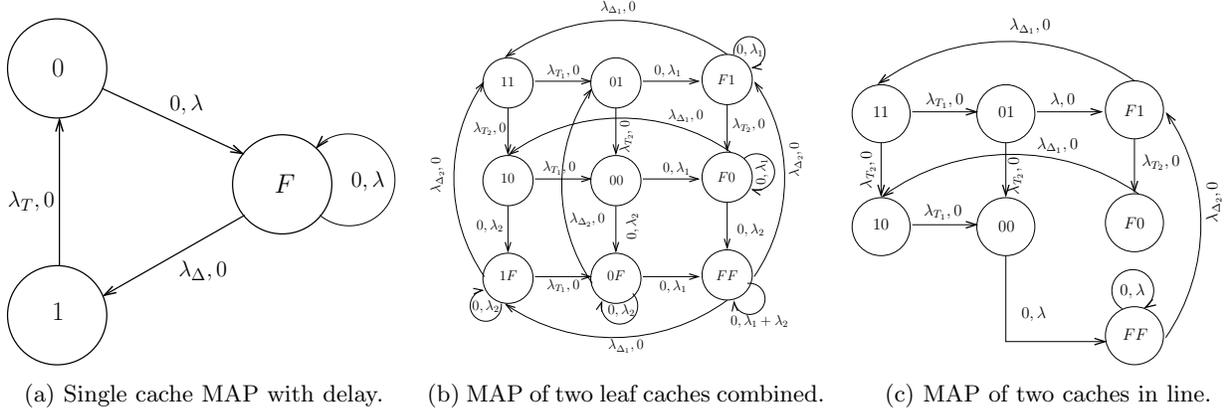}}         		
	\caption{Single cache MAP with delay.}
	\label{Single MAP}
     \end{subfigure}
     \hfill
     \begin{subfigure}[b]{0.32\textwidth}
         \centering
	\resizebox{0.99\linewidth}{!}{\ifx\XFigwidth\undefined\dimen1=0pt\else\dimen1\XFigwidth\fi
\divide\dimen1 by 5251
\ifx\XFigheight\undefined\dimen3=0pt\else\dimen3\XFigheight\fi
\divide\dimen3 by 4890
\ifdim\dimen1=0pt\ifdim\dimen3=0pt\dimen1=4143sp\dimen3\dimen1
  \else\dimen1\dimen3\fi\else\ifdim\dimen3=0pt\dimen3\dimen1\fi\fi
\tikzpicture[x=+\dimen1, y=+\dimen3]
{\ifx\XFigu\undefined\catcode`\@11
\def\temp{\alloc@1\dimen\dimendef\insc@unt}\temp\XFigu\catcode`\@12\fi}
\XFigu4143sp
% Uncomment to scale line thicknesses with the same
% factor as width of the drawing.
%\pgfextractx\XFigu{\pgfqpointxy{1}{1}}
\ifdim\XFigu<0pt\XFigu-\XFigu\fi
\pgfdeclarearrow{
  name = xfiga0,
  parameters = {
    \the\pgfarrowlinewidth \the\pgfarrowlength \the\pgfarrowwidth},
  defaults = {
	  line width=+7.5\XFigu, length=+120\XFigu, width=+60\XFigu},
  setup code = {
    % miter protrusion = thk * sqrt(wd^2 + (tipmv*len)^2) / (2 * wd)
    \dimen7 2.15\pgfarrowlength\pgfmathveclen{\the\dimen7}{\the\pgfarrowwidth}
    \dimen7 2\pgfarrowwidth\pgfmathdivide{\pgfmathresult}{\the\dimen7}
    \dimen7 \pgfmathresult\pgfarrowlinewidth
    \pgfarrowssettipend{+\dimen7}
    \pgfarrowssetbackend{+-\pgfarrowlength}
    \dimen9 -0.5\pgfarrowlinewidth
    \pgfarrowssetvisualbackend{+\dimen9}
    \pgfarrowssetlineend{+-0.5\pgfarrowlinewidth}
    \pgfarrowshullpoint{+\dimen7}{+0pt}
    \pgfarrowsupperhullpoint{+-\pgfarrowlength}{+0.5\pgfarrowwidth}
    \pgfarrowssavethe\pgfarrowlinewidth
    \pgfarrowssavethe\pgfarrowlength
    \pgfarrowssavethe\pgfarrowwidth
  },
  drawing code = {\pgfsetdash{}{+0pt}
    \ifdim\pgfarrowlinewidth=\pgflinewidth\else\pgfsetlinewidth{+\pgfarrowlinewidth}\fi
    \pgfpathmoveto{\pgfqpoint{-\pgfarrowlength}{0.5\pgfarrowwidth}}
    \pgfpathlineto{\pgfqpoint{0pt}{0pt}}
    \pgfpathlineto{\pgfqpoint{-\pgfarrowlength}{-0.5\pgfarrowwidth}}
    \pgfusepathqstroke
  }
}
\clip(614,-5640) rectangle (5865,-750);
\tikzset{inner sep=+0pt, outer sep=+0pt}
\pgfsetlinewidth{+7.5\XFigu}
\draw  (3105,-3172) circle [radius=+335];
\draw  (3105,-4477) circle [radius=+335];
\draw  (3105,-1867) circle [radius=+335];
\draw  (1665,-3172) circle [radius=+335];
\draw  (1665,-4477) circle [radius=+335];
\draw  (1665,-1867) circle [radius=+335];
\pgfsetarrows{[line width=15\XFigu, width=75\XFigu]}
\pgfsetarrowsstart{xfiga0}
\draw (1665,-2835) arc[start angle=+126.23, end angle=+55.51, radius=+2566.3];
\pgfsetarrows{-xfiga0}
\draw (4590,-4770) arc[start angle=+-53.77, end angle=+-124.49, radius=+2566.3];
\draw (1350,-4455) arc[start angle=+214.48, end angle=+145.52, radius=+2305];
\draw (2790,-4545) arc[start angle=+214.48, end angle=+145.52, radius=+2305];
\pgfsetarrows{xfiga0-}
\draw (4950,-1845) arc[start angle=+34.48, end angle=+-34.48, radius=+2305];
\draw (1620,-1530) arc[start angle=+126.23, end angle=+55.51, radius=+2566.3];
\draw  (4590,-3127) circle [radius=+335];
\draw  (4590,-4432) circle [radius=+335];
\draw  (4590,-1822) circle [radius=+335];
\pgfsetarrows{-xfiga0}
\draw (3465,-1845)--(4230,-1845);
\draw (3465,-3150)--(4230,-3150);
\draw (3465,-4455)--(4230,-4455);
\draw (2025,-4455)--(2790,-4455);
\draw (1665,-2205)--(1665,-2835);
\draw (1665,-3510)--(1665,-4140);
\draw (3105,-3510)--(3105,-4140);
\draw (4590,-3465)--(4590,-4095);
\draw (4590,-2160)--(4590,-2790);
\draw (2025,-3150)--(2790,-3150);
\draw (3105,-2205)--(3105,-2835);
\draw (2025,-1845)--(2790,-1845);

\pgfsetarrows{-xfiga0}
\draw (4860,-4545) arc[start angle=+97.6, end angle=+-174.9, radius=+208.2];
\draw (4815,-2880) arc[start angle=+133.5, end angle=+-108.4, radius=+241.9];
\draw (3330,-4680) arc[start angle=+403.5, end angle=+161.6, radius=+241.9];
\draw (1575,-4815) arc[start angle=+367.6, end angle=+95.1, radius=+208.2];
\draw (4680,-1485) arc[start angle=+187.6, end angle=+-84.9, radius=+208.2];
\pgftext[base,left,at=\pgfqpointxy{3005}{-5425}] {\fontsize{12}{14.4}$\lambda_{\Delta_1},0$}
\pgftext[base,left,at=\pgfqpointxy{3690}{-4635}] {\fontsize{12}{14.4}$0,\lambda_1$}
\pgftext[base,left,at=\pgfqpointxy{2185}{-4620}] {\fontsize{12}{14.4}$\lambda_{T_1},0$}
\pgftext[base,left,at=\pgfqpointxy{1200}{-2520}] {\fontsize{12}{14.4}$\lambda_{T_2},0$}
\pgftext[base,left,at=\pgfqpointxy{1245}{-3825}] {\fontsize{12}{14.4}$0,\lambda_2$}
\pgftext[base,left,at=\pgfqpointxy{3620}{-1755}] {\fontsize{12}{14.4}$0,\lambda_1$}
\pgftext[base,left,at=\pgfqpointxy{2185}{-1755}] {\fontsize{12}{14.4}$\lambda_{T_1},0$}
\pgftext[base,left,at=\pgfqpointxy{3805}{-2350}] {\fontsize{12}{14.4}$\lambda_{\Delta_1},0$}
\pgftext[base,left,at=\pgfqpointxy{2490}{-3735}] {\fontsize{12}{14.4}$\lambda_{\Delta_2},0$}
\pgftext[base,left,at=\pgfqpointxy{2130}{-3045}] {\fontsize{12}{14.4}$\lambda_{T_1},0$}
\pgftext[base,left,at=\pgfqpointxy{3735}{-3060}] {\fontsize{12}{14.4}$0,\lambda_1$}
\pgftext[base,left,at=\pgfqpointxy{5100}{-3230}, rotate=+90] {\fontsize{12}{14.4}$0,\lambda_1$}
\pgftext[base,left,at=\pgfqpointxy{4715}{-1480}] {\fontsize{12}{14.4}$0,\lambda_1$}
\pgftext[base,left,at=\pgfqpointxy{4635}{-5100}] {\fontsize{12}{14.4}$0,\lambda_1+\lambda_2$}
\pgftext[base,left,at=\pgfqpointxy{3000}{-4950}] {\fontsize{12}{14.4}$0,\lambda_2$}
\pgftext[base,left,at=\pgfqpointxy{1190}{-4900}] {\fontsize{12}{14.4}$0,\lambda_2$}

\pgftext[base,left,at=\pgfqpointxy{1570}{-1910}] {\fontsize{12}{14.4}$11$}
\pgftext[base,left,at=\pgfqpointxy{3020}{-1910}] {\fontsize{12}{14.4}$01$}
\pgftext[base,left,at=\pgfqpointxy{4460}{-1890}] {\fontsize{12}{14.4}$F1$}
\pgftext[base,left,at=\pgfqpointxy{4460}{-3190}] {\fontsize{12}{14.4}$F0$}
\pgftext[base,left,at=\pgfqpointxy{3020}{-3230}] {\fontsize{12}{14.4}$00$}
\pgftext[base,left,at=\pgfqpointxy{1570}{-3205}] {\fontsize{12}{14.4}$10$}
\pgftext[base,left,at=\pgfqpointxy{1550}{-4500}] {\fontsize{12}{14.4}$1F$}
\pgftext[base,left,at=\pgfqpointxy{2990}{-4500}] {\fontsize{12}{14.4}$0F$}
\pgftext[base,left,at=\pgfqpointxy{4440}{-4500}] {\fontsize{12}{14.4}$FF$}
\pgftext[base,left,at=\pgfqpointxy{2900}{-900}] {\fontsize{12}{14.4}$\lambda_{\Delta_1},0$}
\pgftext[base,left,at=\pgfqpointxy{765}{-3330},rotate=+90] {\fontsize{12}{14.4}$\lambda_{\Delta_2},0$}
\pgftext[base,left,at=\pgfqpointxy{5550}{-3120},rotate=+90] {\fontsize{12}{14.4}$\lambda_{\Delta_2},0$}
\pgftext[base,left,at=\pgfqpointxy{3375}{-3960},rotate=+90] {\fontsize{12}{14.4}$0,\lambda_2$}
\pgftext[base,left,at=\pgfqpointxy{3285}{-2805},rotate=+90] {\fontsize{12}{14.4}$\lambda_{T_2},0$}
\pgftext[base,left,at=\pgfqpointxy{4660}{-2475}] {\fontsize{12}{14.4}$\lambda_{T_2},0$}

\pgftext[base,left,at=\pgfqpointxy{4725}{-3875}] {\fontsize{12}{14.4}$0,\lambda_2$}
\endtikzpicture%
}
    \caption{MAP of two leaf caches combined.}
	\label{Multiple Caches MAP}
     \end{subfigure}
     \hfill
     \begin{subfigure}[b]{0.32\textwidth}
         \centering
	\resizebox{0.99\linewidth}{!}{\ifx\XFigwidth\undefined\dimen1=0pt\else\dimen1\XFigwidth\fi
\divide\dimen1 by 4543
\ifx\XFigheight\undefined\dimen3=0pt\else\dimen3\XFigheight\fi
\divide\dimen3 by 4025
\ifdim\dimen1=0pt\ifdim\dimen3=0pt\dimen1=4143sp\dimen3\dimen1
  \else\dimen1\dimen3\fi\else\ifdim\dimen3=0pt\dimen3\dimen1\fi\fi
\tikzpicture[x=+\dimen1, y=+\dimen3]
{\ifx\XFigu\undefined\catcode`\@11
\def\temp{\alloc@1\dimen\dimendef\insc@unt}\temp\XFigu\catcode`\@12\fi}
\XFigu4143sp
% Uncomment to scale line thicknesses with the same
% factor as width of the drawing.
%\pgfextractx\XFigu{\pgfqpointxy{1}{1}}
\ifdim\XFigu<0pt\XFigu-\XFigu\fi
\pgfdeclarearrow{
  name = xfiga0,
  parameters = {
    \the\pgfarrowlinewidth \the\pgfarrowlength \the\pgfarrowwidth},
  defaults = {
	  line width=+7.5\XFigu, length=+120\XFigu, width=+60\XFigu},
  setup code = {
    % miter protrusion = thk * sqrt(wd^2 + (tipmv*len)^2) / (2 * wd)
    \dimen7 2.15\pgfarrowlength\pgfmathveclen{\the\dimen7}{\the\pgfarrowwidth}
    \dimen7 2\pgfarrowwidth\pgfmathdivide{\pgfmathresult}{\the\dimen7}
    \dimen7 \pgfmathresult\pgfarrowlinewidth
    \pgfarrowssettipend{+\dimen7}
    \pgfarrowssetbackend{+-\pgfarrowlength}
    \dimen9 -0.5\pgfarrowlinewidth
    \pgfarrowssetvisualbackend{+\dimen9}
    \pgfarrowssetlineend{+-0.5\pgfarrowlinewidth}
    \pgfarrowshullpoint{+\dimen7}{+0pt}
    \pgfarrowsupperhullpoint{+-\pgfarrowlength}{+0.5\pgfarrowwidth}
    \pgfarrowssavethe\pgfarrowlinewidth
    \pgfarrowssavethe\pgfarrowlength
    \pgfarrowssavethe\pgfarrowwidth
  },
  drawing code = {\pgfsetdash{}{+0pt}
    \ifdim\pgfarrowlinewidth=\pgflinewidth\else\pgfsetlinewidth{+\pgfarrowlinewidth}\fi
    \pgfpathmoveto{\pgfqpoint{-\pgfarrowlength}{0.5\pgfarrowwidth}}
    \pgfpathlineto{\pgfqpoint{0pt}{0pt}}
    \pgfpathlineto{\pgfqpoint{-\pgfarrowlength}{-0.5\pgfarrowwidth}}
    \pgfusepathqstroke
  }
}
\clip(1322,-4775) rectangle (5865,-750);
\tikzset{inner sep=+0pt, outer sep=+0pt}
\pgfsetarrows{[line width=15\XFigu, width=75\XFigu]}
\pgfsetarrowsstart{xfiga0}
\pgfsetlinewidth{+7.5\XFigu}
\draw (1665,-2835) arc[start angle=+126.23, end angle=+55.51, radius=+2566.3];
\draw (4950,-1845) arc[start angle=+34.48, end angle=+-34.48, radius=+2305];
\draw (1620,-1530) arc[start angle=+126.23, end angle=+55.51, radius=+2566.3];
\draw  (4590,-3127) circle [radius=+335];
\draw  (4590,-4432) circle [radius=+335];
\draw  (4590,-1822) circle [radius=+335];
\draw  (1665,-3172) circle [radius=+335];
\draw  (1665,-1867) circle [radius=+335];
\draw  (3105,-3172) circle [radius=+335];
\draw  (3105,-1867) circle [radius=+335];
\pgfsetarrows{-xfiga0}
\draw (3465,-1845)--(4230,-1845);
\draw (1665,-2205)--(1665,-2835);
\draw (4590,-2160)--(4590,-2790);
\draw (2025,-3150)--(2790,-3150);
\draw (3105,-2205)--(3105,-2835);
\draw (2025,-1845)--(2790,-1845);
\draw (3105,-3510)--(3105,-4500)--(4275,-4500);
\draw (4450,-4120) arc[start angle=+600, end angle=+300.6, radius=+241.9];

\pgftext[base,left,at=\pgfqpointxy{1555}{-2730},rotate=+90] {\fontsize{12}{14.4}$\lambda_{T_2},0$}
\pgftext[base,left,at=\pgfqpointxy{3620}{-1755}] {\fontsize{12}{14.4}$\lambda,0$}
\pgftext[base,left,at=\pgfqpointxy{2180}{-1755}] {\fontsize{12}{14.4}$\lambda_{T_1},0$}
\pgftext[base,left,at=\pgfqpointxy{3465}{-2270}] {\fontsize{12}{14.4}$\lambda_{\Delta_1},0$}
\pgftext[base,left,at=\pgfqpointxy{2160}{-3040}] {\fontsize{12}{14.4}$\lambda_{T_1},0$}
\pgftext[base,left,at=\pgfqpointxy{1570}{-1910}] {\fontsize{12}{14.4}$11$}
\pgftext[base,left,at=\pgfqpointxy{3020}{-1910}] {\fontsize{12}{14.4}$01$}
\pgftext[base,left,at=\pgfqpointxy{4470}{-1890}] {\fontsize{12}{14.4}$F1$}
\pgftext[base,left,at=\pgfqpointxy{4470}{-3190}] {\fontsize{12}{14.4}$F0$}
\pgftext[base,left,at=\pgfqpointxy{3020}{-3230}] {\fontsize{12}{14.4}$00$}
\pgftext[base,left,at=\pgfqpointxy{1570}{-3205}] {\fontsize{12}{14.4}$10$}
\pgftext[base,left,at=\pgfqpointxy{4455}{-4485}] {\fontsize{12}{14.4}$FF$}
\pgftext[base,left,at=\pgfqpointxy{3150}{-900}] {\fontsize{12}{14.4}$\lambda_{\Delta_1},0$}
\pgftext[base,left,at=\pgfqpointxy{5550}{-3120},rotate=+90] {\fontsize{12}{14.4}$\lambda_{\Delta_2},0$}
\pgftext[base,left,at=\pgfqpointxy{3255}{-2800},rotate=+90] {\fontsize{12}{14.4}$\lambda_{T_2},0$}
\pgftext[base,left,at=\pgfqpointxy{4675}{-2475}] {\fontsize{12}{14.4}$\lambda_{T_2},0$}
\pgftext[base,left,at=\pgfqpointxy{3285}{-4230}] {\fontsize{12}{14.4}$0,\lambda$}
\pgftext[base,left,at=\pgfqpointxy{4425}{-3960}] {\fontsize{12}{14.4}$0,\lambda$}

\endtikzpicture%
}
    \caption{MAP of two caches in line.}
	\label{Line caching MAP}
     \end{subfigure}
        \caption{Basic MAPs of different caching setups: \textbf{\textit{State $1$}, \textit{State $F$} and \textit{State $0$} denote object in cache, object being fetched or object out of cache respectively}. Combination of States
        %\textbf{\textit{State $FF$}}
        depends on the semantics of the hierarchy, i.e. whether the MAP is for a combination of leaf caches or child/parent caches. }
        \label{fig:three graphs}
        %\vspace{-10pt}
\end{figure*}

As TTL caches decouple objects it is sufficient to analyze the caching process for one object~\cite{Fofack}.
First, we show how to incorporate the object fetch delay into the MAP representation of a single cache.
Fig.~\ref{Single MAP} shows the MAP for one object at a single M/M/M cache.
Here, the MAP has three states capturing the life cycle of the object in the system.

When an object request arrives to the cache, it can find the object in cache (\textit{State 1}) or being fetched from the origin (\textit{State $F$}) or the object is neither in cache nor being fetched which we denote ``out of cache'' (\textit{State 0}).
% We denoted the corresponding states as follows:  refers to "object is out of cache", \textit{state 1} refers to "object in the cache" and \textit{state $F$} denotes that the object is being fetched.
We define a \textbf{cache miss} as a request arriving when the object \textbf{is not in cache}.
Hence, the \textbf{active transitions} in the MAP \textbf{denote misses} that happen when a request arrives in states $0$ or $F$.
In contrast to the classical cache analysis the object is not instantaneously admitted upon a miss due to the fetching delay between the cache and the origin server.
In this first step we consider exponentially distributed fetch times with parameter $\lambda_\Delta$.
We generalize the delay model in the following sections.
%in state $F$ before being admitted.
Hence, the hidden and active matrices of the MAP are

{\small
\begin{equation}
\label{eq:hidde_active_matrices_single_cache_MMM}
\mathbf{D}_0=
\begin{bmatrix}
-\lambda &      0        & 0    \\
   \lambda_T   & -\lambda_T  & 0    \\
        0        & \lambda_\Delta & -\lambda_\Delta-\lambda
\end{bmatrix}
\;, \;
\mathbf{D}_1=
\begin{bmatrix}
0 & 0 & \lambda \\
0 & 0 &         0       \\
0 & 0 &         \lambda
\end{bmatrix}
\;,
\end{equation}
}
where $\lambda$ is the request arrival rate, $1/\lambda_T$ is the mean TTL and $1/\lambda_\Delta$ is the mean fetch delay.

%\begin{figure}[h!]
%\centering
%\resizebox{0.9\linewidth}{!}{\input{Figs/Single_cache_MAP.tex}}
%\caption{MAP for a single cache with object fetch delay}
%\label{Single MAP}
%\end{figure}

\subsection{Construction and Analysis of Cache Hierarchies}

We first consider one object that is contained in a tree caching hierarchy as depicted in Fig.~\ref{MAP formation}.
The figure shows an ``algorithmic'' overview of the steps of our approach for the incorporation of the delay into the MAP model of the cache hierarchy.
First, caches on the same level, i.e., having the same parent, are grouped together through an operation that we denote \textit{level superposition}.
In the next step, parent and child caches are grouped through an operation denoted \textit{line superposition}.
%In addition, we represent the level superposition of a number of caches $\mathbf{C}$  by a function we denote by $\Phi$ while we represent the line superposition by a function we denote by $\Psi$.
%In the following, we show how to form the MAP for multiple caches in a hierarchy as in Fig.~\ref{MAP formation}.
% figure to be done
Now we can build the overall MAP from the MAPs of the single caches starting from the leaf nodes until the root.
Level and line superposition correspond to non-trivial compounding of the MAPs of the single caches.
%\todo{Don't do this if not needed.. better use only the word cache.. The term cache and node are going to be used interchangeably}
%There are two main operations that govern our approach, i.e., level superposition and line superposition. As depicted in Fig.~\ref{MAP formation} we compound the MAPs of caches on the same hierarchy level  using the level superposition and in a following step combine the MAPs of parent-child caches using the line superposition operation.
We expand the definition of a \textbf{miss to caching hierarchies} to denote object requests where the object is not fetched from caches within the hierarchy but from the origin server.

\subsubsection{Level superposition $(\Phi)$}
The level superposition forms a MAP from two separate sub-trees connected directly to the same parent node.
%Both sub-trees are referred to as being in the same level.
% the words in bracket depends on the definition done in the previous sections that has not be done yet may be changed at the end
%(This is previously illustrated in the exact approach by Daniel)
The level superposition for caches under the zero-delay assumption was shown first in \cite{Daniel}.
Here, the MAP $(M)$ resulting from the superposition is formed by the Kronecker sum $\oplus$  of the two MAPs (${M}_1$, ${M}_2$) of the two sub-trees as
%\vspace{-10pt}
%producing all the possible state combinations and corresponding transitions as
\vspace{-10pt}
\begin{align*}
\vspace{-10pt}
M&= M_1 \oplus M_2
\nonumber \\
\big(\mathbf{D}_0,\mathbf{D}_1\big)&=\big(\mathbf{D}_0^{(1)},\mathbf{D}_1^{(1)}\big) \oplus
 \big(\mathbf{D}_0^{(2)},\mathbf{D}_1^{(2)}\big) \;,
\end{align*}
where
%$\oplus$ denotes the Kronecker sum,
$\mathbf{D}_0^{(i)}$ and $\mathbf{D}_1^{(i)}$ are the hidden and active transition matrices of $M_i$, respectively.
As the two sub-trees receive independent request streams, the Kronecker sum produces the exact states with the right transitions between them.
A basic example of the level superposition is of two separate caches described each by \eqref{eq:hidde_active_matrices_single_cache_MMM} that are connected to a parent cache. Applying the Kronecker sum we obtain the MAP in Fig.~\ref{Multiple Caches MAP}.
% \todo{smaller matrices}
% \begin{equation}
% \mathbf{D}_0=
% \begin{bmatrix}
% -\lambda_1 &      0        & 0    \\
%   \mu_1   & -\mu_1  & 0    \\
%         0        & \lambda_{\Delta_1} & -\lambda_{\Delta_1}-\lambda_1
% \end{bmatrix}
% %
% \oplus
% %
% \begin{bmatrix}
% -\lambda_2 &      0        & 0    \\
%   \mu_2  & -\mu_2  & 0    \\
%         0        & \lambda_{\Delta_2} & -\lambda_{\Delta_2}-\lambda_2
% \end{bmatrix}
% \end{equation}
% \begin{equation}
% \mathbf{D}_1=
% \begin{bmatrix}
% 0&0& \lambda_1    \\
%  0&0&0\\
%  0&0& \lambda_1\\
% \end{bmatrix}
% %
% \oplus
% %
% \begin{bmatrix}
% 0&0& \lambda_2    \\
%  0&0&0\\
%  0&0& \lambda_2\\
% \end{bmatrix}
% \;.
% \end{equation}
% %\begin{figure}[h!]
% %\centering
% %\resizebox{0.7\linewidth}{!}{\input{Figs/Level_sup_MAP.tex}}
% %\caption{MAP of two leaf caches connected to the server
% %\todo{can you center the numbers in the circles?}}
% %\label{Multiple Caches MAP}
% %\end{figure}
% Observe that this corresponds to the transitions in Fig.~\ref{Multiple Caches MAP}.
\subsubsection{Line superposition $(\Psi)$}
\label{subsec:line_superposition}
Line superposition involves constructing the MAP for a parent node and the sub-tree below it.
% Note: This not exactly correct as the sub-tree goes through level and line operations depending on its structure not just level. So I would rather keep it general without any specification
%Every level of this sub-tree goes first through the level superposition.
In contrast to level superposition, the caches considered in line superposition are dependent.
We carry out the line superposition using the following four steps:
% \begin{itemize}
% \item Perform Kronecker sum
% \item Remove invalid states
% \item Remove the hidden transitions from $\mathbf{D}_1$
% \item Update $\mathbf{D}_0$ and $\mathbf{D}_1$ single transition
% \end{itemize}
% \vspace{2mm}
\paragraph{Kronecker sum}
As the parent cache is not associated with a direct request stream the active transition matrix of the MAP representing the added parent is  given by $\mathbf{D}_1^{(2)} = \mathbf{0} $
% \begin{equation}
% 	\mathbf{D}_1^{(2)} = \mathbf{0} \;,
% \end{equation}
where the superscript denotes the cache id along the line.
Performing the Kronecker sum on the MAP of the parent cache and the existing MAP of the sub-tree of children results in all state combinations.
Due to the dependency between the parent cache and the children sub-tree connected to it, not all possible state combinations are valid and need to be removed.
We consider in the following two caches in line for simplicity, however, the MAP of the child cache can well be a superposed caching sub-tree.
%some transitions are not defined as in~\eqref{transitions}.
Note that the result of the Kronecker sum exactly represents the overall MAP only when dealing with two independent MAPs where the transitions are given by
\begin{equation}
M^{(ij \rightarrow i'j')}=
\begin{cases}
  M_1^{(i \rightarrow i')} & \; \; \text{when $j=j'$} \\
  M_2^{(j \rightarrow j')} & \; \; \text{when $i=i'$}  \end{cases}
%\label{transitions}
\;,
\end{equation}
where ${M}^{(ij \rightarrow i'j')}$ represents the transition in the resulting MAP  ${M}$ from state $ij$ to $i'j'$  where $ij$ represents the state superposition of state $i$ from ${M}_1$ and state $j$ from ${M}_2$.

 Fig.~\ref{Line caching MAP}
%to be done
describes the MAP of two caches in line. This example is used in the following to illustrate the changes required after the Kronecker sum to produce a MAP that exactly represents the combined behavior of the caches after line superposition.
% \vspace{2mm}
%\begin{figure}[H]
%\centering
%\resizebox{0.6\linewidth}{!}{\input{Figs/Line_sup_MAP.tex}}
%\caption{MAP of two caches connected in line}
%\label{Line caching MAP}
%\end{figure}
\paragraph{Removal of invalid states and transitions}
Observe that when comparing the MAP resulting from the Kronecker sum in Fig.~\ref{Multiple Caches MAP} to the final  MAP in Fig.~\ref{Line caching MAP} there are two removed invalid states when dealing with two caches in line which are \textit{0$F$} and \textit{1$F$}.
State \textit{0$F$} denotes that the object is out of the child cache and is being fetched by its parent which contradicts our assumption that
%is can never happen since the request can never be
objects are fetched by the parent based on a request by the child cache.
%unless it is being fetched from it by its child.
Further, the state \textit{1$F$} denotes that the object is in cache at the child while being fetched by the parent which obviously violates causality.
%, thus no fetching should be initiated for the parent cache if a request arrives.
Based on this observation, the hidden and active matrices are updated by removing the rows and columns of the invalid states.

Applying this idea to tree hierarchies, we build the MAP for the hierarchy
%when building the MAP step by step starting from the leaves, the
with line superposition applied when $M_2$ represents the parent MAP and $M_1$ represents the superposition of all of its children.
Note that in this case the invalid states set $I$ is defined by %$I=\underset{k=1}{\overset{n_{ch}}{{\bigcap}}} I_k$
\vspace{-10pt}
\begin{equation}
\label{eq:set_of_invalid_states}
  I=\underset{k=1}{\overset{n_{ch}}{{\bigcap}}} I_k \;,
\end{equation}
where $I_k$ is the invalid states set of an added parent cache and its child $k$ as given by the two node example and $n_{ch}$ is the number of children caches.

Further, the hidden transition in the Kronecker sum given by $M^{(FF \rightarrow 1F)}$ is invalid and removed as the object is  admitted in the parent cache before the child.
Note that in the two cache case, this transition is obsolete  as the state \textit{1$F$} is removed as stated above, however, in general for a parent cache and a child sub-tree this type of transition is generally removed.
% %since the state \textit{1$F$} is invalid,
% however when generally dealing with any parent node and its sub-tree, this type of transition should be changed.

%%%%%%%%%%%%%%%%%%%%%%%%%%%%%%%%%%%%%%%%%%%%%%%%%%%%%%%%%%%%%%%%
%% Till here I am confident not perfect but consistent
% \vspace{2mm}
\paragraph{Accounting for system hits by moving transitions from $\mathbf{D}_1$ to $\mathbf{D}_0$}
Since the active transitions are related to requests which induce misses, when performing line superposition some of the requests no longer generate a \textbf{system miss} as the object is contained in the parent cache.
%is no longer needed to be fetched from the origin server.
Hence, the Kronecker sum of the active matrices is adapted to account for that fact.
The following active transitions are moved from $\mathbf{D}_1$ to $\mathbf{D}_0$:
%changed to a hidden one in either a state to a different state transition or a self transition
\begin{equation*}
\label{eq:moving_active_to_hidden_transitions}
	M^{(ij \rightarrow i'j)} \hspace{0.5cm}\text{for} \ i,i' \in \mathcal{S}_1  \text{  and  } j \neq F \; ,
\end{equation*}
where $\mathcal{S} _1$ represents the set of the states of the MAP $M_1$ of the connected sub-tree.
To illustrate this, we first consider the case of two caches in line (Fig.~\ref{Line caching MAP}).
Now it is obvious that the transition $M^{(01 \rightarrow F1)}$ is hidden as it does not denote a \textbf{system miss}.
Moreover, this argument also holds for the self transitions $M^{(F1 \rightarrow F1)}$ and $M^{(F0 \rightarrow F0)}$.
Note that we do not draw the hidden self transitions in Fig.~\ref{Line caching MAP}.
%for clarity of exposition.
%\vspace{2mm}

%\subsubsection{Updating $\mathbf{D}_0$ and $\mathbf{D}_1$ single transition}
\paragraph{Updating $\mathbf{D}_0$ and $\mathbf{D}_1$ for system misses}
Finally, a single transition of the Kronecker sum is adapted to express the \textit{system miss}.
%in both the hidden and the active transition matrices has to be adapted.
Considering first the line superposition of a child and a parent cache we observe that when a request generates a system miss, i.e., when the two caches are in state \textit{0}, the child cache fetches the object from its parent and the parent fetches the object from the origin server.
Thus the transition obtained by the Kronecker sum as $M^{(00 \rightarrow F0)}$ in $\mathbf{D}_1$ is replaced by the corresponding transition $M^{(00 \rightarrow FF)}$.

The four preceding steps to calculate the overall system MAP for a \textbf{ general tree cache hierarchy} with one parent/root an $l-1$ children are given in Algorithm~\ref{MAP Algorithm}.
Here, the caches are labeled from $1$ to $l$ such that $1$ and $l$ represent the first leaf cache and the root, respectively.
%Note that Algorithm~\ref{MAP Algorithm} calculates the MAP for tree hierarchies irrespective of the depth.
The function children$(i)$ returns the indices of the children of cache $i$.
$M_i$ represents the MAP of a single cache $i$ while $M_{t,i}$ represents the MAP of the sub-tree with root $i$.
The functions $\Phi$ and $\Psi$ represent the level and line superposition, respectively.
$M_{\mathrm{lvl}}$ represents the MAP obtained from the level superposition $\Phi$.
Note that $\Phi(0)=0$ and $\Psi(M_i, 0)=M_i$.
\vspace{-5pt}
\begin{algorithm}[h!]
\SetAlgoLined
\textbf{Input:} $M_i \; \; \; \mathrm{i} \in \{1,...., l\} $. \\
\KwResult{$M_{t,l}$}
 \textbf{Initialization}: $i=1$ , $M_{t,i}=0$ $\forall i$  s.t $\text{children}(i)= \emptyset$ \;
 \While{$i\leq l$}{
 	$M_{\mathrm{lvl}}= \Phi(M_{t,{\text{children}(i)}})$ \;
	$M_{t,i}= \Psi(M_i,M_{\mathrm{lvl}})$\;
  }
 \caption{Calculation of the combined MAP for a tree caching topology}
   \label{MAP Algorithm}
   %\vspace{-5pt}
\end{algorithm}
\vspace{-20pt}
\subsection{M/M/PH Caching System}
Next, we generalize the approach illustrated above to M/M/PH caching systems where the delay is PH distributed.
Recall that the PH distribution is dense in the class of positive-valued distributions making it a good approximate for trace-based evaluations.
% Here we show how the MAP is formed in case of having a PH delay distribution.
% In case of exponential distribution, the delay in the MAP of a single cache is represented according to Fig.~\ref{Single MAP} by a single state with exponential holding time.

\begin{figure}[H]
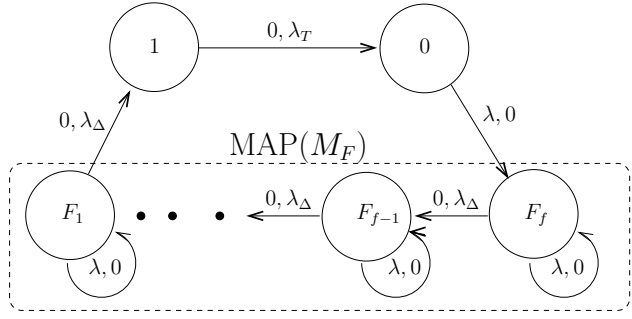

\centering
\resizebox{0.6\linewidth}{!}{\ifx\XFigwidth\undefined\dimen1=0pt\else\dimen1\XFigwidth\fi
\divide\dimen1 by 6381
\ifx\XFigheight\undefined\dimen3=0pt\else\dimen3\XFigheight\fi
\divide\dimen3 by 2655
\ifdim\dimen1=0pt\ifdim\dimen3=0pt\dimen1=4143sp\dimen3\dimen1
  \else\dimen1\dimen3\fi\else\ifdim\dimen3=0pt\dimen3\dimen1\fi\fi
\tikzpicture[x=+\dimen1, y=+\dimen3]
{\ifx\XFigu\undefined\catcode`\@11
\def\temp{\alloc@1\dimen\dimendef\insc@unt}\temp\XFigu\catcode`\@12\fi}
\XFigu4143sp
% Uncomment to scale line thicknesses with the same
% factor as width of the drawing.
%\pgfextractx\XFigu{\pgfqpointxy{1}{1}}
\ifdim\XFigu<0pt\XFigu-\XFigu\fi
\pgfdeclarearrow{
  name = xfiga0,
  parameters = {
    \the\pgfarrowlinewidth \the\pgfarrowlength \the\pgfarrowwidth},
  defaults = {
	  line width=+7.5\XFigu, length=+120\XFigu, width=+60\XFigu},
  setup code = {
    % miter protrusion = thk * sqrt(wd^2 + (tipmv*len)^2) / (2 * wd)
    \dimen7 2.15\pgfarrowlength\pgfmathveclen{\the\dimen7}{\the\pgfarrowwidth}
    \dimen7 2\pgfarrowwidth\pgfmathdivide{\pgfmathresult}{\the\dimen7}
    \dimen7 \pgfmathresult\pgfarrowlinewidth
    \pgfarrowssettipend{+\dimen7}
    \pgfarrowssetbackend{+-\pgfarrowlength}
    \dimen9 -0.5\pgfarrowlinewidth
    \pgfarrowssetvisualbackend{+\dimen9}
    \pgfarrowssetlineend{+-0.5\pgfarrowlinewidth}
    \pgfarrowshullpoint{+\dimen7}{+0pt}
    \pgfarrowsupperhullpoint{+-\pgfarrowlength}{+0.5\pgfarrowwidth}
    \pgfarrowssavethe\pgfarrowlinewidth
    \pgfarrowssavethe\pgfarrowlength
    \pgfarrowssavethe\pgfarrowwidth
  },
  drawing code = {\pgfsetdash{}{+0pt}
    \ifdim\pgfarrowlinewidth=\pgflinewidth\else\pgfsetlinewidth{+\pgfarrowlinewidth}\fi
    \pgfpathmoveto{\pgfqpoint{-\pgfarrowlength}{0.5\pgfarrowwidth}}
    \pgfpathlineto{\pgfqpoint{0pt}{0pt}}
    \pgfpathlineto{\pgfqpoint{-\pgfarrowlength}{-0.5\pgfarrowwidth}}
    \pgfusepathqstroke
  }
}
\clip(2775,-3342) rectangle (9156,-687);
\tikzset{inner sep=+0pt, outer sep=+0pt}
\pgfsetlinewidth{+7.5\XFigu}
\pgfsetarrowsstart{xfiga0}

\draw (8235,-2655) arc[start angle=+67.4, end angle=+-180.0, radius=+292.5];
\draw (6795,-2655) arc[start angle=+67.4, end angle=+-180.0, radius=+292.5];
\draw (4275,-2655) arc[start angle=+67.4, end angle=+-180.0, radius=+292.5];
\draw  (4613,-1074) circle [radius=+380];
\draw  (6431,-2527) circle [radius=+380];
\draw  (3893,-2514) circle [radius=+380];
\filldraw  (5175,-2520) circle [radius=+33];
\filldraw  (4770,-2520) circle [radius=+33];
\filldraw  (4500,-2520) circle [radius=+33];
\draw  (6927,-1085) circle [radius=+380];
\draw  (7863,-2503) circle [radius=+380];
\pgfsetarrows{[line width=15\XFigu, width=75\XFigu]}
\pgfsetarrowsstart{xfiga0}
\draw (6525,-1080)--(4995,-1080);
\draw (6791,-2655)--(6799,-2655);
\pgfsetarrows{-xfiga0}
\draw (7470,-2520)--(6840,-2520);
\draw (6030,-2520)--(5400,-2520);
\draw (7155,-1395)--(7650,-2205);
\draw (4050,-2160)--(4410,-1440);
\pgfsetdash{{+45\XFigu}{+45\XFigu}}{++0pt}
\draw (3375,-3330) [rounded corners=+105\XFigu] rectangle (8685,-2070);

\pgftext[base,left,at=\pgfqpointxy{4050}{-3015},rotate=+360] {\fontsize{12}{14.4}$\lambda,0$}
\pgftext[base,left,at=\pgfqpointxy{6615}{-3015},rotate=+360] {\fontsize{12}{14.4}$\lambda,0$}
\pgftext[base,left,at=\pgfqpointxy{8010}{-3015},rotate=+360] {\fontsize{12}{14.4}$\lambda,0$}
\pgftext[base,left,at=\pgfqpointxy{3820}{-2550}] {\fontsize{12}{14.4}$F_1$}
\pgftext[base,left,at=\pgfqpointxy{6345}{-2550}] {\fontsize{12}{14.4}$F_{f-1}$}
\pgftext[base,left,at=\pgfqpointxy{7785}{-2550}] {\fontsize{12}{14.4}$F_f$}
\pgftext[base,left,at=\pgfqpointxy{3800}{-1755}] {\fontsize{12}{14.4}$ 0,\lambda_\Delta$}
\pgftext[base,left,at=\pgfqpointxy{7425}{-1710}] {\fontsize{12}{14.4}$\lambda,0$}
\pgftext[base,left,at=\pgfqpointxy{5580}{-990}] {\fontsize{12}{14.4}$0,\lambda_T$}
\pgftext[base,left,at=\pgfqpointxy{4570}{-1120}] {\fontsize{12}{14.4}$1$}
\pgftext[base,left,at=\pgfqpointxy{6885}{-1120}] {\fontsize{12}{14.4}$0$}
\pgftext[base,left,at=\pgfqpointxy{5255}{-2000},rotate=+360] {\fontsize{16}{14.4}$\mathrm{MAP}(M_F)$}
\pgftext[base,left,at=\pgfqpointxy{5565}{-2430}] {\fontsize{12}{14.4}$ 0,\lambda_\Delta$}
\pgftext[base,left,at=\pgfqpointxy{7000}{-2430}] {\fontsize{12}{14.4}$ 0,\lambda_\Delta$}
\endtikzpicture%
}
\caption{MAP of M/M/$\text{E}_f$ single cache}
\label{Single Erlang}
\end{figure}

In case of PH distributed delay, the fetching process is given by a multiple state MAP $M_{F}$ with states $F_\alpha$ where $\alpha \in \{1,2,...,f\}$ and $f$ represents the number of the phases.
%%will still define PH
%%will define the D0 D1
Without loss of generality, we use here an Erlang \textit{$\text{E}_f$} delay distributions with $f$ states to illustrate our approach.
%example of the PH distribution, however any PH distribution can be dealt the same way.
For a single M/M/PH cache we provide the MAP in Fig.~\ref{Single Erlang} highlighting the $\text{E}_f$ fetching process.
\vspace{10pt}
The corresponding transition matrices of the single cache MAP are given as
% remove color add zeros%%%%%%%%%%%%%%%%%%%%%%%%%%%%%%%
%do not number an equation not used
{\small
\begin{equation}
\mathbf{D}_0=
\begin{bmatrix}
-\lambda &   0    & 0 & 0 & 0 \\
   \lambda_T   & -\lambda_T  & 0 & 0&0  \\
       0  & \tikzmark{left}{$\lambda_\Delta$} & (-\lambda_\Delta-\lambda)  &0 &0  \\
     0&0 & \ddots & \ddots &0  \\
     0&  0  & 0& \lambda_\Delta &(-\lambda_\Delta-\lambda)\tikzmark{right}{}
     \\
\end{bmatrix}
\;, \
\mathbf{D}_1=
\begin{bmatrix}
0& 0&0  & 0& \lambda\\
0& 0&0 &  0&0 \\
0& 0&\tikzmark{left}{$\lambda$} & 0&0 \\
0&0 &0 & \ddots &  0 \\
0&0 &0 &0 &\lambda\tikzmark{right}{} \\
\end{bmatrix}
\;.
\nonumber
%\Highlight[second]
\end{equation}
}
%Note that bold face values represent vectors or matrices.
%The highlighted part of $\mathbf{D}_0$ and $\mathbf{D}_1$ represent the hidden and active matrices of the fetching MAP
%$\mathbf{D}_{0,F}$ and $\mathbf{D}_{1,F}$
%with the absorbing state represented by state $1$.

% As illustrated in the M/M/M caching system, in order to form a MAP for a hierarchy, level and line superposition operations are performed on the transition matrices of the single caches to group those in the same level or the same line respectively.

Similar to the derivation of the MAP for M/M/M caching hierarchies we apply the same operations of level and line superposition for the M/M/PH caching hierarchy.
To this end, we essentially define the line superposition in a general manner to capture PH delay distributions.

% However, some of the steps performed in order to carry out the Line superposition need to be generally defined to describe any delay distribution.

% As previously defined, the line superposition consists of four main steps. Next,
We show how to generalize the four main steps given in Sect.~\ref{subsec:line_superposition} for PH delay distributions.
The Kronecker sum in the first step is applied without changes.
The second step represented by removing the invalid states is generalized next. As the fetching process is represented by multiple states, the invalid states are defined for all state combinations.
Note that in case of exponential delay distribution, the invalid states are defined by the two states $0F$ and $1F$.
For PH delay distribution, the invalid states due to a child cache $k$ represented by multiple states $F_\alpha$ are given as
\begin{align*}
	I_k=&\{0F_\alpha, \ 1F_\alpha, \ F_\beta F_\alpha\}
	\nonumber
	\\	
	 &\alpha \in\{1,..,f\}, \ \beta \in\{1,...,f-1\} \;.
\end{align*}
The first two state combination sets $0F_\alpha$ and $1F_\alpha$ are analogous to the case of exponential delay distribution.
However, given PH distributed delays the invalid state combinations
$F_\beta F_\alpha$ when $\alpha \in\{1,..,f\}$ and $ \beta \in\{1,...,f-1\}$ arise.
This comes from causality, i.e., object fetching ends at the parent cache before starting at the child.
The third step in Sect.~\ref{subsec:line_superposition} representing the transitions to be changed from active to hidden is extended such that self transitions include all the fetching states.
%instead of one state.
Hence, the states changed from active to hidden are given by
%\begin{align}
%& \mathrm{M}^{(i1 \rightarrow i'1)} \; \; \text{for} \; \;i,i' \in S_1
%\nonumber
%\\
%& \mathrm{M}^{(ij \rightarrow i'j)} \; \; \text{when} \; \; j \neq F_i \ \forall i \;.
%\end{align}
\begin{equation*}
	M^{(ij \rightarrow i'j)} \hspace{0.5cm}\text{for} \ i,i' \in \mathcal{S}_1  \text{  and  } j \neq F_\alpha \; \forall \alpha \;.
\end{equation*}
The final step is analogous to the case of the exponential distribution, i.e. changing the transition $M^{(00 \rightarrow F_f0)}$ to $M^{(00 \rightarrow F_fF_f)}$.
%This is analogical to the same step in case of the exponential distribution.
\vspace{-10pt}
\subsection{PH/M/M Caching System}
\begin{figure}[t]
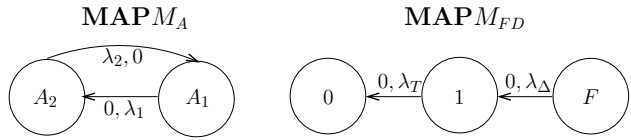

\centering
\resizebox{0.5\linewidth}{!}{\ifx\XFigwidth\undefined\dimen1=0pt\else\dimen1\XFigwidth\fi
\divide\dimen1 by 6584
\ifx\XFigheight\undefined\dimen3=0pt\else\dimen3\XFigheight\fi
\divide\dimen3 by 1439
\ifdim\dimen1=0pt\ifdim\dimen3=0pt\dimen1=4143sp\dimen3\dimen1
  \else\dimen1\dimen3\fi\else\ifdim\dimen3=0pt\dimen3\dimen1\fi\fi
\tikzpicture[x=+\dimen1, y=+\dimen3]
{\ifx\XFigu\undefined\catcode`\@11
\def\temp{\alloc@1\dimen\dimendef\insc@unt}\temp\XFigu\catcode`\@12\fi}
\XFigu4143sp
% Uncomment to scale line thicknesses with the same
% factor as width of the drawing.
%\pgfextractx\XFigu{\pgfqpointxy{1}{1}}
\ifdim\XFigu<0pt\XFigu-\XFigu\fi
\pgfdeclarearrow{
  name = xfiga0,
  parameters = {
    \the\pgfarrowlinewidth \the\pgfarrowlength \the\pgfarrowwidth},
  defaults = {
	  line width=+7.5\XFigu, length=+120\XFigu, width=+60\XFigu},
  setup code = {
    % miter protrusion = thk * sqrt(wd^2 + (tipmv*len)^2) / (2 * wd)
    \dimen7 2.15\pgfarrowlength\pgfmathveclen{\the\dimen7}{\the\pgfarrowwidth}
    \dimen7 2\pgfarrowwidth\pgfmathdivide{\pgfmathresult}{\the\dimen7}
    \dimen7 \pgfmathresult\pgfarrowlinewidth
    \pgfarrowssettipend{+\dimen7}
    \pgfarrowssetbackend{+-\pgfarrowlength}
    \dimen9 -0.5\pgfarrowlinewidth
    \pgfarrowssetvisualbackend{+\dimen9}
    \pgfarrowssetlineend{+-0.5\pgfarrowlinewidth}
    \pgfarrowshullpoint{+\dimen7}{+0pt}
    \pgfarrowsupperhullpoint{+-\pgfarrowlength}{+0.5\pgfarrowwidth}
    \pgfarrowssavethe\pgfarrowlinewidth
    \pgfarrowssavethe\pgfarrowlength
    \pgfarrowssavethe\pgfarrowwidth
  },
  drawing code = {\pgfsetdash{}{+0pt}
    \ifdim\pgfarrowlinewidth=\pgflinewidth\else\pgfsetlinewidth{+\pgfarrowlinewidth}\fi
    \pgfpathmoveto{\pgfqpoint{-\pgfarrowlength}{0.5\pgfarrowwidth}}
    \pgfpathlineto{\pgfqpoint{0pt}{0pt}}
    \pgfpathlineto{\pgfqpoint{-\pgfarrowlength}{-0.5\pgfarrowwidth}}
    \pgfusepathqstroke
  }
}
\clip(3324,-2594) rectangle (9908,-1155);
\tikzset{inner sep=+0pt, outer sep=+0pt}
\pgfsetarrows{[line width=15\XFigu, width=75\XFigu]}
\pgfsetarrowsend{xfiga0}
\pgfsetlinewidth{+7.5\XFigu}
\draw (3735,-1755) arc[start angle=+110.40, end angle=+66.42, radius=+2163.9];
\draw  (5317,-2185) circle [radius=+401];
\draw  (3733,-2167) circle [radius=+401];
\draw  (9499,-2155) circle [radius=+401];
\draw  (8109,-2144) circle [radius=+401];
\draw  (6712,-2154) circle [radius=+401];
\draw (4905,-2160)--(4095,-2160);
\pgfsetarrows{xfiga0-}
\draw (8505,-2160)--(9090,-2160);
\draw (7110,-2160)--(7695,-2160);
\pgftext[base,left,at=\pgfqpointxy{5200}{-2240}]  {\fontsize{15}{14.4}$A_1$} 
\pgftext[base,left,at=\pgfqpointxy{3585}{-2240}]  {\fontsize{15}{14.4}$A_2$} 
\pgftext[base,left,at=\pgfqpointxy{4315}{-1800}] {\fontsize{15}{14.4}$\lambda_2,0$} 
\pgftext[base,left,at=\pgfqpointxy{4335}{-2350}] {\fontsize{15}{14.4}$0,\lambda_1$} 
\pgftext[base,left,at=\pgfqpointxy{8590}{-2070}] {\fontsize{15}{14.4}$0,\lambda_\Delta$} 
\pgftext[base,left,at=\pgfqpointxy{7240}{-2070}] {\fontsize{15}{14.4}$0,\lambda_T$} 
\pgftext[base,left,at=\pgfqpointxy{7520}{-1405}] {\fontsize{16}{14.4}${\mathbf{MAP}M_{FD}}$}
\pgftext[base,left,at=\pgfqpointxy{4100}{-1405}] {\fontsize{16}{14.4}${\mathbf{MAP}M_A}$}
\pgftext[base,left,at=\pgfqpointxy{8050}{-2240}] {\fontsize{15}{14.4}$1$} 
\pgftext[base,left,at=\pgfqpointxy{6660}{-2240}] {\fontsize{15}{14.4}$0$} 
\pgftext[base,left,at=\pgfqpointxy{9410}{-2240}] {\fontsize{15}{14.4}$F$} 
\endtikzpicture%
}
\caption{$\text{E}_2$/M/M: MAPs of the arrivals and of the delay and fetching process.}
\label{PH_input}
\end{figure}

\begin{figure}[t]
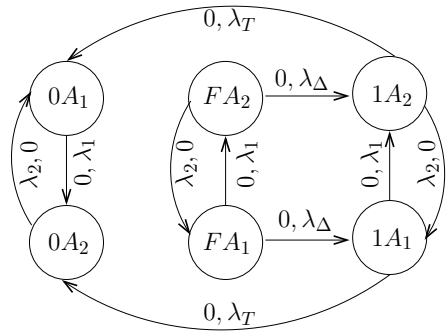

\centering
\resizebox{0.35\linewidth}{!}{\ifx\XFigwidth\undefined\dimen1=0pt\else\dimen1\XFigwidth\fi
\divide\dimen1 by 3934
\ifx\XFigheight\undefined\dimen3=0pt\else\dimen3\XFigheight\fi
\divide\dimen3 by 2946
\ifdim\dimen1=0pt\ifdim\dimen3=0pt\dimen1=4143sp\dimen3\dimen1
  \else\dimen1\dimen3\fi\else\ifdim\dimen3=0pt\dimen3\dimen1\fi\fi
\tikzpicture[x=+\dimen1, y=+\dimen3]
{\ifx\XFigu\undefined\catcode`\@11
\def\temp{\alloc@1\dimen\dimendef\insc@unt}\temp\XFigu\catcode`\@12\fi}
\XFigu4143sp
% Uncomment to scale line thicknesses with the same
% factor as width of the drawing.
%\pgfextractx\XFigu{\pgfqpointxy{1}{1}}
\ifdim\XFigu<0pt\XFigu-\XFigu\fi
\pgfdeclarearrow{
  name = xfiga0,
  parameters = {
    \the\pgfarrowlinewidth \the\pgfarrowlength \the\pgfarrowwidth},
  defaults = {
	  line width=+7.5\XFigu, length=+120\XFigu, width=+60\XFigu},
  setup code = {
    % miter protrusion = thk * sqrt(wd^2 + (tipmv*len)^2) / (2 * wd)
    \dimen7 2.15\pgfarrowlength\pgfmathveclen{\the\dimen7}{\the\pgfarrowwidth}
    \dimen7 2\pgfarrowwidth\pgfmathdivide{\pgfmathresult}{\the\dimen7}
    \dimen7 \pgfmathresult\pgfarrowlinewidth
    \pgfarrowssettipend{+\dimen7}
    \pgfarrowssetbackend{+-\pgfarrowlength}
    \dimen9 -0.5\pgfarrowlinewidth
    \pgfarrowssetvisualbackend{+\dimen9}
    \pgfarrowssetlineend{+-0.5\pgfarrowlinewidth}
    \pgfarrowshullpoint{+\dimen7}{+0pt}
    \pgfarrowsupperhullpoint{+-\pgfarrowlength}{+0.5\pgfarrowwidth}
    \pgfarrowssavethe\pgfarrowlinewidth
    \pgfarrowssavethe\pgfarrowlength
    \pgfarrowssavethe\pgfarrowwidth
  },
  drawing code = {\pgfsetdash{}{+0pt}
    \ifdim\pgfarrowlinewidth=\pgflinewidth\else\pgfsetlinewidth{+\pgfarrowlinewidth}\fi
    \pgfpathmoveto{\pgfqpoint{-\pgfarrowlength}{0.5\pgfarrowwidth}}
    \pgfpathlineto{\pgfqpoint{0pt}{0pt}}
    \pgfpathlineto{\pgfqpoint{-\pgfarrowlength}{-0.5\pgfarrowwidth}}
    \pgfusepathqstroke
  }
}
\clip(1160,-3970) rectangle (5094,-1024);
\tikzset{inner sep=+0pt, outer sep=+0pt}
\pgfsetarrows{[line width=15\XFigu, width=75\XFigu]}
\pgfsetarrowsstart{xfiga0}
\pgfsetlinewidth{+7.5\XFigu}
\draw (1665,-1530) arc[start angle=+126.23, end angle=+55.51, radius=+2566.3];
\pgfsetarrows{-xfiga0}
\draw (4590,-3465) arc[start angle=+-53.77, end angle=+-124.49, radius=+2566.3];
\draw (2790,-1890) arc[start angle=+146.95, end angle=+213.05, radius=+1114];
\pgfsetarrows{xfiga0-}
\draw (4905,-3150) arc[start angle=+-33.05, end angle=+33.05, radius=+1114];
\pgfsetarrows{-xfiga0}
\draw (1350,-3015) arc[start angle=+213.05, end angle=+146.95, radius=+1114];
\draw  (4590,-1822) circle [radius=+335];
\draw  (1665,-3172) circle [radius=+335];
\draw  (1665,-1867) circle [radius=+335];
\draw  (3105,-3172) circle [radius=+335];
\draw  (3105,-1867) circle [radius=+335];
\draw  (4590,-3127) circle [radius=+335];
\draw (3465,-1845)--(4230,-1845);
\draw (1665,-2205)--(1665,-2835);
\draw (3465,-3150)--(4230,-3150);
\draw (3105,-2835)--(3105,-2205);
\draw (4590,-2790)--(4590,-2160);
%\draw (2000,-3150)--(2780,-3150);

\pgftext[base,left,at=\pgfqpointxy{1500}{-1920}] {\fontsize{15}{14.4}$0A_1$} 
\pgftext[base,left,at=\pgfqpointxy{2900}{-1920}] {\fontsize{15}{14.4}$FA_2$} 
\pgftext[base,left,at=\pgfqpointxy{4425}{-1900}] {\fontsize{15}{14.4}$1A_2$} 
\pgftext[base,left,at=\pgfqpointxy{4425}{-3200}] {\fontsize{15}{14.4}$1A_1$} 
\pgftext[base,left,at=\pgfqpointxy{2900}{-3240}] {\fontsize{15}{14.4}$FA_1$} 
\pgftext[base,left,at=\pgfqpointxy{1500}{-3240}] {\fontsize{15}{14.4}$0A_2$} 
\pgftext[base,left,at=\pgfqpointxy{4485}{-2675},rotate=90] {\fontsize{15}{14.4}$0,\lambda_1$} 
\pgftext[base,left,at=\pgfqpointxy{3580}{-3030}] {\fontsize{15}{14.4}$0,\lambda_\Delta$} 
%\pgftext[base,left,at=\pgfqpointxy{2070}{-3030}] {\fontsize{15}{14.4}$\lambda_2,0$} 

\pgftext[base,left,at=\pgfqpointxy{3360}{-2675},rotate=+90] {\fontsize{15}{14.4}$0,\lambda_1$} 
\pgftext[base,left,at=\pgfqpointxy{2810}{-2675},rotate=+90] {\fontsize{15}{14.4}$\lambda_2,0$} 
\pgftext[base,left,at=\pgfqpointxy{5000}{-2675},rotate=+90] {\fontsize{15}{14.4}$\lambda_2,0$}
\pgftext[base,left,at=\pgfqpointxy{3550}{-1755}] {\fontsize{15}{14.4}$0,\lambda_\Delta$} 
\pgftext[base,left,at=\pgfqpointxy{2905}{-1215}] {\fontsize{15}{14.4}$0,\lambda_T$} 
\pgftext[base,left,at=\pgfqpointxy{2905}{-3875}] {\fontsize{15}{14.4}$0,\lambda_T$} 
\pgftext[base,left,at=\pgfqpointxy{1400}{-2685},rotate=+90] {\fontsize{15}{14.4}$\lambda_2,0$} 
\pgftext[base,left,at=\pgfqpointxy{1885}{-2660},rotate=90] {\fontsize{15}{14.4}$0,\lambda_1$} 
\endtikzpicture%
}
\caption{$\text{E}_2$/M/M: Combination of the arrival MAP with the delay and fetching MAP.}
\label{PH_input_total}
\end{figure}

%So far we considered Poisson request processes.
%is only considered to be a Poisson process. In this section,
Next, we generalize the Poisson request processes to show how to form the MAP of a caching hierarchy for PH distributed inter-arrival times.
%First, we show how to form the MAP of a single cache.
%In case of Poisson arrivals, the single cache MAP is easily formed.
%However, forming the single cache MAP for PH arrivals needs to be generally shown.
First, considering a single PH/M/M cache we show two required steps to form the MAP.
%which can be also applied to Poisson arrivals.
We regard two separate MAPs; one representing the arrival process which we denote as $M_A$ and the other representing the TTL and the fetching processes denoted by $M_{FD}$.
We form the MAP by \textit{(i)} applying the superposition between $M_A$ and $M_{FD}$ and  \textit{(ii)} changing the transitions resulting in misses.
For illustration the MAPs $M_A$ and $M_{FD}$ for an $\text{E}_2$/M/M system are shown in Fig.~\ref{PH_input}.
%The corresponding $M_A$ and $M_{FD}$
$M_A$ can be shown to have transition matrices
\vspace{-5pt}
\begin{equation*}
{\small
\mathbf{D}_0=
\begin{bmatrix}
-\lambda_1 &   \lambda_1 \\
0 & -\lambda_2
\end{bmatrix}
\;, \;
\mathbf{D}_1=
\begin{bmatrix}
0 & 0 \\
 \lambda_2   & 0
\end{bmatrix}
\;,
}
\end{equation*}
%\vspace{-5pt}
while $M_{FD}$ is represented by
\vspace{-5pt}
\begin{equation*}
{\small
\mathbf{D}_0=
\begin{bmatrix}
0 &      0        & 0    \\
   \lambda_T   & -\lambda_T  & 0    \\
        0        & \lambda_\Delta & -\lambda_\Delta
\end{bmatrix}
\;, \;
\mathbf{D}_1= \mathbf{0}
\;.
}
\end{equation*}
%\vspace{-5pt}
The second step, i.e. changing the transitions that result in misses, requires the modification of the two active transitions $M^{(0A_2 \rightarrow 0A_1)}$ and $M^{(1A_2 \rightarrow 1A_1)}$.
The first represents an  arrival when the object is not in the cache, therefore the transition is changed as $M^{(0A_2 \rightarrow FA_1)}$.
% \begin{equation}
%  M^{(0A_2 \rightarrow FA_1)}
% \end{equation}
The latter transition should not be an active transition as the object is not in the cache.
The resulting MAP of the superposition before the modification of the two active transitions is shown in Fig.~\ref{PH_input_total}.

%The question to be asked now is whether forming the MAPs in case of caching systems needs modifications or not.
The MAP formation of the caching system requires performing level and line superposition according to the topology.
%We have shown when forming the MAP in case of PH delay distribution that the MAP formation for Poisson delay has to be generalized.
Note that the previously discussed MAP extension depends on the delay MAP.
Changing the request input process influences the MAP formation as in
%The only difference when carrying the same operation is
that every transition or state to be changed exists a number of times as many as the states of the input process.
We do not show the derivation of PH/M/PH as it is straightforward given the analysis in this section.
%\vspace{-5pt}
%\vspace{-10pt}

%\vspace{-5pt}
\subsection{Computational Complexity Discussion}{\label{sec:complexity}}
In this subsection, we discuss the computational complexity of our MAP approach for caching trees.
%We show the complexity of each step in our MAP approach as well as the total complexity.
Further, we show the computational complexity for calculating the hit probability.

Our MAP approach is based on the level superposition and the line superposition operations.
First, the computational complexity of the level superposition operation is that of the Kronecker sum.
The Kronecker sum of two square matrices of size $m_1$ and $m_2$ results in a square matrix of size $m_1 m_2$.
The only operation involved between the elements of both matrices are addition operations.
Thus, the level superposition affects first the memory requirement and assignment.
%, however, no major mathematical operations (multiplication or division) are involved.

On the other hand, the computational complexity of line superposition operation is that of its four steps.
Except for the first step (Kronecker sum), the three other steps have a complexity of a search function in the state set.
For example, the identification of the invalid states requires a search through the state set.
Let the first step be a Kronecker sum between the MAP of a parent cache with $m_p$ number of states and the MAP of the children sub-trees with size $m_{ch}$ number of states.
The computational complexity of the Line superposition, i.e., the search in the state set with size $m_p m_{ch}$ in a tree is $\mathcal{O}(\log(m_p m_{ch}))$.
Here, $m_p=m_c$, i.e. the number of states of a single cache MAP which is given for a defined caching system with stationary stochastic parameters (input, TTL and delay).
The complexity of our MAP approach is then determined by the search complexity $\mathcal{O}(\log(m_p m_{ch}))$ in the number of states.
At each step of constructing the system MAP starting from the leaves to the root, $m_{ch}$ increases exponentially depending on the number of caches $n$ represented.
Therefore, we know for $m_{ch}$ that
\begin{equation*}
     m_{ch} \sim \mathcal{O}({m_c}^{n})\;.
\end{equation*}
Although each step of our approach has a complexity corresponding to the number of the caches involved, overall  we find the complexity of system MAP construction to be $\mathcal{O}( \log({m_c}^{n_c}))=\mathcal{O}( n_c\log({m_c}))$, where $n_c$ denotes the total number of caches of the hierarchy.
Therefore, the complexity of constructing the overall MAP grows linearly with the number of caches in the system.
%This means that we represent the complexity by that of the last step (highest complexity step) of line superposition with the root cache as a parent.
For example, a binary M/M/M tree with $L$ levels has a complexity of $ \mathcal{O}( {2^{L-1}}\log({3}))$.

The calculation of the hit probability depends on the steady state probability of the states denoted $\mathbf{\pi}$ in \eqref{eq:object_hit_probability}.
The calculation of the steady state vector has a complexity of solving a  system of ${m_c}^{n_c}$ linear equations.
Thus, it has a time complexity of $\mathcal{O}( {m_c}^{\xi n_c})$ with $\xi \in(2.4,3]$~\cite{Inv_complexity}.

\section{Approximation for single caches and cache hierarchies under non-zero delay}
\label{sec:approx_hierarchies}
%As discussed in Sect.~\ref{sect:background} the CDF of the inter miss times as well as the hit probability for a single cache under non-zero delay can be calculated according to \eqref{hit rate delay}-\eqref{miss process delay}.
\subsection{Approximation for single caches under non-zero delay}
\label{subsec:approx_single_cache}
Motivated by the high computational complexity of the calculation of the CDF of the inter miss times for a single cache in \eqref{hit rate delay}-\eqref{miss process delay} we provide a computable approximation for the Laplace-Stieltjes transform (LST) of the inter-miss times $F_Y^*(s)$ that also permits calculating the hit probability given renewal inter-request times, and arbitrary distributions of the TTL and the delay\footnote{as long as the corresponding Laplace transform exists.}.
Note that \eqref{miss process delay} contains iterative convolutions making it practically only computable for simple inter-arrival time distributions such as the exponential distribution.
%Next, we provide a computable approximation for the Laplace transform of the CDF of the inter-miss times $F_Y'(s)$ that also permits calculating the hit probability given arbitrary distributions
%which results in less computational complexity for the miss process and the hitrate given the
%of the inter-request times, the TTL and the delay\footnote{as long as their Laplace transform exists}.
In contrast to \eqref{hit rate delay}-\eqref{miss process delay} the following approximation can readily be incorporated in analytical methods to model entire caching hierarchies.
Our approximation is based on the following:
\begin{assumption}
\label{assumption1}
Given a miss there exists a request arrival at the instant the object is admitted to the cache, i.e., exactly at the end of the object fetch delay duration.
\label{Analytical assum}
\end{assumption}

% \noindent Fig.~\ref{Single_approx} illustrates this assumption.
% %, where the latter represents the single cache model using the assumption unlike the other.
% The transparent arrow represent a request arrival time that is per Assumption~\ref{Analytical assum} shifted to the moment the TTL starts.
% The approximated model equates the excess time $\Gamma_\Delta$ to the inter-arrival time $X_2$ in Fig.~\ref{Single_approx}.

% \begin{figure}[H]
% \vspace{-5pt}
% \centering
% \resizebox{0.4\textwidth}{!}{\input{Figs/Cache_TTL_D _approx.tex}}
% \caption{Approximation of the TTL cache model under fetch delay under Assumption~\ref{assumption1}.}
% \label{Single_approx}
% \vspace{-5pt}
% \end{figure}
% %

\begin{proposition}
\label{prop:LST_inter_miss_time}
Under Assumption~\ref{Analytical assum} and given the object fetch delay LST $F_\Delta^*(s)$, the LST of the renewal inter-request time $F_X^*(s)$ and the LST of the joint probability $L^*(s)$ from  Sect.~\ref{subsec:single_cache_analysis} at a single cache, the LST of the inter-miss time is
\begin{equation}
F_Y^*(s)= \frac{F_\Delta^*(s) [F_X^*(s) -L^*(s)]}{1- L^*(s)} .
\label{eq:Approx}
\end{equation}
\end{proposition}

\begin{proof}
%\renewcommand{\qedsymbol}{}
% \begin{flalign}
%F_Y(t)&=\mathsf{P}(X+\Delta<t,X>T)  &
%		\nonumber
%		\\	
%	&+\mathsf{P}(Y<t,x<t,X<t,X<T) &
%\end{flalign}
We calculate the probability that the inter-miss time $Y<t$ for inter-request time $X>T$ as $\mathbb{P}(X+\Delta<t)$.
%and when $X<T$ is expressed in general as $\mathsf{P}(Y<t)$.
Hence, the CDF of the inter-miss times is given by
\begin{equation}
F_Y(t) =\mathbb{P}(T<X<t-\Delta) +\mathbb{P}(Y<t,X<t,X<T)\; , \nonumber
\end{equation}
We manipulate the terms as
\begin{flalign}
&\mathbb{P}(T\!<\!X\!<\!t-\Delta)= \mathbb{P}(X\!<\!t-\Delta) -\mathbb{P}(X\!<\!T,X\!<\!t-\Delta)
		 = \E[F_X(t-\Delta)]				
		 -\E\bigg[\int_0^{t-\Delta} (1-F_T(x)) dF_X(x)\bigg] \; , &
	%	 \label{term 1}
		 \nonumber
\end{flalign}
%and
\begin{flalign*}
\mathbb{P}(Y\!<\!t,X\!<\!t,X\!<\!T) &= \E[F_Y(t) \ 1_{X<t,X<T}] =\! \int_0^t \!\! F_Y(t) (1\!\!-\!\!F_T(x))dF_X\!(x) \;,
			   %\!\!=\! \!\!\int_0^t \!\!\! F_Y(t\!\!-\!\!x) (1\!\!-\!\!F_T(x))dF_X\!(x)	
			   \nonumber
			   %\label{term_2}	   				 	
\end{flalign*}

% \begin{flalign}
% &\mathsf{P}(Y\!<\!t,X\!<\!t,X\!<\!T)=\! \int_0^t \! F_Y(t) (1-F_T(x))dF_X(x)
%  \nonumber \\
%   &= \!\int_0^t \! F_Y(t-x) (1-F_T(x))dF_X(x)
%   \nonumber
% \end{flalign}
% %
%From Equation.~\eqref{term 1} and Equation.~\eqref{term 2}  in Equation.~\eqref{main eqn},
\noindent
to express $F_Y(t)$ in terms of the PDFs and CDFs of the TTL and the inter-request time as
\begin{flalign}
F_Y(t)=\E[F_X(t-\Delta)]-\E\bigg[\int_0^{t-\Delta} (1-F_T(x)) dF_X(x)\bigg]	+\int_0^t F_Y(t) (1-F_T(x))dF_X(x)	\;,
    \label{CDF_y_approx}		
\end{flalign}
%
%The expectation represents the average with respect to the delay.
% \begin{fleqn}
% \[L(t)=\mathsf{P}(X<t,X<T)=\int_{0}^{t}(1-F_T(x)) dF_X(x)\]
% \end{fleqn}
Note that we marginalized here over the random variable $\Delta$.
Using $L(t)$ from Sect.~\ref{subsec:single_cache_analysis} and by substitution in \eqref{CDF_y_approx} as well as expressing the CDF of the inter-miss times in terms of the delay we obtain
\begin{flalign}
F_Y(t)  =\int_0^\infty F_X(t-\delta) dF_\Delta(\delta) -\int_0^\infty L(t-\delta) dF_\Delta(\delta)
	  + \int_0^t F_Y(t-x) dL(x) \;.
\end{flalign}
Note that $F_Y(t)$ is replaced with $F_Y(t-x)$ in the last term since $F_Y(t)=F_Y(t-x)$. This is a property of the renewal process \cite{renewal_property}.
As a result, $F_Y(t)$ is expressed in a recursive form.
Taking the LSTs we obtain
\begin{fleqn}
\[F_Y^*(s)=F_X^*(s) F_\Delta^*(s) -L^*(s) F_\Delta^*(s) + F_Y^*(s) L^*(s) \;,\]
\end{fleqn}
% \begin{fleqn}
% \[F_Y'(s)= F_\Delta'(s) \frac{ [F_X'(s) -L'(s)]}{1- L'(s)}\]
% \end{fleqn}
\noindent
which completes the proof.
\end{proof}

%\vspace{-10pt}
We observe that the closed form above is similar to the result of the zero delay expression in~\eqref{Miss_process laplace}. Both can be further related by $F_Y^*(s)= F_\Delta^*(s) F_Y^*(s|\Delta=0)$.
% % \begin{equation}
% % F_Y'(s)= F_\Delta'(s) F_Y'(s|\Delta=0) \;.
% % \label{intuitive proof}
% % \end{equation}
% This can be viewed as the inter-miss time random variable in the presence of the delay is only the addition of the delay and the inter-miss time given zero delay.
This is intuitive in light of Assumption~\ref{Analytical assum}.
%
%The approximation error is negligible as ${\E[\Delta]}/{\E[X]}$ becomes very high.
%This is the case we deal with in our modeling since the delay effect becomes more significant as long as ${\E[\Delta]}/{\E[X]}$ increases,  i.e., the request rate is significantly high.
%
%%%%%%%%%%%%%%%%%%%%%%%%%%%%%%%%%%%%%%%%%%%%%%%%%%%%%%%%
Next, we use Assumption~\ref{Analytical assum} to approximate the object hit probability.
Note that the probability of the number of hits $N$ between two misses being $n$
%$\mathbb{P}(N=n)$
%is $n$ is independent of the delay and is
can be represented by
\begin{equation*}
%\vspace{-10pt}
\mathbb{P}(N=n)=
\begin{cases}
  \mathbb{P}(X>T) & \; \; \text{for $n=0$} \\
  \mathbb{P}(X<T)^n   \mathbb{P}(X>T)& \; \; \text{for $n>0$}
  \end{cases}
  \;.
\end{equation*}
Letting $q=\mathbb{P}(X<T)$, $\E(N)$ is given from the geometric distribution as
\begin{equation}
\E(N)=\frac{q}{(1-q)}\;.
\label{Approx2}
\end{equation}
Next we compare the hit probability given non-zero object fetch delay to that under the zero delay assumption.
To this end, we use the following definition:
%define the {delay impairment} $\eta$ as
\begin{definition}
The delay impairment $\eta$ under random delay $\Delta$ is defined as
\begin{equation}
\eta := 1-\frac{P_h|_{\Delta\geq0}}{P_h|_{\Delta=0}} \;,
\end{equation}
with $P_h$ denoting the object hit probability.
\label{def_delay_impairment}
\end{definition}%
%as $\eta=1-\frac{P_h}{P_h|_{\Delta=0}}$.
This ratio $\eta \in [0,1]$ describes the reduction of the hit probability due to the object fetch delay, i.e., for $\eta=0$ there is no reduction in the hit probability, whereas $\eta=1$ denotes that the hit probability under delay is reduced to 0.
\begin{lemma}
\label{lem:delay_impairment}
Given Assumption~\ref{Analytical assum} the delay impairment for a single M/M/M cache  is
\begin{equation}
\eta = \frac{\bar{\tau}_\Delta}{\bar{\tau}_T+\bar{\tau}_\Delta +1} \;.
\label{reduction ratio}
\end{equation}
\end{lemma}%
Here, the ratio of the expected delay $\E[\Delta]$ to the expected inter-request time $\E[X]$ is defined as $\bar{\tau}_\Delta:=\E[\Delta] / \E[X]$.
%$\bar{\tau}_\Delta = \bar{\Delta} / \bar{X}$
%with $\bar{\Delta}:=\mathsf{E}[\Delta]$ and $\bar{X}:=\mathsf{E}[X]$ denoting the expected delay and  expected inter-request time, respectively.
Further, the ratio of the expected TTL $\E[T]$ to the expected inter-request time is defined as $\bar{\tau}_T := \E[T]/\E[X]$.
%with the expected TTL given as $\bar{T}:=\mathsf{E}[T]$.
Obviously, the delay impairment $\eta$ in \eqref{reduction ratio} increases monotonically from $0$ to $1$ as $\bar{\tau}_\Delta$ increases from $0$ to $\infty$.
%This means that as $\bar{\tau}_\Delta \rightarrow \infty$ the hit Pr. under zero delay is reduced by $100\%$.
In addition, we observe that the delay impairment $\eta$ decreases with the expected TTL
%.
%as concluded from \eqref{reduction ratio}.
%As $\bar{\tau}_T$ increases, $Ph_r$ decreases.
%This
which suggests that the reduction in hit probability due to larger delays can be compensated by larger TTL.
However, considering
%the decaying of $Ph_r$ due to the increase of the TTL is slow where the
the rate of change of the delay impairment with respect to the expected TTL $\frac{\partial \eta}{\partial \bar{\tau}_T}= \frac{-\bar{\tau}_\Delta}{(\bar{\tau}_T+\bar{\tau}_\Delta +1)^2}$
we conclude that compensating the object fetch delays using longer TTLs is disproportionately expensive.
% As a result, the change of $Ph_r$ is a small fraction of the change in $\bar{\tau}_T$.
% We conclude that expense of compensating the loss of the hit Pr. by increasing the TTL is not worthy.

%Proof
%\begin{align}
%Ph_z &= \frac{\lambda_x}{\lambda_x+\lambda_T} = \frac{1}{1+\bar{X}/\bar{T}}=
%		\frac{1}{1+1/\bar{\tau}_T}=\frac{\bar{\tau}_T}{\bar{\tau}_T+1}
%		\nonumber		
%		\\
%Ph_nz  & \approx \frac{\lambda_x}{\lambda_x+\lambda_T+\lambda_x \lambda_T /  	
%		\lambda_D} = \frac{1}{\bar{X}/\bar{T}+\bar{D}/\bar{T}+1}
%		\nonumber
%		\\
%		&\approx \frac{1}{1+1/\bar{\tau}_T+\bar{\tau}_D /\bar{\tau}_T}
%		=\frac{\bar{\tau}_T}{\bar{\tau}_T+\bar{\tau}_D +1}
%		\nonumber
%		\\
%Ph_r&\approx 1- \frac{1+\bar{\tau}_T}{\bar{\tau}_T+\bar{\tau}_D +1}
%		\nonumber
%		\\
%		&= \frac{\bar{\tau}_D}{\bar{\tau}_T+\bar{\tau}_D +1}
%\end{align}
%
\subsection{Approximations for caching hierarchies under non-zero delay}

Next, we extend the approximate caching hierarchy model in~\cite{Fofack} to consider random object fetching delays.
We will compare this approximation to our exact MAP based model in the following evaluation section. %Moreover, we illustrate the challenges facing the extension of the model. Furthermore, we state the drawbacks of the model and similar approximation models.

%We first briefly describe the main steps of the approximation in~\cite{Fofack}.
Similar to the level superposition and line superposition operations defined in the context of MAPs in Sect.~\ref{sec:exact_models}, the two main operations required in \cite{Fofack} to model a caching system are \textit{(i)} Input-output operation and \textit{(ii)} Superposition of the miss process of multiple caches.
% \begin{itemize}
% \item Input-output operation.
% \item Superposition of the miss process of multiple caches.
% \end{itemize}
Assuming zero delay, the input-output operation is used to exactly calculate the miss process of a given cache given the input process according to \eqref{Miss_process laplace}.
Further, \cite{Fofack} approximates the superposition of renewal miss processes as a renewal process.
%as in~\eqref{Superposition Assumption}.
Finally, the input request rate into each parent cache in the system is calculated recursively as the sum of input rates to the children caches each weighted by their respective miss probabilities.
%give~\eqref{hit rate no delay}.

Here, we extend this approximation to incorporate non-zero random object fetch delay.
We use our approximation from Proposition~\ref{prop:LST_inter_miss_time} to carry out the input-output operation according to \eqref{eq:Approx} instead of \eqref{Miss_process laplace}.
In addition, we calculate the hit probability of each cache according to~(\ref{hit rate delay}) and (\ref{Approx2}) using $m(\Delta)$ from \cite{Mostafa}.
%\todo[inline]{this is not correct as \eqref{Approx2} does not show a hit probability. Added equation (9) also}
Note that the superposition operation as defined by \cite{Fofack} remains unaffected by the delay.
%is the same because the delay has no influence on the superposition of multiple processes.
%Now, there are \textbf{two challenges} facing the straightforward extension of the model from \cite{Fofack} to include the object fetch delay, i.e., \textit{(i)} the need to estimate the delay PDF from each cache to the server and \textit{(ii)} that \cite{Fofack} lacks a metric for calculating the system hit probability.
Now, a {challenge} facing the straightforward extension of the model from \cite{Fofack} to incorporate the object fetch delay, is that this model lacks a metric for calculating the system hit probability.
% \begin{itemize}
% \item The need to have an estimation to the delay PDF from each cache to the server.
% \item The lack of an appropriate metric for calculating the system hit Pr.
% \end{itemize}
%The first challenge is simply put the need for a probabilistic description
%of the delay from every cache up to the origin server. This stands in contrast to our MAP based model that only requires PDF of the delay a cache and its parent.
%Further, in \cite{Fofack} the system hit probability is not proposed nor used to evaluate the cache hierarchy.
There the system hit probability is not proposed nor used to evaluate the cache hierarchy.
We opt here to use the classical method from~\cite{Che} to calculate the system hit probability, i.e.,
%According to Che the system hit rate in is calculated by
\begin{equation*}
	P_{h,sys}=1-\frac{M_{r,out}}{\sum_{i=1}^K \lambda_i}\;,
	\label{eq:approx2}
\end{equation*}
where $M_{r,out}$ is the output miss rate, i.e the miss rate of the root cache and $\lambda_i$ represents the input rate to the leaf cache $i$ while $K$ denotes the number of leaf caches in the system.
%\todo[inline]{why is K only used? for which topology does this work? for all or just for a flat one with K leaf nodes? K is the number of all nodes and work for any topology. the input $\lambda$ for non-leaf caches is zero}
Here, the system miss probability is obtained as the ratio of the output to the input rate of the cache hierarchy.

\section{Computable performance metrics for large Cache hierarchies through MAP Lumping}
\label{sec:lumpability}
In this section, we show how to reduce the state set size of the caching tree MAP to reduce the computational complexity of the cache performance metric computation.
Here, we use graph automorphisms to define equivalent states which are then lumped together and denoted as a lumpable partition.

The equivalent states within a MAP are the states that have the same description of the object status within the caching system.
To illustrate this, we use the example of a cache system consisting of two leaf caches as represented by the MAP in Fig.~\ref{Multiple Caches MAP}.
For example, state "10" in Fig.~\ref{Multiple Caches MAP} denotes that the object is stored in the first cache only while state "01" denotes that the object is stored in the second cache only.
If the two caches are symmetric in terms of the request processes, the TTL and the delay distributions, the order of the caches having the object is \textit{irrelevant} and the system description is \textit{identical} when having the object in either of the caches.
Here, it is clear that states "01" and "10" are hence equivalent states and can be lumped together.
In general, we denote each group of equivalent states within a MAP as a lumpable partition while the number of lumpable partitions is given by $N_p$.
To reduce the state set of a given MAP, we find all lumpable partitions and represent each of them using only one of their states, thus we  only use $N_p$ states for the final performance metric computation.

We observe that the problem of finding lumpable partitions within a MAP that represents a caching tree is equivalent to first finding the automorphisms of a graph featuring the caching tree. Then, we can let the automorphism group act on the states set~\cite{wasiur, bio}.
%Using the previous definition of the problem,
We focus in the following on finding the lumpable partitions.

Given a caching hierarchy with depth $L$ (denoted the number of levels) and defining the set $[L] = \{1,\dots,L\}$.
We call a caching tree \textit{homogeneous} when the caches on the same level have the same input, TTL and delay distributions, i.e.,
\begin{align*}
    &F_{X_{i,j}}(t)= F_{X_{i,k}(t)},   F_{T_{i,j}}(t)= F_{T_{i,k}(t)},
    F_{\Delta_{i,j}}(t)= F_{\Delta_{i,k}(t)}
\forall  j,k\in [n_i] \;, \forall i \in [L]  \;,
\end{align*}
where $F_{X_{i,j}}(t)$, $F_{T_{i,j}}(t)$ and $F_{\Delta_{i,j}}(t)$ are respectively the input (inter-request time), TTL and delay distribution of the $j$th cache in the $i$th level.
Here $n_i$ is the number of caches in level $i$ and $[n_i] :=\{1,\dots,n_i\}$.
%\todo[inline]{The sentence above can be written as formula.}

We represent the homogeneous cache tree with a graph $G(V,E)$ whose vertices $V\in[n_c]$ are the caches, where $n_c$ is the total number of caches in the hierarchy, and the edges in $E$ denote their connections.
\begin{definition}
A bijection $f\ : V \rightarrow V$ is an automorphism of G if for $(i,j) \in E$ $(f(i),f(j))\in E$ $\forall \ i,j \in V$, i.e., it is a permutation of the graph vertices that retains the edges.
The set of all the automorphisms $f$ that are defined on the graph $G$ forms a group $\text{Aut}(G)$ under function composition~\cite{Book:Godsil}.
\label{auto def}
\end{definition}

Given the set of states of a MAP of a single cache $\gamma$, as depicted for example for an M/M/M or an M/M/$\text{E}_f$ cache in Fig.~\ref{Single MAP} or Fig.~\ref{Single Erlang}, respectively.
In order to find the lumpable partitions, we let $\text{Aut}(G)$ act on $\gamma^{n_c}-I$ where $I$ is the set of the invalid states defined in \eqref{eq:set_of_invalid_states}.
Note that for a state $u \in \gamma^{n_c}$, $u_i \in \gamma$ for $i\in[n_c]$. For example, for Fig.~\ref{Multiple Caches MAP} we have $u\in\gamma^2$ with $u_i\in \{0,1,F\}$, i.e. when $u=(1,F)$ (that is the depicted state $1F$ in the figure) then $u_1=1, u_2=F$.
The group action is a map $\text{Aut}(G) \times \gamma^{n_c}-I \rightarrow \gamma^{n_c}-I$ defined by
\begin{align}
\label{group_action}
     f.u = v   \leftrightarrow u_i=v_{f(i)}\forall i\in [{n_c}], \forall f\in \text{Aut}(G), u \ \text{and} \ v \in \gamma^{n_c}-I \;,
\end{align}
using the equivalence relation $\leftrightarrow$.
This group action on $u$ is the permutation of the elements of $u$ each representing a state of one of the tree caches according to $f$.
Using this concept, we define next the equivalence between two states $u,v$ in $\gamma^{n_c}-I$.
\begin{definition}
States $u,v \in \gamma^{n_c}-I$ are equivalent if and only if there exist $f \in \text{Aut}(G):\ f.u=v$ \;.
\label{Def: equivalence}
\end{definition}

\subsection{Iterative state lumping}
In this subsection, we describe our approach to reducing the state set using lumpability first by a solution to the problem of finding automorphisms then by iteratively lumping the state set.
The problem of finding the automorphisms of any graph is NP complete and can be solved using McKay's algorithm~\cite{morphism_alg}.
However, given the symmetric and recursive structure of the tree graph and based on Def.~\ref{auto def}, we define the automorphisms in a caching tree graph using Theorem \ref{Lem:tree_auto}.
\begin{lemma}
A permutation of a tree graph is an automorphism if it is a composition of permutations between sibling sub-trees that are symmetric.
\label{Lem:tree_auto}
\end{lemma}
\vspace{-5pt}
\begin{proof}
This lemma imposes three conditions on a permutation to be identified as an automorphism.
We prove these conditions in the following:
First, only siblings can be permuted such that it preserves the vertex-edge connectivity.
Let $i,j$  represent the indices of two \emph{leaf} nodes in a tree graph and $i_p,j_p$ represent the corresponding parent.
As a result $(i,i_p)$ as well as $(j,j_p) \in E$.
A permutation between $i$ and $j$ is an automorphism if and only if $(i,j_p)$ and $(j,i_p) \in E$.
This is only true if $j_p=i_p$, thus $i$ and $j$ are siblings.
The second condition is that the permutation of non-leaf nodes requires that of their entire sub-trees.
Hence, let $i,j$  represent the indices of two \emph{non-leaf} nodes in a tree graph and $i_c,j_c$ represent the indices of one of their corresponding children.
As a result $(i,i_c)$ as well as $(j,j_c) \in E$.
A permutation between $i$ and $j$ is an automorphism if and only if $(i,j_c)$ and $(j,i_c) \in E$ which is impossible to happen.
Hence, the simultaneous permutation of the vertices and their corresponding children is indispensable to keep the same vertex-edge connectivity.
The third condition is that the sibling sub-trees must be symmetric.
We define two sub-trees to be symmetric if, for each vertex of a sub-tree, there exists a corresponding symmetric one in the other located in the same level and has the same number of children.
This condition is a straight-forward result to the second condition.
\end{proof}

Note that the lemma above implies that a composition between two permutations that fulfill the three conditions is also an automorphism.
This is due to the fact that the set of all automorphisms forms a group $\text{Aut}(G)$ under the function composition.
Graphically, this can be seen from the recursive structure of the tree where one of the permutable sub-trees can contain permutable sub-trees.

Recall that in our MAP approach in Sect.~\ref{sec:exact_models}, we build the MAP of the caching tree iteratively by applying the level superposition operation on MAPs modelling the sibling cache sub-trees in each level starting from the leaves followed by the line superposition operation between the parent caches and their superposed children.
To lump states one may form the group $\text{Aut}(G)$ of the overall tree MAP according to Lem.~\ref{Lem:tree_auto}.
Then, we let $\text{Aut}(G)$ act on the state set to determine the lumpable partitions.
As a result, the complexity of solving the MAP for the steady state probability $\mathbf{\pi}$ is reduced.
However, we can further reduce the complexity of solving the MAP by iterative lumping.
Our approach to reducing the state set is to lump the equivalent states for each MAP resulting from the level superposition instead of applying the lumping approach for the overall MAP.
Both approaches result in having exactly the same MAP with the same number of states, thus they have the same complexity of the hit probability calculation.
However, applying the lumping approach in each level superposition operation has the following advantages:
First, it is beneficial for speeding up the construction of the overall MAP.
As a result of lumping per level superposition operation, the size of the state set in every step is reduced (given there are equivalent states), hence, the complexity of each step is reduced.
On the other hand, the complexity related to finding the automorphisms and their actions on the states is significantly reduced.
Given that we apply the level superposition on sibling sub-trees, Lem.~\ref{Lem:tree_auto} implies that all the permutations between the symmetric sub-trees are automorphisms.
%Thus, finding the automorphisms almost has no influence on the computational complexity.
%In addition, the number of the tree automorphisms that result in more than one equivalent state is reduced.
%Using the lumpability per level superposition is equivalent to the action of the tree automorphisms which only result in the permutation of the corresponding sub-trees in every level.
%Assuming that there are $k$ sibling sub-trees with the tree, thus $k$ level superposition operation are applied.
%Let $f^k \in \text{Aut(G)}$ such that it only includes the permutation of the $k$-th sibling sub-trees.
%Applying lumpability iteratively does not only account for the application of $f^k$ at the $k$-th operation, but also the application of $f^k.f^{k-1}$, i.e, the automorphisms equivalent to the composition between the ones that got applied
\subsection{The Size of the State set}
Next, we derive the size of the reduced state set as a result of state lumping for each level superposition operation in comparison to the original state set size.
We define some important notations and definitions that we will use later.
\begin{Corollary}
The individual MAPs modelling the permutable sibling sub-trees are the same.
\label{cor:same_MAPs}
\end{Corollary}
Corollary~\ref{cor:same_MAPs} follows directly from Lem.~\ref{Lem:tree_auto}.
Permutable sibling sub-trees denote the sub-trees whose permutations are considered automorphisms.
This comes from the fact that an automorphism only exists between symmetric sub-trees according to Lem.~\ref{Lem:tree_auto}.

As a result of Corollary~\ref{cor:same_MAPs}, the number of the states of $M_{\text{lvl}}$ resulting from the level superposition of $n$ permutable sub-trees is $m_\mathcal{S}^n$ where $m_\mathcal{S}$ is the number of the states of the MAP modelling any of the sub-trees.
Let the state of a subtree be defined as $S=(S_1,S_2,.....,S_n) \in \mathcal{S}^n$, i.e. it represents a state in $M_{\text{lvl}}$ such that $S_j \in \mathcal{S} \ \forall j \in [n]$ where $\mathcal{S}$ denotes the set of states of the MAP of one sub-tree.
In a lumpable partition $\rho$, we denote the frequency of the presence of a state $s^{(i)} \in \mathcal{S}$ as an element in a state $S$ by $\nu_i$ where $i \in [m_\mathcal{S}]$.
For example, for three binary M/M/M sub-trees each of one parent and two child caches we have for $S=(001,10F,001)$ and $s^{(i)}= (001)$ the frequency $\nu_i=2$.
We represent $\nu_i$ by
\begin{equation*}
\vspace{-5pt}
  \nu_i= \sum_{\forall j \in [n]} \mathbf{1}_{S_j=s^{(i)}}  \;,
\end{equation*}
where $\mathbf{1_{(\cdot)}}$ is the indicator function.
According to Lem.~\ref{Lem:tree_auto}, the states in each lumpable partition are considered permutations with respect to their elements $S_j$.
This is exactly the result of the action of the automorphisms of $G$ that include only the permutations of the intended sub-trees on the state set $\mathcal{S}^n$.
As a result, the frequency $\nu_i$ of all the states in one partition is the same.
Hence, each lumpable partition has a \textit{unique frequency vector} $\nu$ of length $m_\mathcal{S}$ which defines the different lumpable partitions in $M_\text{lvl}$.
%We denote a lumpable partition by $\rho_\nu$.
Since the length of $S$ with respect to its elements $S_i$ is $n$, we have
\vspace{-5pt}
\begin{equation*}
\sum_{b=1}^{m_\mathcal{S}} \nu_b =n \;.
\end{equation*}
\begin{theorem}
\label{thm:nr_lumpabe_partitions}
The number of lumpable partitions within the MAP representing the level superposition of $n$ symmetric sub-trees with refined individual MAPs is
\begin{equation*}
N_p= {{n+m_\mathcal{S}-1}\choose{m_\mathcal{S}-1}} \leq {m_\mathcal{S}}^n \hspace{0.5cm} m_\mathcal{S}\geq3,n\geq1 \;.
\end{equation*}
\end{theorem}
\noindent where $m_\mathcal{S}$ is the number of the states of the MAP of one sub-tree.
We denote a MAP as \textit{refined} if there exists no equivalence between its states.
Note that this is the case after lumping.
The number of the lumpable partitions corresponds to the number of all possible values of $\nu$.
Therefore, the number of lumpable partitions is the number of $m_\mathcal{S}$-tuples of non-negative integers whose sum is $n$.
According to the ``stars and bars'' theorem in \cite{stars_bars}, the number of the tuples is given by~${{n+m_\mathcal{S}-1}\choose{m_\mathcal{S}-1}}$.
The proof of the relation in Theorem~\ref{thm:nr_lumpabe_partitions} is given in the appendix~\ref{appendix:proof_of_}.
\begin{Corollary}
\label{cor:nr_lumpabe_partitions_polynomial}
The number of partitions $N_p$ is bounded by a polynomial function of $n$
\begin{equation*}
    N_p \sim \mathcal{O}(n^{m_\mathcal{S}})
\end{equation*}
\end{Corollary}
The proof of Cor.~\ref{cor:nr_lumpabe_partitions_polynomial} is in the appendix in Sect.~\ref{appendix:derivation_poly_Np}. Essentially, Cor.~\ref{cor:nr_lumpabe_partitions_polynomial} states that the growth of the number of states with the number of symmetric sub-trees in the cache hierarchy is broken down from originally an exponential growth (see the right hand side of Theorem~\ref{thm:nr_lumpabe_partitions}) to a polynomial one under the lumpability approach.
%%%%%%%%%%%%%%%%%%%%%%%%%%%%%%%%%%%%%%

\begin{figure}[t]
\centering
\begin{tikzpicture}

\begin{axis}[%
axis line style = thick,
width=2.1in,
height=1.625in,
at={(0in,0in)},
scale only axis,
xmin=1,
xmax=10,
xlabel style={font=\color{white!15!black}},
xlabel={Number of symmetric sub-trees $(n)$},
ymode=log,
ymin=1,
ymax=1e+15,
xlabel style={yshift=0.15cm},
ylabel style={yshift=-0.3cm},
yminorticks=true,
ylabel style={font=\color{white!15!black}},
ylabel={Number of states},
axis background/.style={fill=white},
xmajorgrids,
ymajorgrids,
yminorgrids,
legend style={legend cell align=left, align=left, draw=white!15!black},
legend pos=outer north east
]
\draw[line width=0.005 \linewidth](current axis.south west)rectangle(current axis.north east);
\begin{comment}

\addplot [color=blue, mark=x, mark options={solid, blue}]
  table[row sep=crcr]{%
1	3\\
2	9\\
3	27\\
4	81\\
5	243\\
6	729\\
7	2187\\
8	6561\\
9	19683\\
10	59049\\
};
\addlegendentry{1-lvl}

\addplot [color=red, mark=x, mark options={solid, red}]
  table[row sep=crcr]{%
1	3\\
2	6\\
3	10\\
4	15\\
5	21\\
6	28\\
7	36\\
8	45\\
9	55\\
10	66\\
};
\addlegendentry{1-lvl lump}

\addplot [color=black, only marks, mark=x, mark options={solid, black}]
  table[row sep=crcr]{%
1	3\\
2	6\\
3	10\\
4	15\\
5	21\\
6	28\\
7	36\\
8	45\\
9	55\\
10	66\\
};
%\addlegendentry{1-lvl lump all}
\end{comment}

\addplot [color=blue, mark=asterisk, mark options={solid, blue}]
  table[row sep=crcr]{%
1	27\\
2	729\\
3	19683\\
4	531441\\
5	14348907\\
6	387420489\\
7	10460353203\\
8	282429536481\\
9	7625597484987\\
10	205891132094649\\
};
\addlegendentry{Original}

\addplot [color=red, mark=asterisk, mark options={solid, red}]
  table[row sep=crcr]{%
1	27\\
2	378\\
3	3654\\
4	27405\\
5	169911\\
6	906192\\
7	4272048\\
8	18156204\\
9	70607460\\
10	254186856\\
};
\addlegendentry{Lump}

\addplot [color=black, mark=asterisk, mark options={solid, black}]
  table[row sep=crcr]{%
1	18\\
2	171\\
3	1140\\
4	5985\\
5	26334\\
6	100947\\
7	346104\\
8	1081575\\
9	3124550\\
10	8436285\\
};
\addlegendentry{Lump+}
\end{axis}

\end{tikzpicture}%
\caption{The number of states of the level superposition of $n$ $2$-lvl M/M/M binary symmetric sub-trees: The reduced state set size due to lumpability grows significantly slower (polynomially) for an increasing number of sub-trees compared to the original MAP that grows exponentially in $n$. Lumping the states of inner caches (permutation between leaves) shows an additional improvement compared to only lumping with respect to the permutations between the sub-trees.}
\label{fig:Lumpability}
\end{figure}
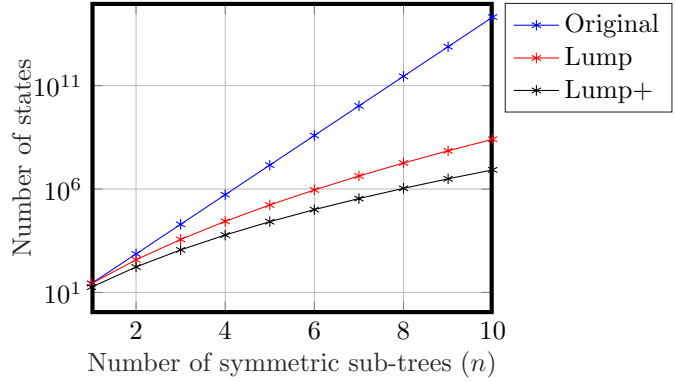
We show in Fig.~\ref{fig:Lumpability} the significant effect of the lumpability on the number of the states of the MAP representing $n\in\{1,\dots,10\}$ 2-lvl M/M/M symmetric binary sub-trees.
We compare the size of the state set for three cases: the original MAP, the MAP after applying the lumpability approach with respect to the sub-tree permutations only (Lump) and the MAP applying lumpability approach  with respect to all the possible automorphisms (Lump+).
Here we use 2-lvl binary sub-trees as an example, but the qualitative behavior of Fig.~\ref{fig:Lumpability} remains unchanged when having more levels or different number of children per node.
As described above, the figure shows that applying the lumpability approach results in a remarkable decrease in the growth of the number of states with the number of symmetric sub-trees, i.e., from an originally exponential growth to a polynomial one.
In addition, lumping all the possible states taking into account not only the sub-trees permutations but also the inner nodes permutations leads to an additional reduction in the state set size (Lump+).

%Recall that in our approach we lump the states after each level superposition operation which results in lumping all the possible states.
\subsection{MAP Lumpability}
In this subsection, we prove that the MAP within each level superposition operation is lumpable with respect to the lumpable partitions $\mathcal{\rho}=\{\rho_1, \rho_2, ...., \rho_{N_p}\}$ formed according to the equivalence relation between the states based on Def.~\ref{Def: equivalence}.

\begin{definition}
We define $\mathbf{{Q}}=\mathbf{D_0}+\mathbf{D_1}$ as the total transition matrix of the MAP resulting from the level superposition of $n$ symmetric sub-trees.
Elements of $\mathbf{{Q}}$, i.e., $Q_{A,B}$ denote the transition rate from state $A$ to state $B$.
In addition, we define $\mathbf{q}$ as the total transition matrix of each of the $n$ symmetric sub-trees.
Elements of $\mathbf{{q}}$, i.e., $q_{a,b}$ denote the transition rate from state $a$ to state $b$.
\end{definition}
\vspace{-5pt}
\begin{lemma}
There exists no transition from a state $A$ to a state $B$ in the MAP $M_{\text{lvl}}$ resulting from the level superposition of the MAPs $M_j$ of $n$ symmetric trees if $A_j \neq B_j $ for more than one element $j \in [n]$.
\begin{equation*}
Q_{A,B}=0 \ \text{if} \ \sum_{\forall j \in [n]} \mathbf{1}_{A_j \neq B_j}>1, \  A,B \in \mathcal{S}^n
\end{equation*}
\label{lemma:zero trans}
\end{lemma}
\begin{proof}
We note that only one event (active or hidden transition) takes place at a time.
The level superposition is applied on MAPs of independent trees, i.e. a transition between different states in one MAP is independent of any other transitions in other MAPs.
Therefore, having two states $A$ and $B$ that have a difference of at least two pairs of the corresponding superposed states $A_j$ and $B_j$ implies the occurrence of at least two independent events.
For these events no single transition exists.
\end{proof}
%A direct result of Lemma \ref{zero trans} is that transition rate from state $S_i$ to state $S_k$ is given by $Q_{S_i,S_k}=0$ such that $i\neq k$ if $s_{\{1,j\}}\neq s_{\{2,j\}} $ for more than one element $j$ where  $j \in \{1,2,...,n\}$.
Given a MAP $M_\text{lvl}$ resulting from  the level superposition of $n$ independent symmetric sub-trees in a graph $G_\text{lvl}$,
we consider states $A,B$ with non-zero transition rates $Q_{A,B}$ as defined by Lem.~\ref{lemma:zero trans}. Now we can represent the transition rate $Q_{A,B}$ as the sum of the corresponding sub-tree transition rates as
%\vspace{-10pt}
\begin{equation}
    Q_{A,B}=  \sum_{j=1}^n \left. q_{A_j,B_j} \right|_{A_j \neq B_j}, \ \ A\neq B \;.
    \label{eq:total_trans_element}
\end{equation}
%\vspace{-10pt}
Note that for $Q_{A,B} \neq 0$ there is only one value of $j$ for which $A_j \neq B_j$.

Now, we show that under the equivalence relation from Def.~\ref{Def: equivalence} for an automorphism $f \in Aut(G_\text{lvl})$, i.e. a permutation of the $n$ symmetric sub-trees, the total transition rates between states in different lumpable partitions remain unchanged.
\begin{lemma}
\label{lemma:same_trans}
\begin{align*}
    & \hspace{2cm} Q_{A^*,B^*}=Q_{A,B}: \nonumber \\
    & A \in \rho_1, \ B \in \rho_2, \ A \neq B, \ A^*=f.A, \ B^*=f.B \nonumber \\
    & \hspace{2cm} \forall f \in \mathrm{Aut}(G_\text{lvl}) \;.
\end{align*}
\label{same transitions}
\end{lemma}
\noindent The proof of Lem. \ref{lemma:same_trans} is given in the appendix in Sect.~\ref{appendix:proof_of_thm4}.
In the following, based on the equivalence relation from Def.~\ref{Def: equivalence} we obtain lumpable partitions of the MAP of the level superposition.
\begin{theorem}
The MAP $M_\text{lvl}$ resulting from the level superposition of $n$ independent symmetric sub-trees is lumpable with respect to the partitions $\mathcal{\rho}=\{\rho_1, \rho_2, ...., \rho_{N_p}\}$ formed based on the equivalence relation between the states according to Def.~\ref{Def: equivalence}.
\end{theorem}
\begin{proof}
A necessary and sufficient condition for a MAP to be lumpable with respect to $\mathcal{\rho}$ is that for every pair of partition sets $\rho_i$ and $\rho_j$, $Q_{A,\rho_j}:=\sum_{B\in \rho_j} Q_{A,B}$ remains unchanged $\forall A \in \rho_i$~\cite{Book:Lump_cond} .
From Lem.~\ref{lemma:same_trans} we know that any pair of states $A,A^*$ from one partition $\rho_i$ has the same transition rate to a state $B$ in another partition $\rho_j$ and its permutation $B^*$ in the same partition respectively.
This implies that the $Q_{A,\rho_j}=Q_{A^*,\rho_j}\  \forall A,A^* \in \rho_i$ and $j\in[N_p] \setminus i$.
The only remaining task is to prove that $Q_{A,\rho_i}=Q_{A^*,\rho_i}  \ \forall A,A^* \in  \rho_i $ , i.e, the sum of the transition rates from one state to its equivalent ones remains constant for all the states in the same partition.
As every two states $A,A^*$ within the same partition are permutations, the corresponding sub-tree states $A_j,A^*_j$ are not equal for at least two values of $j \in [n]$.
Therefore, by Lem.~\ref{lemma:zero trans} there exists no transition between $A$ and $A^*$ $\forall A,A^* \in \rho_i : A\neq A^*$, i.e., there exist no transitions between states from the same partition.
Therefore, the condition in \cite{Book:Lump_cond} holds for partitions defined under the equivalence relation Def.~\ref{Def: equivalence}.
\end{proof}
Note that the transition matrix of the lumped MAP is an $N_p \times N_p$ matrix with elements representing the transition rate from one partition to another, i.e, an element $Q_{\rho_i,\rho_j}:=\sum_{B\in \rho_j} Q_{A,B}$ where $A$ is any of the states in $\rho_i$ since they all have the same value of the sum.
%$[Q_{\rho_i,\rho_j}]: \ i,j \in [N_p]$.
% \vspace{-10p}

\section{Evaluations}
\label{sec:evaluations}
%- Delay has a  substantial impact on the cache hitrate (single cache but also hierarchy)

%- MAP and simulation coincide: For PH/M/M and M/M/PH: show that our framework provides MAP results equal to simulations for PH/M/M and M/M/PH

%- What is the impact of the delay in we add more levels to the hierarchy? (binary tree)

%- What is the impact of the TTL on the hitrate for a fixed delay. Fix the expected delay D = X = 1 (the expected interarrival time) and change the TTL T. Plot the  hitrate vs. D/T.

%- what is the comparable result obtained from the SOTA (Fofack+Che with system hitrate)

%- Microbenchmark: Fofack for root is bad (similar to Daniel who does IRM while Fofack does renewal assumption)

%- Benchmark of the our miss process Approximation (with delay) in (32) - show here the hitrate with the approx vs the hit rate of the simulation for a single cache.

%- (check approximation vs parameter of gamma(alpha,alpha*beta) distribution of X for $\alpha \in(0,\infty))$

\def\biggg{% This file was created by matlab2tikz.
\definecolor{mycolor1}{rgb}{1.00000,0.00000,1.00000}%
\begin{tikzpicture}

\begin{axis}[%
axis line style = thick,
width=2.1in,
height=1.625in,
at={(0in,0in)},
scale only axis,
xmin=0,
xmax=20,
xlabel style={font=\color{white!15!black}},
xlabel={Delay to inter-request time ratio $\bar{\tau}_\Delta$},
xlabel style={yshift=0.15cm},
ylabel style={yshift=-0.4cm},
ymin=0,
ymax=1,
ylabel style={font=\color{white!15!black}},
ylabel={Object hit probability $P_h$},
axis background/.style={fill=white},
axis x line*=bottom,
axis y line*=left,
xmajorgrids,
ymajorgrids,
legend style={legend cell align=left, align=left, draw=white!15!black, line width= 0.002\linewidth},
legend pos=outer north east]
\draw[line width=0.005 \linewidth](current axis.south west)rectangle(current axis.north east);
  
\addplot [color=red, mark=o, mark options={solid, red},mark size= 2.3 pt]
  table[row sep=crcr]{%
0	0.666664444451852\\
1	0.5\\
2	0.4\\
3	0.333333333333333\\
4	0.285714285714286\\
5	0.25\\
6	0.222222222222222\\
7	0.2\\
8	0.181818181818181\\
9	0.166666666666667\\
10	0.153846153846154\\
11	0.142857142857143\\
12	0.133333333333333\\
13	0.125\\
14	0.117647058823529\\
15	0.111111111111111\\
16	0.105263157894736\\
17	0.0999999999999999\\
18	0.0952380952380951\\
19	0.0909090909090908\\
20	0.0869565217391299\\
};
\addlegendentry{MAP}

\addplot [color=blue, mark=asterisk, mark options={solid, blue}]
  table[row sep=crcr]{%
0	0.666372666372666\\
1	0.499133499133499\\
2	0.399471399471399\\
3	0.333251333251333\\
4	0.285823285823286\\
5	0.248638248638249\\
6	0.222678222678223\\
7	0.2001532001532\\
8	0.181379181379181\\
9	0.165174165174165\\
10	0.153832153832154\\
11	0.142970142970143\\
12	0.133079133079133\\
13	0.124874124874125\\
14	0.117812117812118\\
15	0.110585110585111\\
16	0.105634105634106\\
17	0.0999480999480999\\
18	0.0960940960940961\\
19	0.0905910905910906\\
20	0.0866620866620866\\
};
\addlegendentry{Sim}

\addplot [color=black, mark=diamond, mark options={solid, black},mark size=2.2pt]
  table[row sep=crcr]{%
0	0.666666666666667\\
1	0.5\\
2	0.4\\
3	0.333333333333333\\
4	0.285714285714286\\
5	0.25\\
6	0.222222222222222\\
7	0.2\\
8	0.181818181818182\\
9	0.166666666666667\\
10	0.153846153846154\\
11	0.142857142857143\\
12	0.133333333333333\\
13	0.125\\
14	0.117647058823529\\
15	0.111111111111111\\
16	0.105263157894737\\
17	0.1\\
18	0.0952380952380952\\
19	0.0909090909090909\\
20	0.0869565217391304\\
};
\addlegendentry{Approx. \\  \ \ (\ref{eq:Approx}-\ref{Approx2})}

\end{axis}

\end{tikzpicture}%
}
\def\smalll{% This file was created by matlab2tikz.
\definecolor{mycolor1}{rgb}{1.00000,0.00000,1.00000}%
\begin{tikzpicture}

\begin{axis}[%
axis line style = thick,
width=1.2in,
height=0.9in,
at={(0in,0in)},
scale only axis,
xmin=0,
xmax=2,
xlabel style={font=\color{white!15!black}},
ymin=0.35,
ymax=0.7,
ylabel style={font=\color{white!15!black}},
axis background/.style={fill=white},
axis x line*=bottom,
axis y line*=left,
xmajorgrids,
ymajorgrids,
legend style={legend cell align=left, align=left, draw=white!15!black}
]
\draw[line width=0.005 \linewidth](current axis.south west)rectangle(current axis.north east);
  
\addplot [color=red, mark=o, mark options={solid, red},mark size= 2.3 pt]
  table[row sep=crcr]{%
0	0.666664444451852\\
0.2	0.625\\
0.4	0.588235294117647\\
0.6	0.555555555555556\\
0.8	0.526315789473684\\
1	0.5\\
1.2	0.476190476190476\\
1.4	0.454545454545454\\
1.6	0.434782608695652\\
1.8	0.416666666666667\\
2	0.4\\
};
%\addlegendentry{MAP}

\addplot [color=blue, mark=asterisk, mark options={solid, blue}]
  table[row sep=crcr]{%
0	0.666627666627667\\
0.2	0.624642624642625\\
0.4	0.588101588101588\\
0.6	0.554526554526555\\
0.8	0.525868525868526\\
1	0.501031501031501\\
1.2	0.476986476986477\\
1.4	0.454109454109454\\
1.6	0.434023434023434\\
1.8	0.415819415819416\\
2	0.3995723995724\\
};
%\addlegendentry{Sim}

\addplot [color=black, mark=diamond, mark options={solid, black},mark size=2.2pt]
  table[row sep=crcr]{%
0	0.666666666666667\\
0.2	0.625\\
0.4	0.588235294117647\\
0.6	0.555555555555556\\
0.8	0.526315789473684\\
1	0.5\\
1.2	0.476190476190476\\
1.4	0.454545454545455\\
1.6	0.434782608695652\\
1.8	0.416666666666667\\
2	0.4\\
};
%\addlegendentry{theo}

\end{axis}

\end{tikzpicture}%
}
\begin{figure}[t]
\centering
\stackinset{l}{72pt}{t}{5pt}{\input{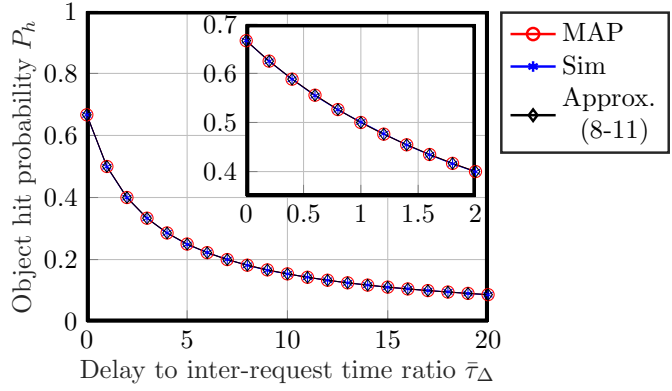}}{% This file was created by matlab2tikz.
\begin{tikzpicture}

\begin{axis}[%
axis line style = thick,
width=2.1in,
height=1.625in,
at={(0in,0in)},
scale only axis,
xmin=0,
xmax=20,
xlabel style={font=\color{white!15!black}},
ylabel style={font=\color{white!15!black}},
xlabel={Delay to inter-request time ratio $\bar{\tau}_\Delta$},
ylabel={Object hit probability $P_h$},
xlabel style={yshift=0.15cm},
ylabel style={yshift=-0.38cm},
ymin=0.1,
ymax=0.9,
axis background/.style={fill=white},
xmajorgrids,
ymajorgrids,
legend style={legend cell align=left, align=left, draw=white!15!black, line width= 0.002\linewidth},
legend pos=outer north east]

\addplot [color=red, mark=square, mark options={solid, red}]
  table[row sep=crcr]{%
0	0.809423046219045\\
1	0.67492541669421\\
2	0.581865827861812\\
3	0.509292277184041\\
4	0.45232200650395\\
5	0.406644327412684\\
6	0.369272977757274\\
7	0.338156694536657\\
8	0.31185750101\\
9	0.289342601941163\\
10	0.269852888651046\\
11	0.252818653203286\\
12	0.237804317101406\\
13	0.224471312309755\\
14	0.212552581268351\\
15	0.201834688922339\\
16	0.192145033174001\\
17	0.183342540990419\\
18	0.175310792921176\\
19	0.167952868926468\\
20	0.161187433753734\\
};
\addlegendentry{MAP}

\addplot [color=blue, dashed, mark=o, mark options={solid, blue}]
  table[row sep=crcr]{%
0	0.792795094532448\\
1	0.683731476750128\\
2	0.566013030148186\\
3	0.496423096576392\\
4	0.42076520183955\\
5	0.382505109862034\\
6	0.355518650996423\\
7	0.309753449156873\\
8	0.278008431272356\\
9	0.270918497700562\\
10	0.245305314256515\\
11	0.249233520694941\\
12	0.225983648441492\\
13	0.207141032192131\\
14	0.2177120592744\\
15	0.186158661216147\\
16	0.170605518650996\\
17	0.155084312723556\\
18	0.176226366888094\\
19	0.13196218702095\\
20	0.146525293817067\\
};
\addlegendentry{Trace}

\end{axis}

\end{tikzpicture}%
}
\caption{Single M/M/M Cache: Object hit probability ${P_h}$ for increasing ratio $\bar{\tau}_\Delta$ of the expected delay to the expected inter-request time. Note that the MAP approach from Sect.~\ref{sec:exact_models} exactly matches the simulation. The analytical approximation for a single cache is accurate.}
\label{Single Cache fig}
\end{figure}
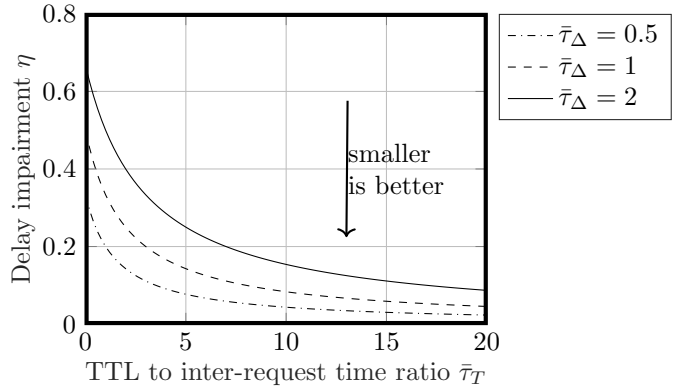
\begin{figure}[t]
\centering
\begin{tikzpicture}

\begin{axis}[%
axis line style = thick,
width=2.1in,
height=1.625in,
at={(0in,0in)},
scale only axis,
xmin=0,
xmax=20,
xlabel style={font=\color{white!15!black}},
xlabel={TTL to inter-request time ratio $\bar{\tau}_T$},
xlabel style={yshift=0.15cm},
ylabel style={yshift=-0.4cm},
ymin=0,
ymax=0.8,
ylabel style={font=\color{white!15!black}},
ylabel={Delay impairment $\eta$},
axis background/.style={fill=white},
xmajorgrids,
ymajorgrids,
legend style={legend cell align=left, align=left, draw=white!15!black},
legend pos=outer north east
]
\draw[line width=0.005 \linewidth](current axis.south west)rectangle(current axis.north east);

\addplot [color=black, dashdotted, mark options={solid, black},mark size=2.2pt]
  table[row sep=crcr]{%
0	0.333333333333333\\
0.2	0.294117647058824\\
0.4	0.263157894736842\\
0.6	0.238095238095238\\
0.8	0.217391304347826\\
1	0.2\\
1.2	0.185185185185185\\
1.4	0.172413793103448\\
1.6	0.161290322580645\\
1.8	0.151515151515152\\
2	0.142857142857143\\
2.2	0.135135135135135\\
2.4	0.128205128205128\\
2.6	0.121951219512195\\
2.8	0.116279069767442\\
3	0.111111111111111\\
3.2	0.106382978723404\\
3.4	0.102040816326531\\
3.6	0.0980392156862745\\
3.8	0.0943396226415094\\
4	0.0909090909090909\\
4.2	0.087719298245614\\
4.4	0.0847457627118644\\
4.6	0.0819672131147541\\
4.8	0.0793650793650794\\
5	0.0769230769230769\\
5.2	0.0746268656716418\\
5.4	0.072463768115942\\
5.6	0.0704225352112676\\
5.8	0.0684931506849315\\
6	0.0666666666666667\\
6.2	0.0649350649350649\\
6.4	0.0632911392405063\\
6.6	0.0617283950617284\\
6.8	0.0602409638554217\\
7	0.0588235294117647\\
7.2	0.0574712643678161\\
7.4	0.0561797752808989\\
7.6	0.0549450549450549\\
7.8	0.0537634408602151\\
8	0.0526315789473684\\
8.2	0.0515463917525773\\
8.4	0.0505050505050505\\
8.6	0.0495049504950495\\
8.8	0.0485436893203883\\
9	0.0476190476190476\\
9.2	0.0467289719626168\\
9.4	0.0458715596330275\\
9.6	0.045045045045045\\
9.8	0.0442477876106195\\
10	0.0434782608695652\\
10.2	0.0427350427350427\\
10.4	0.0420168067226891\\
10.6	0.0413223140495868\\
10.8	0.040650406504065\\
11	0.04\\
11.2	0.0393700787401575\\
11.4	0.0387596899224806\\
11.6	0.0381679389312977\\
11.8	0.037593984962406\\
12	0.037037037037037\\
12.2	0.0364963503649635\\
12.4	0.0359712230215827\\
12.6	0.0354609929078014\\
12.8	0.034965034965035\\
13	0.0344827586206897\\
13.2	0.0340136054421769\\
13.4	0.0335570469798658\\
13.6	0.033112582781457\\
13.8	0.0326797385620915\\
14	0.032258064516129\\
14.2	0.0318471337579618\\
14.4	0.0314465408805031\\
14.6	0.031055900621118\\
14.8	0.0306748466257669\\
15	0.0303030303030303\\
15.2	0.029940119760479\\
15.4	0.029585798816568\\
15.6	0.0292397660818713\\
15.8	0.0289017341040462\\
16	0.0285714285714286\\
16.2	0.0282485875706215\\
16.4	0.0279329608938547\\
16.6	0.0276243093922652\\
16.8	0.0273224043715847\\
17	0.027027027027027\\
17.2	0.0267379679144385\\
17.4	0.0264550264550265\\
17.6	0.0261780104712042\\
17.8	0.0259067357512953\\
18	0.0256410256410256\\
18.2	0.0253807106598985\\
18.4	0.0251256281407035\\
18.6	0.0248756218905473\\
18.8	0.0246305418719212\\
19	0.024390243902439\\
19.2	0.0241545893719807\\
19.4	0.0239234449760766\\
19.6	0.023696682464455\\
19.8	0.0234741784037559\\
20	0.0232558139534884\\
};
\addlegendentry{$\bar{\tau}_\Delta=0.5$}
\addplot [color=black,dashed, mark options={solid, black},mark size=2.2pt]
  table[row sep=crcr]{%
0	0.5\\
0.2	0.454545454545455\\
0.4	0.416666666666667\\
0.6	0.384615384615385\\
0.8	0.357142857142857\\
1	0.333333333333333\\
1.2	0.3125\\
1.4	0.294117647058824\\
1.6	0.277777777777778\\
1.8	0.263157894736842\\
2	0.25\\
2.2	0.238095238095238\\
2.4	0.227272727272727\\
2.6	0.217391304347826\\
2.8	0.208333333333333\\
3	0.2\\
3.2	0.192307692307692\\
3.4	0.185185185185185\\
3.6	0.178571428571429\\
3.8	0.172413793103448\\
4	0.166666666666667\\
4.2	0.161290322580645\\
4.4	0.15625\\
4.6	0.151515151515152\\
4.8	0.147058823529412\\
5	0.142857142857143\\
5.2	0.138888888888889\\
5.4	0.135135135135135\\
5.6	0.131578947368421\\
5.8	0.128205128205128\\
6	0.125\\
6.2	0.121951219512195\\
6.4	0.119047619047619\\
6.6	0.116279069767442\\
6.8	0.113636363636364\\
7	0.111111111111111\\
7.2	0.108695652173913\\
7.4	0.106382978723404\\
7.6	0.104166666666667\\
7.8	0.102040816326531\\
8	0.1\\
8.2	0.0980392156862745\\
8.4	0.0961538461538462\\
8.6	0.0943396226415094\\
8.8	0.0925925925925926\\
9	0.0909090909090909\\
9.2	0.0892857142857143\\
9.4	0.087719298245614\\
9.6	0.0862068965517241\\
9.8	0.0847457627118644\\
10	0.0833333333333333\\
10.2	0.0819672131147541\\
10.4	0.0806451612903226\\
10.6	0.0793650793650794\\
10.8	0.078125\\
11	0.0769230769230769\\
11.2	0.0757575757575758\\
11.4	0.0746268656716418\\
11.6	0.0735294117647059\\
11.8	0.072463768115942\\
12	0.0714285714285714\\
12.2	0.0704225352112676\\
12.4	0.0694444444444444\\
12.6	0.0684931506849315\\
12.8	0.0675675675675676\\
13	0.0666666666666667\\
13.2	0.0657894736842105\\
13.4	0.0649350649350649\\
13.6	0.0641025641025641\\
13.8	0.0632911392405063\\
14	0.0625\\
14.2	0.0617283950617284\\
14.4	0.0609756097560976\\
14.6	0.0602409638554217\\
14.8	0.0595238095238095\\
15	0.0588235294117647\\
15.2	0.0581395348837209\\
15.4	0.0574712643678161\\
15.6	0.0568181818181818\\
15.8	0.0561797752808989\\
16	0.0555555555555556\\
16.2	0.0549450549450549\\
16.4	0.0543478260869565\\
16.6	0.0537634408602151\\
16.8	0.0531914893617021\\
17	0.0526315789473684\\
17.2	0.0520833333333333\\
17.4	0.0515463917525773\\
17.6	0.0510204081632653\\
17.8	0.0505050505050505\\
18	0.05\\
18.2	0.0495049504950495\\
18.4	0.0490196078431373\\
18.6	0.0485436893203883\\
18.8	0.0480769230769231\\
19	0.0476190476190476\\
19.2	0.0471698113207547\\
19.4	0.0467289719626168\\
19.6	0.0462962962962963\\
19.8	0.0458715596330275\\
20	0.0454545454545455\\
};
\addlegendentry{$\bar{\tau}_\Delta=1$}
\addplot [color=black, mark options={solid, black},mark size=2.2pt]
  table[row sep=crcr]{%
0	0.666666666666667\\
0.2	0.625\\
0.4	0.588235294117647\\
0.6	0.555555555555556\\
0.8	0.526315789473684\\
1	0.5\\
1.2	0.476190476190476\\
1.4	0.454545454545455\\
1.6	0.434782608695652\\
1.8	0.416666666666667\\
2	0.4\\
2.2	0.384615384615385\\
2.4	0.37037037037037\\
2.6	0.357142857142857\\
2.8	0.344827586206897\\
3	0.333333333333333\\
3.2	0.32258064516129\\
3.4	0.3125\\
3.6	0.303030303030303\\
3.8	0.294117647058824\\
4	0.285714285714286\\
4.2	0.277777777777778\\
4.4	0.27027027027027\\
4.6	0.263157894736842\\
4.8	0.256410256410256\\
5	0.25\\
5.2	0.24390243902439\\
5.4	0.238095238095238\\
5.6	0.232558139534884\\
5.8	0.227272727272727\\
6	0.222222222222222\\
6.2	0.217391304347826\\
6.4	0.212765957446809\\
6.6	0.208333333333333\\
6.8	0.204081632653061\\
7	0.2\\
7.2	0.196078431372549\\
7.4	0.192307692307692\\
7.6	0.188679245283019\\
7.8	0.185185185185185\\
8	0.181818181818182\\
8.2	0.178571428571429\\
8.4	0.175438596491228\\
8.6	0.172413793103448\\
8.8	0.169491525423729\\
9	0.166666666666667\\
9.2	0.163934426229508\\
9.4	0.161290322580645\\
9.6	0.158730158730159\\
9.8	0.15625\\
10	0.153846153846154\\
10.2	0.151515151515152\\
10.4	0.149253731343284\\
10.6	0.147058823529412\\
10.8	0.144927536231884\\
11	0.142857142857143\\
11.2	0.140845070422535\\
11.4	0.138888888888889\\
11.6	0.136986301369863\\
11.8	0.135135135135135\\
12	0.133333333333333\\
12.2	0.131578947368421\\
12.4	0.12987012987013\\
12.6	0.128205128205128\\
12.8	0.126582278481013\\
13	0.125\\
13.2	0.123456790123457\\
13.4	0.121951219512195\\
13.6	0.120481927710843\\
13.8	0.119047619047619\\
14	0.117647058823529\\
14.2	0.116279069767442\\
14.4	0.114942528735632\\
14.6	0.113636363636364\\
14.8	0.112359550561798\\
15	0.111111111111111\\
15.2	0.10989010989011\\
15.4	0.108695652173913\\
15.6	0.10752688172043\\
15.8	0.106382978723404\\
16	0.105263157894737\\
16.2	0.104166666666667\\
16.4	0.103092783505155\\
16.6	0.102040816326531\\
16.8	0.101010101010101\\
17	0.1\\
17.2	0.099009900990099\\
17.4	0.0980392156862745\\
17.6	0.0970873786407767\\
17.8	0.0961538461538462\\
18	0.0952380952380952\\
18.2	0.0943396226415094\\
18.4	0.0934579439252336\\
18.6	0.0925925925925926\\
18.8	0.0917431192660551\\
19	0.0909090909090909\\
19.2	0.0900900900900901\\
19.4	0.0892857142857143\\
19.6	0.0884955752212389\\
19.8	0.087719298245614\\
20	0.0869565217391304\\
};
\addlegendentry{$\bar{\tau}_\Delta=2$}
%\node[label={0:{(2,0.4)}},circle,fill,inner sep=2pt] at (axis cs:2,0.4) {};
%\node[label={0:{(2,0.75)}},circle,fill,inner sep=2pt] at (axis cs:2,0.75) {};
 \node[anchor=west] (source) at (axis cs:12.6,0.6){};
       \node[] (destination) at (axis cs: 13,0.2){ };
       \node [ align=left] at (axis cs: 15.5,0.4){smaller \\is better };
       \draw[->, black, thick](source)--(destination);

\end{axis}

\end{tikzpicture}%
\caption{Single M/M/M cache: The delay impairment $\eta \in [0,1]$ for different TTL to  inter-request time ratios $\bar{\tau}_T$. Note that compensating the impact of the delay on the hit probability using increased object TTL has diminishing returns.}
\label{TTL effect on delay}
\vspace{10pt}
\end{figure}
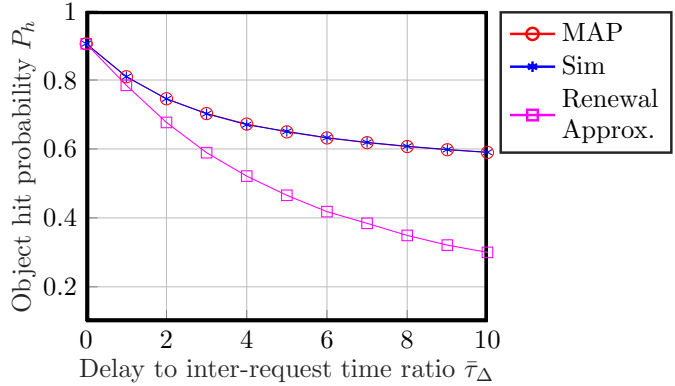
\begin{figure}[t]
\centering
\definecolor{mycolor1}{rgb}{1.00000,0.00000,1.00000}%
\begin{tikzpicture}

\begin{axis}[%
axis line style = thick,
width=2.1in,
height=1.625in,
at={(0in,0in)},
scale only axis,
xmin=0,
xmax=10,
xlabel style={font=\color{white!15!black}},
xlabel={Delay to inter-request time ratio $\bar{\tau}_\Delta$},
ymin=0.1,
ymax=1,
xlabel style={yshift=0.15cm},
ylabel style={yshift=-0.4cm},
ylabel style={font=\color{white!15!black}},
ylabel={Object hit probability $P_h$},
legend style={legend cell align=left, align=left, draw=white!15!black, line width= 0.002\linewidth},
axis background/.style={fill=white},
axis x line*=bottom,
axis y line*=left,
xmajorgrids,
ymajorgrids,
legend style={legend cell align=left, align=left, draw=white!15!black},
legend pos=outer north east
]
\draw[line width=0.005 \linewidth](current axis.south west)rectangle(current axis.north east);

\addplot [color=red, mark=o, mark options={solid, red},mark size= 2.3 pt]
  table[row sep=crcr]{%
0	0.905113530948298\\
1	0.80949260052494\\
2	0.745665304679627\\
3	0.702647546286501\\
4	0.672129866968702\\
5	0.649471499551127\\
6	0.632023100243996\\
7	0.618189835091452\\
8	0.606961627982455\\
9	0.597670114613988\\
10	0.58985635794757\\
11	0.583195191660367\\
12	0.577449987861042\\
13	0.57244454367396\\
14	0.568045005945132\\
15	0.564147900533655\\
16	0.560671995334068\\
17	0.557552640385452\\
18	0.554737749486093\\
19	0.552184894447292\\
20	0.549859168966039\\
};
\addlegendentry{MAP}

\addplot [color=blue, mark=asterisk, mark options={solid, blue}]
  table[row sep=crcr]{%
0	0.905373055892694\\
1	0.809575956256804\\
2	0.745195380438183\\
3	0.702130854218541\\
4	0.670409529410293\\
5	0.650802167140051\\
6	0.632598691207567\\
7	0.619064826516185\\
8	0.607858260031781\\
9	0.59832034133707\\
10	0.590475594649717\\
11	0.585653833160246\\
12	0.578330601468951\\
13	0.574217539820573\\
14	0.567877779307066\\
15	0.5684108056706\\
16	0.562498467236396\\
17	0.558493512343439\\
18	0.551812915253813\\
19	0.553834911975576\\
20	0.550124748188836\\
};
\addlegendentry{Sim}

\addplot [color=mycolor1, mark=square, mark options={solid, mycolor1}]
  table[row sep=crcr]{%
0	0.905114638447972\\
1	0.784614928686515\\
2	0.676911015204863\\
3	0.589158105823818\\
4	0.52120197395012\\
5	0.465561625676368\\
6	0.41815204313774\\
7	0.384253228119195\\
8	0.349069268414663\\
9	0.321043207952258\\
10	0.29992402924879\\
11	0.281290042094229\\
12	0.261445785872179\\
13	0.247416419670185\\
14	0.227664705581083\\
15	0.218499660981943\\
16	0.206391956322356\\
17	0.194172666403634\\
18	0.19291686901183\\
19	0.177884583176477\\
20	0.173629265832381\\
};
\addlegendentry{Renewal \\ Approx.}

\end{axis}

\end{tikzpicture}%
\caption{Two level M/M/M binary caching tree: The object hit probability $P_h$ of the MAP approach and the simulation coincide. State-of-the-art approximations deviate significantly under object fetch delays.}
\label{2-lvl-MMM}
\vspace{10pt}
\end{figure}

 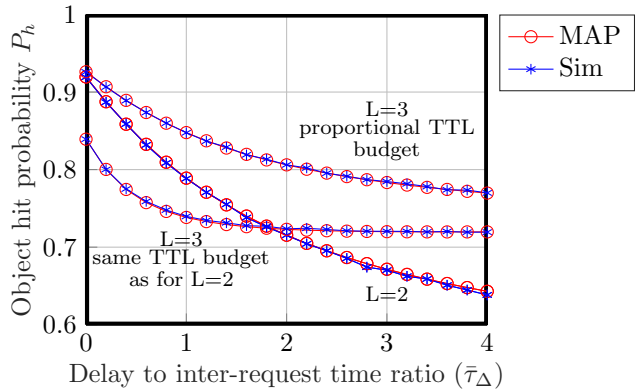
\begin{figure}[t]
\definecolor{mycolor1}{rgb}{1.00000,0.00000,1.00000}%
\begin{tikzpicture}

\begin{axis}[%
axis line style = thick,
width=2.1in,
height=1.625in,
at={(0in,0in)},
scale only axis,
xmin=0,
xmax=4,
xlabel style={font=\color{white!15!black}},
xlabel={Delay to inter-request time ratio ($\bar{\tau}_\Delta$)},
xlabel style={yshift=0.15cm},
ylabel style={yshift=-0.4cm},
ymin=0.6,
ymax=1,
ylabel style={font=\color{white!15!black}},
ylabel={Object hit probability $P_h$},
axis background/.style={fill=white},
axis x line*=bottom,
axis y line*=left,
xmajorgrids,
ymajorgrids,
legend style={legend cell align=left, align=left, draw=white!15!black},
legend pos=outer north east
]
\draw[line width=0.005 \linewidth](current axis.south west)rectangle(current axis.north east);

\addplot [color=red, mark=o, mark options={solid, red},mark size= 2.3 pt]
  table[row sep=crcr]{%
0	0.919504505860105\\
0.2	0.887521374568563\\
0.4	0.858521294808042\\
0.6	0.832573021246023\\
0.8	0.809491858148913\\
1	0.788995653084095\\
1.2	0.770781279803188\\
1.4	0.754559845938965\\
1.6	0.74007082810852\\
1.8	0.727085868758921\\
2	0.715407787932268\\
2.2	0.704867617205718\\
2.4	0.695321034018166\\
2.6	0.686644840717127\\
2.8	0.678733761438649\\
3	0.671497646582217\\
3.2	0.664859087924587\\
3.4	0.658751409837024\\
3.6	0.653116988958217\\
3.8	0.647905853271596\\
4	0.643074515306568\\
};
\addlegendentry{MAP}

\addplot [color=blue, mark=asterisk, mark options={solid,blue }]
  table[row sep=crcr]{%
0	0.919185183182588\\
0.2	0.886907318792474\\
0.4	0.858819889396547\\
0.6	0.831813862579766\\
0.8	0.808647816217634\\
1	0.788469645129733\\
1.2	0.770733641533068\\
1.4	0.7553076340013\\
1.6	0.738087299227097\\
1.8	0.726094744884115\\
2	0.715420900908478\\
2.2	0.704260425772227\\
2.4	0.6954530025443\\
2.6	0.685507102712218\\
2.8	0.674262518486477\\
3	0.670361458812079\\
3.2	0.662459209594845\\
3.4	0.658965678484995\\
3.6	0.650970308490112\\
3.8	0.644742230673183\\
4	0.638409016494393\\
};
\addlegendentry{Sim}

\addplot [color=red, mark=o, mark options={solid, red},mark size= 2.3 pt]
  table[row sep=crcr]{%
0	0.919504505860105\\
0.2	0.887521374568563\\
0.4	0.858521294808042\\
0.6	0.832573021246023\\
0.8	0.809491858148913\\
1	0.788995653084095\\
1.2	0.770781279803188\\
1.4	0.754559845938965\\
1.6	0.74007082810852\\
1.8	0.727085868758921\\
2	0.715407787932268\\
2.2	0.704867617205718\\
2.4	0.695321034018166\\
2.6	0.686644840717127\\
2.8	0.678733761438649\\
3	0.671497646582217\\
3.2	0.664859087924587\\
3.4	0.658751409837024\\
3.6	0.653116988958217\\
3.8	0.647905853271596\\
4	0.643074515306568\\
};
%\addlegendentry{MAP}

\addplot [color=blue, mark=asterisk, mark options={solid,blue }]
  table[row sep=crcr]{%
0	0.919185183182588\\
0.2	0.886907318792474\\
0.4	0.858819889396547\\
0.6	0.831813862579766\\
0.8	0.808647816217634\\
1	0.788469645129733\\
1.2	0.770733641533068\\
1.4	0.7553076340013\\
1.6	0.738087299227097\\
1.8	0.726094744884115\\
2	0.715420900908478\\
2.2	0.704260425772227\\
2.4	0.6954530025443\\
2.6	0.685507102712218\\
2.8	0.674262518486477\\
3	0.670361458812079\\
3.2	0.662459209594845\\
3.4	0.658965678484995\\
3.6	0.650970308490112\\
3.8	0.644742230673183\\
4	0.638409016494393\\
};
%\addlegendentry{Sim}

\addplot [color=red, mark=o, mark options={solid, red},mark size= 2.3 pt]
  table[row sep=crcr]{%
0	0.839221387427233\\
0.2	0.800238482989351\\
0.4	0.774419047953822\\
0.6	0.757425254448949\\
0.8	0.746074852263066\\
1	0.7383373843687\\
1.2	0.732956115213961\\
1.4	0.729148319692803\\
1.6	0.726416778556823\\
1.8	0.724437950338973\\
2	0.722996103542177\\
2.2	0.721944031743743\\
2.4	0.721179141523664\\
2.6	0.72062855842289\\
2.8	0.720239631333945\\
3	0.71997374149552\\
3.2	0.719802180380906\\
3.4	0.719703351282257\\
3.6	0.719660835351739\\
3.8	0.719662033085573\\
4	0.719697195706067\\
};
%\addlegendentry{MAP}

\addplot [color=blue, mark=asterisk, mark options={solid,blue }]
  table[row sep=crcr]{%
0	0.839274726464086\\
0.2	0.800188503775086\\
0.4	0.77502510377228\\
0.6	0.759024831188406\\
0.8	0.747853903060355\\
1	0.739673380380776\\
1.2	0.734542890872248\\
1.4	0.731403165580674\\
1.6	0.727725788639265\\
1.8	0.726912546674978\\
2	0.723679621749788\\
2.2	0.72438463188149\\
2.4	0.722788212350856\\
2.6	0.721123646851556\\
2.8	0.720524363185944\\
3	0.720444692531402\\
3.2	0.720333454259023\\
3.4	0.719686067556706\\
3.6	0.720228228866231\\
3.8	0.719302746482966\\
4	0.719365881718641\\
};
%\addlegendentry{data1}

\addplot [color=red, mark=o, mark options={solid, red},mark size= 2.3 pt]
  table[row sep=crcr]{%
0	0.926185045679688\\
0.2	0.906832390012559\\
0.4	0.889269791522832\\
0.6	0.87362392880897\\
0.8	0.859829229912878\\
1	0.847727090686001\\
1.2	0.837125815348745\\
1.4	0.827832801577359\\
1.6	0.819669791192601\\
1.8	0.81247870600593\\
2	0.806122631169899\\
2.2	0.800484517330158\\
2.4	0.795464975064734\\
2.6	0.790979856967678\\
2.8	0.786957954206733\\
3	0.783338941147107\\
3.2	0.780071604472823\\
3.4	0.777112347403362\\
3.6	0.77442394047241\\
3.8	0.771974484614382\\
4	0.769736552943892\\
};
%\addlegendentry{MAP}

\addplot [color=blue, mark=asterisk, mark options={solid, blue}]
  table[row sep=crcr]{%
0	0.925928895628629\\
0.2	0.906502306985704\\
0.4	0.888950974236029\\
0.6	0.873657265163288\\
0.8	0.859931614799788\\
1	0.847534479501228\\
1.2	0.83671442338387\\
1.4	0.828991395973161\\
1.6	0.819111473489083\\
1.8	0.81340702724834\\
2	0.806147551868967\\
2.2	0.802169022596783\\
2.4	0.795389135643343\\
2.6	0.791413112057815\\
2.8	0.787077271868506\\
3	0.784599147765855\\
3.2	0.781676514850453\\
3.4	0.778066321516682\\
3.6	0.773950981753089\\
3.8	0.773610208367891\\
4	0.769969445656918\\
};
%\addlegendentry{data1}
\node [ align=left, font=\fontsize{11}{7}] at (axis cs: 3.0,0.85){$\substack{\text{L=3} \\ \text{proportional TTL} \\ \text{budget}}$};
\node [ align=left, font=\fontsize{11}{7}] at (axis cs: 0.95,0.685){$\substack{\text{L=3} \\ \text{same TTL budget} \\ \text{as for L=2}}$};
\node [ align=left, font=\fontsize{8}{7}] at (axis cs: 3,0.64){$\text{L=2}$};

\end{axis}
\end{tikzpicture}%
      \centering
       \caption{Binary M/M/$\E_{2}$ \textbf{caching trees} with $2^L\!\!-\!\!1$ caches: Increasing the depth $L$ of the caching tree under the same TTL budget compensates for large delays. For zero or small delays the TTL overlap is, however, detrimental.}
        \label{fig:add_one_more_level}
\end{figure}
\begin{figure}[t]
\centering
  \captionsetup[subfigure]{oneside,margin={1cm,0cm}}
%      \hspace{1cm}
      \begin{subfigure}[b]{0.4\textwidth}
            \centering
\definecolor{mycolor1}{rgb}{1.00000,0.00000,1.00000}%
\begin{tikzpicture}

\begin{axis}[%
axis line style = thick,
width=2.1in,
height=1.625in,
at={(0in,0in)},
scale only axis,
xmin=0,
xmax=2,
xlabel style={font=\color{white!15!black}},
xlabel={Delay to inter-request time ratio $\bar{\tau}_\Delta$},
xlabel style={yshift=0.15cm},
ylabel style={yshift=-0.4cm},
ymin=0.4,
ymax=0.7,
ylabel style={font=\color{white!15!black}},
ylabel={Object hit probability $P_h$},
axis background/.style={fill=white},
extra x ticks={0.25},
extra x tick labels={$\bar{\tau}_{\delta^*}$},
xmajorgrids,
ymajorgrids,
legend style={nodes={scale=0.7}},
legend style={legend cell align=left, align=left, draw=white!15!black, line width= 0.002\linewidth}
]
\draw[line width=0.005 \linewidth](current axis.south west)rectangle(current axis.north east);
\draw[dashed] (axis cs: 0.25,0.4) -- (axis cs: 0.25,0.631416998107195);
\draw[dashed] (axis cs: 0.55,0.4) -- (axis cs: 0.55,0.606111503151209);

\addplot [color=red, mark=o, mark options={solid, red},mark size= 2.3 pt]
  table[row sep=crcr]{%
0	0.610272132062742\\
0.05	0.616275084601769\\
0.1	0.62228344977465\\
0.15	0.627428989355102\\
0.2	0.630587746327697\\
0.25	0.631416998107195\\
0.3	0.630163647897025\\
0.35	0.627253998970873\\
0.4	0.623096661345048\\
0.45	0.618025907818234\\
0.5	0.612299447816774\\
0.55	0.606111503151209\\
0.6	0.599607425762442\\
0.65	0.592895989961519\\
0.7	0.586058827469155\\
0.75	0.579157420488753\\
0.8	0.572238222580898\\
0.85	0.565336403424522\\
0.9	0.558478598103626\\
0.95	0.551684938926475\\
1	0.544970568837178\\
1.05	0.53834677808065\\
1.1	0.531821865024859\\
1.15	0.525401793350541\\
1.2	0.519090697632767\\
1.25	0.512891275088745\\
1.3	0.50680509114724\\
1.35	0.500832819257593\\
1.4	0.494974430139164\\
1.45	0.489229341880362\\
1.5	0.483596539518047\\
1.55	0.478074670675631\\
1.6	0.472662122310015\\
1.65	0.467357082470889\\
1.7	0.462157590109119\\
1.75	0.457061575311285\\
1.8	0.45206689183172\\
1.85	0.447171343403381\\
1.9	0.442372705006121\\
1.95	0.437668740034676\\
2	0.433057214123053\\
};
\addlegendentry{{ MAP $\E_{20}$/M/M}}

\addplot [color=blue, mark=x, mark options={solid, blue}]
  table[row sep=crcr]{%
0	0.610871610871611\\
0.05	0.615613615613616\\
0.1	0.622318622318622\\
0.15	0.628242628242628\\
0.2	0.630559630559631\\
0.25	0.631298631298631\\
0.3	0.630615630615631\\
0.35	0.627312627312627\\
0.4	0.622793622793623\\
0.45	0.618983618983619\\
0.5	0.612774612774613\\
0.55	0.605838605838606\\
0.6	0.598882598882599\\
0.65	0.593621593621594\\
0.7	0.586379586379586\\
0.75	0.578832578832579\\
0.8	0.571145571145571\\
0.85	0.566593566593567\\
0.9	0.556775556775557\\
0.95	0.550977550977551\\
1	0.544131544131544\\
1.05	0.538229538229538\\
1.1	0.532582532582533\\
1.15	0.524453524453524\\
1.2	0.519504519504519\\
1.25	0.512503512503512\\
1.3	0.506602506602507\\
1.35	0.501692501692502\\
1.4	0.494380494380494\\
1.45	0.489319489319489\\
1.5	0.483459483459483\\
1.55	0.478858478858479\\
1.6	0.473247473247473\\
1.65	0.466415466415466\\
1.7	0.462304462304462\\
1.75	0.457367457367457\\
1.8	0.452513452513453\\
1.85	0.447125447125447\\
1.9	0.441522441522442\\
1.95	0.437761437761438\\
2	0.432384432384432\\
};

\addlegendentry{{ Sim $\E_{20}$/M/M}}

\addplot [color=red, mark=square, mark options={solid, red},mark size= 2.3 pt]
  table[row sep=crcr]{%
0	0.666664444451852\\
0.05	0.655737704918033\\
0.1	0.645161290322581\\
0.15	0.634920634920635\\
0.2	0.625\\
0.25	0.615384615384615\\
0.3	0.606060606060606\\
0.35	0.597014925373134\\
0.4	0.588235294117647\\
0.45	0.579710144927536\\
0.5	0.571428571428572\\
0.55	0.563380281690141\\
0.6	0.555555555555555\\
0.65	0.547945205479452\\
0.7	0.540540540540541\\
0.75	0.533333333333333\\
0.8	0.526315789473684\\
0.85	0.519480519480519\\
0.9	0.512820512820513\\
0.95	0.506329113924051\\
1	0.5\\
1.05	0.493827160493827\\
1.1	0.48780487804878\\
1.15	0.481927710843374\\
1.2	0.476190476190476\\
1.25	0.470588235294118\\
1.3	0.465116279069768\\
1.35	0.459770114942529\\
1.4	0.454545454545454\\
1.45	0.449438202247191\\
1.5	0.444444444444444\\
1.55	0.439560439560439\\
1.6	0.434782608695652\\
1.65	0.43010752688172\\
1.7	0.425531914893617\\
1.75	0.421052631578948\\
1.8	0.416666666666667\\
1.85	0.412371134020619\\
1.9	0.408163265306122\\
1.95	0.404040404040404\\
2	0.4\\
};

\addlegendentry{{MAP M/M/M}}
\addplot [color=blue, mark=asterisk, mark options={solid, blue}]
  table[row sep=crcr]{%
0	0.666458666458666\\
0.05	0.655827655827656\\
0.1	0.644606644606645\\
0.15	0.634944634944635\\
0.2	0.625216625216625\\
0.25	0.615240615240615\\
0.3	0.606419606419606\\
0.35	0.597499597499598\\
0.4	0.588180588180588\\
0.45	0.57985257985258\\
0.5	0.570795570795571\\
0.55	0.563065563065563\\
0.6	0.554763554763555\\
0.65	0.547509547509548\\
0.7	0.54031054031054\\
0.75	0.532612532612533\\
0.8	0.527330527330527\\
0.85	0.520713520713521\\
0.9	0.511970511970512\\
0.95	0.506824506824507\\
1	0.500742500742501\\
1.05	0.494578494578495\\
1.1	0.487590487590488\\
1.15	0.482613482613483\\
1.2	0.476733476733477\\
1.25	0.471168471168471\\
1.3	0.465556465556466\\
1.35	0.459316459316459\\
1.4	0.454876454876455\\
1.45	0.449432449432449\\
1.5	0.444892444892445\\
1.55	0.43960743960744\\
1.6	0.435739435739436\\
1.65	0.431081431081431\\
1.7	0.426206426206426\\
1.75	0.421946421946422\\
1.8	0.416775416775417\\
1.85	0.411715411715412\\
1.9	0.407139407139407\\
1.95	0.404104404104404\\
2	0.399034399034399\\
};
\addlegendentry{{ Sim M/M/M}}
\node [ align=right] at (axis cs: 0.65,0.42){$\bar{\tau}_\Delta^+$};
\end{axis}

\end{tikzpicture}%
       \caption{}
        \label{Single MAP no D}
     \end{subfigure}
      %\hspace{0.1cm}
     \begin{subfigure}[b]{0.4 \textwidth}
     \centering
\definecolor{mycolor1}{rgb}{0.00000,0.44700,0.74100}%
\definecolor{mycolor2}{rgb}{0.85000,0.32500,0.09800}%
\begin{tikzpicture}
\pgfplotsset{
    scale only axis,
    xmin=0, xmax=4
}
\begin{axis}[%
axis x line*=bottom,
axis y line*=right,
axis line style = thick,
width=2.1in,
height=1.625in,
at={(0in,0in)},
xlabel style={yshift=0.15cm},
ylabel style={rotate=180,yshift=7.4cm},
xlabel style={font=\color{white!15!black}},
every outer y axis line/.append style={mycolor2},
every y tick label/.append style={font=\color{mycolor2}},
every y tick/.append style={mycolor2},
ymin=0,
ymax=1,
ylabel style={font=\color{mycolor2}},
xlabel={TTL to inter-request time ratio $\bar{\tau}_T$},
ylabel={Ratio of hit probability $\kappa$},
axis background/.style={fill=white},
xmajorgrids,
ymajorgrids
]
\draw[line width=0.005 \linewidth](current axis.south west)rectangle(current axis.north east);
\addplot [color=mycolor2]
  table[row sep=crcr]{%
0.1	0.00737896228856931\\
0.3	0.326255874653656\\
0.5	0.616088623397184\\
0.7	0.766296984956838\\
0.9	0.846552223415543\\
1.1	0.892944486032918\\
1.3	0.921691778794928\\
1.5	0.940588860152415\\
1.7	0.953558096860974\\
1.9	0.962826135601149\\
2.1	0.969647076233254\\
2.3	0.974799534132253\\
2.5	0.978784754521545\\
2.7	0.981915516865219\\
2.9	0.984416244011138\\
3.1	0.986454128530643\\
3.3	0.98811913107787\\
3.5	0.989507504536656\\
3.7	0.990675761693758\\
3.9	0.991655972222929\\
};
 \label{plot_one}
\end{axis}

\begin{axis}[%
  axis y line*=left,
      scale only axis,
      at={(0in,0in)},
  width=1.9in,
  height=1.525in,
  axis x line=none,
  ymin=0,
  ymax=0.8,
ylabel style={yshift=-0.4cm},
legend style={nodes={scale=0.7}},
  every outer y axis line/.append style={mycolor1},
every y tick label/.append style={font=\color{mycolor1}},
every y tick/.append style={mycolor1},
ylabel style={font=\color{mycolor1}},
ylabel={Delay $\bar{\tau}_{\delta^*}$ maximizing $P_h$}
]
%\addlegendimage{/pgfplots/refstyle=plot_one}\addlegendentry{plot 1}
  
\addplot [color=mycolor1]
  table[row sep=crcr]{%
0.1	0.83\\
0.3	0.59\\
0.5	0.46\\
0.7	0.39\\
0.9	0.35\\
1.1	0.31\\
1.3	0.29\\
1.5	0.27\\
1.7	0.26\\
1.9	0.25\\
2.1	0.24\\
2.3	0.23\\
2.5	0.22\\
2.7	0.22\\
2.9	0.21\\
3.1	0.21\\
3.3	0.2\\
3.5	0.2\\
3.7	0.19\\
3.9	0.19\\
4.0 0.19\\
};
%\addlegendentry{plot 2}
\end{axis}
\end{tikzpicture}%
        \caption{}
        \label{Single_phit_deltastar}
     \end{subfigure}
        \caption{(a) Single $\E_{20}$/M/M cache: For E$_k$ input request processes approaching a \textit{periodic behavior}, the delay results into a non-trivial maximum of the object hit probability $P_h$. For $\bar{\tau}_{\Delta} \leq \bar{\tau}_{\Delta}^+$ the object hit probability is larger or equal to $P_h|_{\Delta=0}$ (b) Single $\E_{20}$/M/M cache: For  nearly \textit{periodic requests} the figure shows the delay $\bar{\tau}_{\delta^*}$ that maximizes the hit probability for different TTLs $ \bar{\tau}_{T}$ (cf. Fig.~\ref{Single MAP no D} for comparison).
        %For larger expected  TTL $P_h$ is maximized at lower delays $\bar{\tau}_{\delta^*}$.
        The ratio of the hit probability $P_h$ at zero delay to $P_h$ at the optimum delay is depicted as $\kappa$
        %$ := P_h|_{\Delta=0}/ P_h|_{\Delta=\delta^*}$
        showing that for small expected TTL the hit probability is significantly increased in presence of delays. It can be shown that this is due to the nearly periodic behavior of the request stream.}
        \label{}
    \vspace{10pt}
\end{figure}
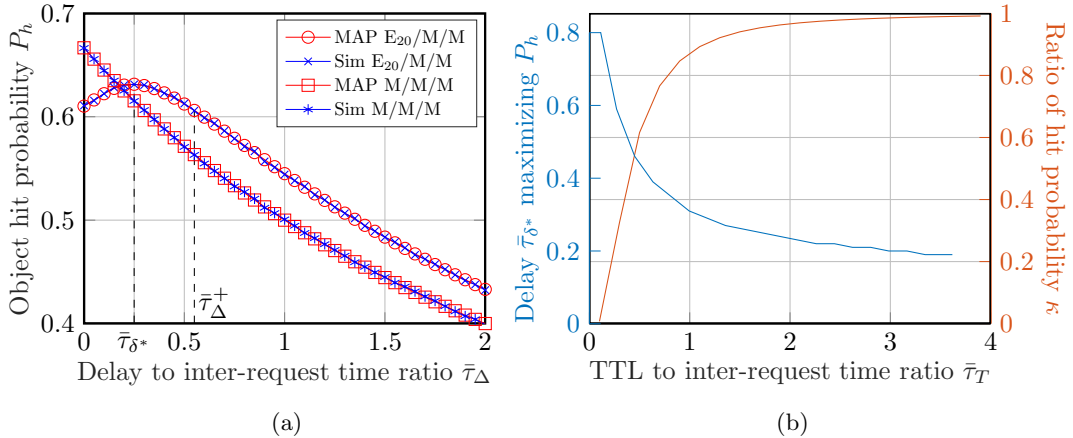

In this section, we show analytical, as well as, simulation-based evaluation results for TTL cache hierarchies under object fetch delays.
%evaluate our proposed approach.
We first show the impact of non-zero fetch delay on the object hit probability using the M/M/M caching system before validating the MAP approach from Sect.~\ref{sec:exact_models} using simulation results.
We then evaluate the accuracy of the approximations from Sect.~\ref{sec:approx_hierarchies} compared to the MAP approach from Sect.~\ref{sec:exact_models}, as well as, the simulation results.
Further, we provide numerical results and comparison with simulations for specific PH/M/M and M/M/PH systems.
Then, in light of the evaluation results we discuss the TTL as means to compensate for the delay impairment.
Finally, we present simulation results using traces obtained from a network file system in a cluster deployment~\cite{Trace}.
The traces contain the request times for file related events in a production storage system.
% numerically show the influence of changing the TTL on the effect of the delay. Then we show the accuracy of our approximation in \ref{eq:Approx}-\ref{eq:approx2}.
%
\subsection{Impact of network delays on the object hit probability}
If not stated otherwise, we use in the following full binary tree caching hierarchies of $L$ levels with $L \in \{1,2,3\}$, i.e. the hierarchies possess $2^{L}-1$ caches.
Note that the provided analytical methods hold for any tree caching hierarchy.
For the illustration of the following results we repeat the definitions from Sect.\ref{subsec:approx_single_cache} of the ratio of the expected delay $\E[\Delta]$ to the expected inter-request time $\E[X]$, i.e., $\bar{\tau}_\Delta:=\E[\Delta] / \E[X]$, as well as, the the ratio of the expected TTL $\E[T]$ to $\E[X]$ defined as $\bar{\tau}_T := \E[T]/\E[X]$.
% In general, we show hit probabilities for any chosen single object in a TTL cache.

%\begin{figure}[t]
%\input{Figs/Evaluations/Three_level_tree_Erlang.tex}
%\caption{Binary $M/M/E_{2}$ caching trees with $2^L\!\!-\!\!1$ caches: Increasing the depth %of the caching tree under the same TTL budget compensates for large delays. For zero or small %delays the TTL overlap is detrimental.}
%\label{fig:add_one_more_level}
%\end{figure}

Fig.~\ref{Single Cache fig} shows a comparison of the object hit probability $P_h$ at an M/M/M single cache from simulations, the MAP as well as the approximations~(\ref{eq:Approx}-\ref{Approx2}). The object hit probability declines with increasing delay for a given expected inter-request time.
%s ratio $\bar{\tau}_\Delta$ is varied on the x-axis.
The ratio  of the TTL to the inter-request time is fixed to $\bar{\tau}_T=2$.	
The figure shows that our MAP approach, as well as, the analytical approximation coincide with the simulation.
Moreover, an increasing fetch delay (even of the order of the expected inter-request time $\mathsf{E}[X]$) is shown to significantly decrease the hit probability.
%Qualitatively, the figure shows how the object hit probability for a single M/M/M cache diminishes with the object fetch delay.
%The figure shows that as $\bar{\tau}_\Delta  \rightarrow \infty$, Ph $\rightarrow 0$.
%From the figure, we can see that when $\bar{\tau}_\Delta=2$, $Ph$ decreases from $2/3$ to $0.4$ having a delay impairment $\eta=0.4$ which is very significant.
%at $\bar{\tau}_\Delta=1$ the delay impairment is $\eta=0.25$.

In Lem.~\ref{lem:delay_impairment} we showed that the TTL can counteract the delay impairment, however, at a diminishing impact.
%and a large increase in the TTL results only in a small change in the hit Pr.
Fig.~\ref{TTL effect on delay} illustrates this for a single M/M/M cache for given expected delays.
Note that for a given expected TTL doubling the expected delay leads to an over-proportional increase in delay impairment $\eta$.

Fig.~\ref{2-lvl-MMM} depicts the decrease in object hit probability with increasing object fetch delay for a two level binary tree cache hierarchy under the M/M/M setting.
The figure shows results obtained by simulation that match our MAP approach in comparison to the renewal approximation.
We set the expected TTL as $\bar{\tau}_T\!\!=\!\!2$ for the first level containing the leaf caches and $\bar{\tau}_T\!\!=\!\!4$ for the root cache.
In contrast to Fig.~\ref{Single Cache fig}
%which shows that the analytical approximation for a \textit{single cache} under delay is accurate,
this result shows that the approximation deteriorates quickly for hierarchies.
%The MAP approach still coincides with the simulation.

% The figure shows that our MAP approach almost have the same result as the simulation in the multiple case scenario.
% On the other hand, the renewal approximation is significantly inaccurate especially in the presence of delay.

%\begin{figure}[t]
%\input{Figs/Evaluations/E1vsE20_input.tex}
%\caption{Single $E_{k}/M/M$ cache: For Erlang$_k$ input request processes approaching a periodic behavior the delay results into a non-trivial optimum of the object hit probability $P_h$ for small TTL. For larger TTL this effect vanishes.}
%\label{Single MAP no D}
%\end{figure}

Fig.~\ref{fig:add_one_more_level} illustrates the gain of adding one level of leaf caches to a hierarchy in terms of the object hit probability.
We compare an $L\!\!=\!\!2$-level caching binary tree with an $L\!\!=\!\!3$-level binary tree.
We keep the aggregate input request process to the system fixed for comparison.
Obviously, increasing the TTL budget proportionally to the additional level of caches results in a hit probability curve ($L\!\!=\!\!3$-proportional in Fig.~\ref{fig:add_one_more_level}) that dom\-i\-nates the $L\!\!=\!\!2$ curve.
More interestingly, is how the behavior shifts when we keep the overall \textit{system TTL budget} fixed and distribute it evenly.
%, i.e., the caches in both systems possess the same aggregate TTL.
As delays are per link random variables an additional level of caches stochastically increases the object fetch delay.
In Fig.~\ref{fig:add_one_more_level}  we observe that an extra level of leaf caches compensates for relatively large object fetch delays (albeit the fixed system TTL budget).
For zero or small delays under fixed system TTL budget the added layer of caches is detrimental as the (per cache) smaller TTLs overlap between children and parent caches. %while the link delays accumulate.

%(not shown for clarity in Fig.~\ref{})

Fig.~\ref{Single MAP no D} shows the object hit probability $P_h$ using $\bar{\tau}_T=2$ for increasing delay for an $\E_{20}$/M/M cache in comparison to an M/M/M cache.
Here, we sample the input inter-request times from an Erlang E$_k$ distribution with a fixed mean such that for larger $k$ the request process becomes periodic.\footnote{A special case of $\Gamma(\alpha,\alpha\beta)$ distributed inter-request times with $\alpha\rightarrow\infty$.}
Interestingly, the figure shows that
%for a given ratio of the expected TTL  to the inter-request times $\bar{\tau}_T$
%\todo[inline]{@Karim: what is the TTL here?}
%and
for request streams becoming periodic there is a non-trivial optimum of the object hit probability at an expected delay denoted $\bar{\tau}_{\delta^*}$.
To give an intuition for this behavior, consider deterministic inter-arrival time $x$, delay $\delta$ and TTL $t$.
In case of zero delay and if $x>t$ each request results in a miss thus zero hit probability.
%$0$ hit probability as shown in Fig.\ref{fig:good_delay_0}.
Having a delay that shifts the TTL start such that $\delta<x<\delta+t$ results in a positive hit probability.
Fig.~\ref{fig:good_delay_0} illustrates this in the appendix.

%For the case of Erlang arrivals we prove in the appendix
{For the case of periodic arrivals we prove in the appendix}
(Sect.~\ref{appendix:derivation_delay_interval}) that there exists an upper bound $\bar{\tau}_\Delta^+$ on the expected delay to inter-request ratio $\bar{\tau}_{\Delta}$ as a function of the expected TTL to inter-request ratio $\bar{\tau}_{T}$ below which the object hit probability is higher than the zero delay case {(For example see $\bar{\tau}_\Delta^+$ in Fig.~\ref{Single MAP no D})}.

Accordingly, Fig.~\ref{Single_phit_deltastar} shows for varying expected TTL $\bar{\tau}_{T}$  the corresponding expected delay $\bar{\tau}_{\delta^*}$ which maximizes the hit probability $P_h$. The figure also shows the ratio $\kappa$ of the hit probability for zero delay to the maximum hit probability at $\Delta=\delta^*$, i.e., $\kappa := P_h|_{\Delta=0}/ P_h|_{\Delta=\delta^*}$.
%\todo[inline]{@Karim: put the formula here}
\begin{comment}
Using an argument on the form of $P_h$ one can show that for $\bar{\tau}_{\Delta}\leq -1/ W_{-1}\big(-\bar{\tau}_{T}^{-1} \mathrm{exp}(-\bar{\tau}_{T}^{-1} ) \big)$
the hit probability $P_h$ is higher than $P_h|_{\Delta=0}$. The function $W_{-1}(.)$ is the Lambert-W function.
%using the -1 branch.
\end{comment}
We observe that for request processes approaching periodicity
there is a non trivial impact of the TTL on the hit probability given non-zero object fetch delay, which is more significant for small TTLs.

\subsection{Trace-based simulation results}
Next, we use trace data from a production system~\cite{Trace} as input to a caching scenario with a single cache that is simulated to show the impact of different average delays on the object hit probabilities.
In addition, we compare the simulation results to analytical results obtained from modelling the caching system based on our MAP approach.
%This is analogous to the simulation results above, e.g. Fig.~\ref{Single Cache fig}.

%%%% Estimation of the data
%Our approach as discussed in Section III models input, delay and the TTL distributions of a cache using a MAP.
Given the MAP model of the input to a single cache, in addition to the TTL and delay, we form a cache MAP as in Fig.~\ref{PH_input}.
Hence, in the first step we fit the inter-request times in the trace to a corresponding density function $\hat{f}_X(t)$ that is represented by a MAP.
% to  the modelling of the trace data of an unknown distribution requires fitting the data to a PDF  .
To optimize the estimation of the density, we first prepare the trace data to ensure its consistency.
We remove statistical outliers based on the $z$-score of the samples~\cite{outliers}, where
%
% Trace samples inconsistent with the whole sample set of the trace and have a relatively significant difference are excluded.
% In statistics, these points are called outliers.
% The outliers especially the large valued ones result in the inflation of the mean and the standard deviation of the samples, thus influence the accuracy of the PDF estimation.
% We identify the outliers by judging a statistical score of each sample called $z$-score\cite{outliers}. %
the $z$-score ($z=\frac{s-\mu_s}{\sigma_s}$) denotes the number of standard deviations a sample deviates from the mean of the trace data.
Here $\mu_s$ and $\sigma_s$ are the mean and the standard deviation of the trace data, respectively.
In this work we use a cutoff value of 2 and a power transform to map our trace data such that we can calculate the $z$-score.
%Note that as this method strictly applies to Gaussian densities we utilize a power transform to map our trace data to the Normal distribution.
% For a sample $s$ the $z$-score is calculated as $z=\frac{s-\mu_s}{\sigma_s}$, where .

% A sample being an outlier has the absolute value of its $z$-score above a certain cut-off value that is manually chosen.
% However, the drawback of the $z$-score is its limitation to only Gaussian distributed samples, therefore, it does not directly fit our trace data.
% To avoid such a limitation the $z$-score is indirectly calculated by using a transformation of the trace data instead of the original one.
% The data transformation is based on the power transform function which maps the original data to transformed data that are normally distributed, thus the outliers are identified using $z$-score.
% For the trace samples, we use a cut-off value of 2. As a result, only $5\%$ of the samples are identified as outliers.

\begin{figure}[t]
 \centering
 \input{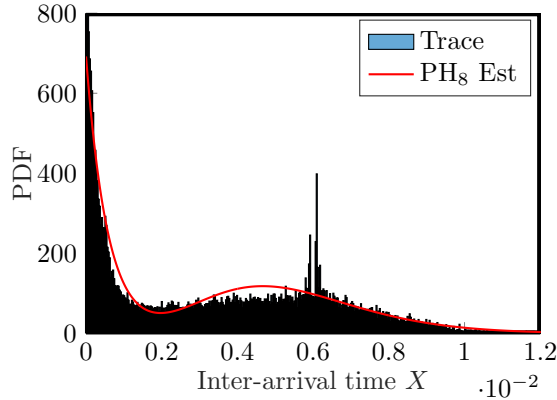}
 \caption{Trace data probability density function estimation: The estimated PDF using PH$_8$ distribution after removing the outliers is very close to the histogram of the trace data.}
 \label{histogram}
 \end{figure}

We carry out a parametric estimation of the density that best represents the trace.
Here, we represent the parameters of the density by $\theta$.
%,i.e., for an exponential distribution, $\theta$ is the rate.
We formulate the parameter estimation problem as a maximum likelihood optimization problem of the form
\begin{equation*}
    \hat{\theta}=  \underset{\theta}{\argmax} f_\Theta(\mathbf{\theta}|\mathcal{T})\;,
\end{equation*}
where $\hat{\theta}$ represents the estimate of the density parameters, $\mathcal{T}$ represents the trace data and $f_\Theta$ is the conditional density of the parameters given the trace.
We apply the expectation maximization (EM) algorithm to solve this optimization problem \cite{PH_modelling}.
We use a PH Coxian distribution $\text{PH}_k$ for the PDF estimation of the trace data.
The Coxian distribution is a generalization of the Erlang distribution where the transition to the absorption state is not limited to the last state $k$, however, it can be reached from any other state~\cite{coxian}.
The larger the number of states used, the more degrees of freedom are available, thus we achieve a better estimation.
%, thus the better the estimation.
On the other hand, the increase in the states comes at the expense of the computational complexity and overfitting.
% \textcolor{red}{Recall in Section \ref{sec:complexity} that the complexity is given by $\mathcal{O}({m_c}^{n_c})$.
% Having more states to represent the input density function, $m_c$ increases.
% There are some tools designed specifically to solve the fitting problem of the data to a MAP.
% An overview about these tools is found in \cite{Tools}.
% }
Several approaches exist to determine the minimum number of phases required for fitting the data \cite{Tools}.
One approach is to use distant metrics such as AIC (Akaike information criterion) \cite{AIC} and BIC (Bayesian information criterion) \cite{BIC} to quantify the distance between the estimation and the data.
Based on these metrics, the minimum number of phases is determined by a tolerance value.
%defined for each application.
Fig.~\ref{histogram} shows the result of the density estimation of the trace data where $\text{PH}_8$ is used.
Although, the histogram of the trace is shown to be non-monotonic, using $\text{PH}_8$ allows for capturing such a behavior.
Here we find $8$ phases enough to capture the statistical characteristics of the data.
% Since the trace data does not exactly coincide with the estimated PDF and due to the presence of the outliers, the result of our MAP approach is expected to have some error to the trace simulation.

\begin{figure}[t]
\centering
\begin{tikzpicture}

\begin{axis}[%
axis line style = thick,
width=2.1in,
height=1.625in,
at={(0in,0in)},
scale only axis,
xmin=0,
xmax=20,
xlabel style={font=\color{white!15!black}},
ylabel style={font=\color{white!15!black}},
xlabel={Delay to inter-request time ratio $\bar{\tau}_\Delta$},
ylabel={Object hit probability $P_h$},
xlabel style={yshift=0.15cm},
ylabel style={yshift=-0.38cm},
ymin=0.1,
ymax=0.9,
axis background/.style={fill=white},
xmajorgrids,
ymajorgrids,
legend style={legend cell align=left, align=left, draw=white!15!black, line width= 0.002\linewidth},
legend pos=outer north east]

\addplot [color=red, mark=square, mark options={solid, red}]
  table[row sep=crcr]{%
0	0.809423046219045\\
1	0.67492541669421\\
2	0.581865827861812\\
3	0.509292277184041\\
4	0.45232200650395\\
5	0.406644327412684\\
6	0.369272977757274\\
7	0.338156694536657\\
8	0.31185750101\\
9	0.289342601941163\\
10	0.269852888651046\\
11	0.252818653203286\\
12	0.237804317101406\\
13	0.224471312309755\\
14	0.212552581268351\\
15	0.201834688922339\\
16	0.192145033174001\\
17	0.183342540990419\\
18	0.175310792921176\\
19	0.167952868926468\\
20	0.161187433753734\\
};
\addlegendentry{MAP}

\addplot [color=blue, dashed, mark=o, mark options={solid, blue}]
  table[row sep=crcr]{%
0	0.792795094532448\\
1	0.683731476750128\\
2	0.566013030148186\\
3	0.496423096576392\\
4	0.42076520183955\\
5	0.382505109862034\\
6	0.355518650996423\\
7	0.309753449156873\\
8	0.278008431272356\\
9	0.270918497700562\\
10	0.245305314256515\\
11	0.249233520694941\\
12	0.225983648441492\\
13	0.207141032192131\\
14	0.2177120592744\\
15	0.186158661216147\\
16	0.170605518650996\\
17	0.155084312723556\\
18	0.176226366888094\\
19	0.13196218702095\\
20	0.146525293817067\\
};
\addlegendentry{Trace}

\end{axis}

\end{tikzpicture}%
%\stackinset{l}{72pt}{t}{3pt}{\smalll}{\biggg}
\caption{Trace-driven simulation of a single cache system with delay: The MAP model accurately provides the object hit probability $P_h$ under varying expected delays.}
\label{Trace sim}
\vspace{10pt}
\end{figure}
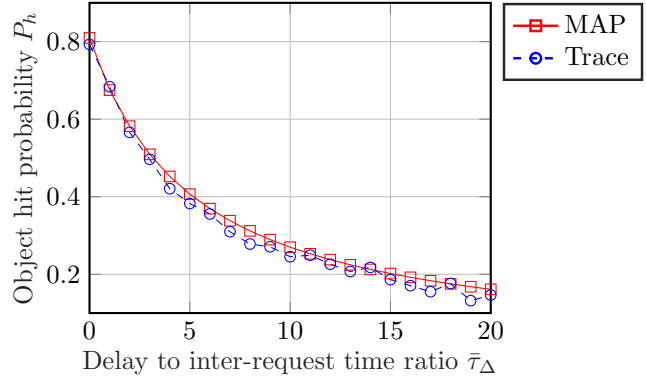

Fig.~\ref{Trace sim} shows the result of our MAP approach using a $\text{PH}_8$/M/M model in comparison to the trace simulation.
Observe that the MAP approach provides a very close result to the trace simulation.
As expected, the object hit probability declines with increasing delay.
The figure shows that if the average delay is in the order of the inter-request times the loss in object hit probability compared to the classical and idealized ''no-delay'' model is substantial.

% , with an error between $10^-4$ and $0.018$ for $\bar{\tau}_\Delta \in [0,1]$.

% \begin{comment}
% In the second step we estimate the distribution to represent the trace samples without the outliers.
% From the histogram of the observations shown in Fig..~\ref{Trace histo}, it is most likely that the observations are exponentially distributed.
% However, for optimal estimation, we use general $\text{Erlang}_k$ and estimate the value of $k$ as well as the rate $\lambda$ of the distribution.
% Maximum likelihood (ML) estimation is used to get the value of $\lambda$ for a defined value of $k$.
% Given the value of $k$, the rate is estimated by \cite{ML:est:Erlang}
% \begin{equation}
%     \hat{\lambda}=\frac{Nk}{\sum_{i=1}^N s_i} \;,
% \end{equation}
% where $N$ is the number of samples.
% In order to estimate the best value of $k$, we use the Kullback-Leibler divergence $D_{KL}$, which measure the difference between two distributions.
% We measure the difference between the samples histogram and the estimated distribution for different values of $k$ after estimating the corresponding $\hat{\lambda}$. The optimal value of $k$ is the one which has the minimum value of $D_{KL}$.
% \end{comment}

\def\biggg{% This file was created by matlab2tikz.
\begin{tikzpicture}

\begin{axis}[%
axis line style = thick,
width=2.1in,
height=1.625in,
at={(0in,0in)},
scale only axis,
xmin=0,
xmax=20,
xlabel style={font=\color{white!15!black}},
ylabel style={font=\color{white!15!black}},
xlabel={Delay to inter-request time ratio $\bar{\tau}_\Delta$},
ylabel={Object hit probability $P_h$},
xlabel style={yshift=0.15cm},
ylabel style={yshift=-0.38cm},
ymin=0.1,
ymax=0.9,
axis background/.style={fill=white},
xmajorgrids,
ymajorgrids,
legend style={legend cell align=left, align=left, draw=white!15!black, line width= 0.002\linewidth},
legend pos=outer north east]

\addplot [color=red, mark=square, mark options={solid, red}]
  table[row sep=crcr]{%
0	0.809423046219045\\
1	0.67492541669421\\
2	0.581865827861812\\
3	0.509292277184041\\
4	0.45232200650395\\
5	0.406644327412684\\
6	0.369272977757274\\
7	0.338156694536657\\
8	0.31185750101\\
9	0.289342601941163\\
10	0.269852888651046\\
11	0.252818653203286\\
12	0.237804317101406\\
13	0.224471312309755\\
14	0.212552581268351\\
15	0.201834688922339\\
16	0.192145033174001\\
17	0.183342540990419\\
18	0.175310792921176\\
19	0.167952868926468\\
20	0.161187433753734\\
};
\addlegendentry{MAP}

\addplot [color=blue, dashed, mark=o, mark options={solid, blue}]
  table[row sep=crcr]{%
0	0.792795094532448\\
1	0.683731476750128\\
2	0.566013030148186\\
3	0.496423096576392\\
4	0.42076520183955\\
5	0.382505109862034\\
6	0.355518650996423\\
7	0.309753449156873\\
8	0.278008431272356\\
9	0.270918497700562\\
10	0.245305314256515\\
11	0.249233520694941\\
12	0.225983648441492\\
13	0.207141032192131\\
14	0.2177120592744\\
15	0.186158661216147\\
16	0.170605518650996\\
17	0.155084312723556\\
18	0.176226366888094\\
19	0.13196218702095\\
20	0.146525293817067\\
};
\addlegendentry{Trace}

\end{axis}

\end{tikzpicture}%
}
\def\smalll{\input{Figs/Trace_single_monte_all_upd}}

\section{Conclusions}
\label{sec:conclusions}
In this paper, we provided an exact model for TTL caches under non-zero object fetch  delays.
We considered tree-based cache hierarchies with random fetch delays, as well as, a wide class of object request processes.
Our analysis provides the object hit probability under delays and allows understanding the impact of delays and TTLs on cache performance.
We provided a rigorous computational speedup method for calculating exact cache performance metrics based on MAP lumpability.
We also show shortcomings of state-of-the-art approximations of the hit probability given object fetch delays.
We provided numerical and trace-based evaluation results validating the accuracy of our model and quantifying the delay impairment, as well as, the benefit of adding caches to a hierarchy under delay constraints. We also discuss the diminishing return of using TTLs as means to compensate object fetch delays.

\section{Appendix}
\label{sec:appendix}

\subsection{Proof of Theorem~\ref{thm:nr_lumpabe_partitions}}
\label{appendix:proof_of_}
Here, we prove the proposition in Theorem~\ref{thm:nr_lumpabe_partitions} that is $ m_\mathcal{S}^n\geq N_p $ for which we use the shorthand notation $P(m_\mathcal{S},n)$. The proof goes by induction which involves three steps.
\begin{proof}
First, we show that the proposition $P(3,1)$ holds:  ${3\choose{2}}= 3^1$.
% \vspace{-10pt}
% \begin{equation*}
%     P(3,1): {3\choose{2}}\geq 3^1
% \end{equation*}
Then we show that $P(m_\mathcal{S},1)$ implies~$P(m_\mathcal{S}+1,1)$.
\begin{align*}
\small
     P(m_\mathcal{S}+1,1):
     &{m_\mathcal{S}+1}\geq{{m_\mathcal{S}+1}\choose{m_\mathcal{S}}} \nonumber \\
     &{m_\mathcal{S}+1}={{m_\mathcal{S}+1}}
\end{align*}
The last step is to show that $P(m_\mathcal{S},n)$ implies $P(m_\mathcal{S},n+1)$
\begin{align*}
\small
     P(m_\mathcal{S},n+1):
     &{m_\mathcal{S}}^{n+1}\geq{{m_\mathcal{S}+n}\choose{m_\mathcal{S}-1}} \nonumber \\
     &{m_\mathcal{S}}^{n+1}\geq\frac{(m_\mathcal{S}+n)!}{(m_\mathcal{S}-1)!(n+1)!} \nonumber \\
     &{m_\mathcal{S}}^{n+1}\geq\frac{(m_\mathcal{S}+n)(m_\mathcal{S}+n-1)!}{(m_\mathcal{S}-1)!(n+1)n!} \nonumber
\end{align*}
Given $P(m_\mathcal{S},n): {m_\mathcal{S}}^{n}\geq\frac{(m_\mathcal{S}+n-1)!}{(m_\mathcal{S}-1)!n!}$, $P(m_\mathcal{S},n+1)$ exists if $m_\mathcal{S}\geq\frac{m_\mathcal{S}+n}{n+1}$, i.e.,
\begin{align*}
    (n+1)m_\mathcal{S}&\geq m_\mathcal{S}+n  \nonumber \\
    m_\mathcal{S} &\geq1
\end{align*}
\end{proof}

\subsection{Proof of Cor.~\ref{cor:nr_lumpabe_partitions_polynomial}}
\label{appendix:derivation_poly_Np}
\begin{proof}
Next, we prove that the number of lumpable partitions grows polynomially with the number of symmetric sub-trees $n$ for a given number of states $m_\mathcal{S}$ of any of the sub-trees.
% \begin{equation}
%     N_p={{n+m_\mathcal{S}-1}\choose{m_\mathcal{S}-1}} \leq \frac{(n+m_\mathcal{S}-1)^{m_\mathcal{S}-1}}{(m_\mathcal{S}-1)!}
% \end{equation}
From Theorem~\ref{thm:nr_lumpabe_partitions} we know that
\begin{align*}
    N_p&={{n+m_\mathcal{S}-1}\choose{m_\mathcal{S}-1}}= \frac{(n+m_\mathcal{S}-1)!}{(m_\mathcal{S}-1)! n!}
    \nonumber \\
    &=\frac{(n+m_\mathcal{S}-1) (n+m_\mathcal{S}-2) .... (n+1) n!}{(m_\mathcal{S}-1)! \ n!}
    \nonumber \\
    &= \frac{\prod_{k=0}^{m_\mathcal{S}-2}(n+m_\mathcal{S}-1-k)}{(m_\mathcal{S}-1)!}
    \;.
\end{align*}
Since $(n+m_\mathcal{S}-1)\geq (n+m_\mathcal{S}-1-k) \ \forall k \in \mathbb{N}$,
\begin{equation*}
    N_p=\frac{\prod_{k=0}^{m_\mathcal{S}-2}(n+m_\mathcal{S}-1-k)}{(m_\mathcal{S}-1)!}\leq \frac{(n+m_\mathcal{S}-1)^{m_\mathcal{S}-1}}{(m_\mathcal{S}-1)!}\;.
\end{equation*}
\end{proof}
\subsection{Proof of Lem.~\ref{lemma:same_trans}}
\label{appendix:proof_of_thm4}
\begin{proof}
Using \eqref{eq:total_trans_element} we can write
\vspace{-5pt}
\begin{align*}
         Q_{A,B}&= \sum_{j=1}^n\left. q_{A_j,B_j}\right|_{A_j \neq B_j} \;,
                \nonumber
                \\
    Q_{A^*,B^*} &=\sum_{j=1}^n \left. q_{A^*_j,B^*_j} \right|_{A^*_j \neq B^*_j} \;.
\end{align*}
\vspace{-5pt}
Together with \eqref{group_action} we have
\begin{equation}
    Q_{A^*,B^*}= \sum_{j=1}^n \left. q_{A_{f^{-1}(j)},B_{f^{-1}(j)}} \right|_{A_{f^{-1}(j)} \neq B_{f^{-1}(j)}} \;.
    \label{equal transitions}
\end{equation}
\noindent
Given that $f$ is any permutation of the $n$ sub-trees, $A^*$ is a permutation of the elements $A_j$ of $A$ according to $f$.
Therefore, $\sum_{j=1}^n \Upsilon(A_j,B_j)= \sum_{j=1}^n\Upsilon(A_{f^{-1}}(j),B_{f^{-1}(j)})$ where $\Upsilon$ represents any non-random test function.
Using that in \eqref{equal transitions} we obtain
\begin{equation*}
    Q_{A^*,B^*}= \sum_{j=1}^n\left. q_{A_j,B_j}\right|_{A_j \neq B_j} = Q_{A,B} \;.
\end{equation*}
\end{proof}
\subsection{Derivation of improved hit probability under delays}
\label{appendix:derivation_delay_interval}
In this part, we derive a bound on the network delay that improves the hit probability for a periodic request process with  inter-request times $x$ and exponentially distributed TTL and exponentially distributed delay with parameters $\lambda_T, \lambda_{\Delta}$, respectively.
In this case, the hit probability improves if for $x>T=t$ the delay shifts the TTL such that $\delta<x<\delta+t$.
The hit probability then improves as
\begin{equation}
    \mathbb{P}(\Delta<x<\Delta+T)\geq\mathbb{P}(x<T)  \;,
    \label{good_delay_ineq}
\end{equation}
with $\mathbb{P}(\Delta<x<\Delta+T)= F_\Delta(x)-F_{\Delta+T}(x)$,  $\mathbb{P}(T>x)=1-F_T(x)$, and finally $F_{\Delta+T}(x)=\int_0^x f_\Delta(t)*f_T(t) dt$.
Now we can compute for the model at hand
\begin{align}
    \mathbb{P}(\Delta<x<\Delta+T)&=\frac{\lambda_\Delta}{\lambda_\Delta-\lambda_T}(\e^{-x\lambda_T }-\e^{-x\lambda_\Delta})
    \nonumber \;
\end{align}
and $\mathbb{P}(x<T) =\e^{-x\lambda_\T}$. Substituting  in \eqref{good_delay_ineq} we obtain
\begin{equation}
\frac{\lambda_\Delta}{\lambda_\Delta-\lambda_T}\e^{-x\lambda_\Delta }\leq\frac{\lambda_T}{\lambda_\Delta-\lambda_T}\e^{-x\lambda_T }
\label{eq:main_ineq}
\end{equation}
Based on $\lambda_\Delta-\lambda_T$, we derive a bound on $\lambda_\Delta$ in two cases:

\noindent\textbf{\textit{Case 1}}: $\mathit{\bar{\tau}_\Delta<\bar{\tau}_T, \text{ i.e.}, \lambda_\Delta>\lambda_T}$ we have
\begin{equation}
-\frac{1}{\bar{\tau}_\Delta}\e^{-1/\bar{\tau}_\Delta}\geq-\frac{1}{\bar{\tau}_T}\e^{-1/\bar{\tau}_T}
\label{eq:reduced_ineq}
\end{equation}
In order to derive the delay interval that satisfies the inequality, we first solve the equation $-\frac{1}{\bar{\tau}_\Delta}\e^{-1/\bar{\tau}_\Delta}=-\frac{1}{\bar{\tau}_T}\e^{-1/\bar{\tau}_T} $.
Let $u=-\frac{1}{\bar{\tau}_\Delta}$ and $h=-\frac{1}{\bar{\tau}_T}\e^{-1/\bar{\tau}_T} $.
For the transcendental equation $u\e^u=h$ we know that
\begin{equation}
u=
\begin{cases}
W_{-1}(h), W_{0}(h) \; \; -1/\e \leq h <0 \\
W_{0}(h) \hspace{1cm} h\geq  0 \;,
\end{cases}
\label{lambert-w_sol}
\end{equation}
where $W_{a}(.)$ is the $a$-th branch of the Lambert-$W$ function.
Note that the possible solutions of \eqref{lambert-w_sol} depend on the value of $h$.
We know that $h=-\frac{1}{\bar{\tau}_T}\e^{-1/\bar{\tau}_T}$ is a convex function over $\bar{\tau}_T \in [0, \infty[$ where $-1/\e\leq h <0$.
Therefore, the solution to the \eqref{lambert-w_sol} is $u=W_{-1}(h),W_{0}(h)$.

Given that $u\e^u$ is decreasing on $u \leq -1$  and $W_{-1}(h)<-1$ the inequality \eqref{eq:reduced_ineq} holds for
\begin{equation}
u\leq W_{-1}(h) .
\label{eq:compact_sol}
\end{equation}
Recall that $\bar{\tau}_\Delta<\bar{\tau}_T$, i.e., $u<-1/\bar{\tau}_T$.
For the interval of $-1<u<-1/\bar{\tau}_T$,
$u\e^u$ is an increasing function such that $-1/e <u\e^u<h$.
As a result, \eqref{eq:reduced_ineq} does not hold for $-1<u<-1/\bar{\tau}_T$.
Now, from \eqref{eq:compact_sol} we finally obtain,
\begin{equation*}
\bar{\tau}_\Delta \leq -1/W_{-1}(-\frac{1}{\bar{\tau}_T}\e^{-1/\bar{\tau}_T})
 \;  \text{if} \; \bar{\tau}_\Delta\leq\bar{\tau}_T
\end{equation*}

\noindent\textbf{\textit{Case 2:}}  $\mathit{\bar{\tau}_\Delta\geq\bar{\tau}_T, \text{ i.e.}, \lambda_\Delta\leq\lambda_T}$ \\
From \eqref{eq:main_ineq} we have
\begin{equation*}
-\frac{1}{\bar{\tau}_\Delta}\e^{-1/\bar{\tau}_\Delta}\leq-\frac{1}{\bar{\tau}_T}\e^{-1/\bar{\tau}_T}
\label{eq:reduced_ineq2}
\end{equation*}
Analogous to the first case the solution to the inequality is
%\vspace{-10pt}
\begin{align*}
 W_{-1}(h)\leq&u<-1 ,  \; -1\leq u \leq W_{0}(h)   \nonumber \\
  W_{-1}(h)\leq&u\leq W_{0}(h)
\end{align*}
\noindent
Note that the following solution is only valid if $\bar{\tau}_\Delta\geq\bar{\tau}_T$.
%\vspace{-5pt}
\begin{equation}
-1/W_{-1}(-\frac{1}{\bar{\tau}_T}\e^{-1/\bar{\tau}_T})
\leq\bar{\tau}_\Delta\leq
-1/W_{0}(-\frac{1}{\bar{\tau}_T}\e^{-1/\bar{\tau}_T})
\label{eq:compact_sol2}
\end{equation}
Since
\begin{equation*}
    W_{0}(-\frac{1}{\bar{\tau}_T}\e^{-1/\bar{\tau}_T})
    \begin{cases}
        =-1/\bar{\tau}_T, \ \text{for} \bar{\tau}_T\geq1 \\
        >-1/\bar{\tau}_T,  \ \text{otherwise} \;,\\
    \end{cases}
\end{equation*}
%\vspace{-10pt}
using that in \eqref{eq:compact_sol} and \eqref{eq:compact_sol2}, we represent bounds for the expected delay for both cases depending on $\bar{\tau}_T$ as $\bar{\tau}_\Delta \leq \bar{\tau}_\Delta^+$,
%\vspace{-10pt}
\begin{equation*}
    \bar{\tau}_\Delta^+ =
    \begin{cases}
        -1/W_{-1}(-\frac{1}{\bar{\tau}_T}\e^{-1/\bar{\tau}_T}),    \text{ for} \; \bar{\tau}_T\geq1 \\
        -1/W_{0}(-\frac{1}{\bar{\tau}_T}\e^{-1/\bar{\tau}_T}),   \text{ otherwise} \;,\\
    \end{cases}
\end{equation*}
%\vspace{-10pt}
% \begin{align}
%         &\leq-1/W_{-1}(-\frac{1}{\bar{\tau}_T}\e^{-1/\bar{\tau}_T}) \; \text{for} \; \bar{\tau}_T\geq1 \nonumber\\
%         &\bar{\tau}_\Delta\leq-1/W_{0}(-\frac{1}{\bar{\tau}_T}\e^{-1/\bar{\tau}_T}) \; \text{otherwise} \;.
%         \label{eq:delay_bound}
% \end{align}
%
Now, for the considered caching system when the expected network delays lie below these bounds this results in an object hit probability that is larger or equal to the object hit probability at zero delay.
We conclude that there exist at least one delay value $\bar{\tau}_{\delta^*}$ that maximizes the hit probability.
For example, recall that in Fig.~\ref{Single MAP no D} we use a nearly periodic $\E_{20}$ input and $\bar{\tau}_T=2$. The figure shows that the hit probability for $\bar{\tau}_\Delta \leq -1/W_{-1}(-\frac{1}{2}\e^{-1/2}) = \bar{\tau}_\Delta^+$ is larger or equal to the hit probability under zero delay.
Moreover, we can show from the equation that as $\bar{\tau}_T \rightarrow \infty$, $\bar{\tau}_\Delta^+\rightarrow 0$, thus $\bar{\tau}_{\delta^*} \rightarrow 0$ (see Fig.~\ref{Single_phit_deltastar}).
An additional remark is that for small values of TTL ($\bar{\tau}_T<1$), the bound on the delay that improves the hit probability exceeds the TTL $\bar{\tau}_\Delta^+>\bar{\tau}_T$.
\vspace{-5pt}
\subsection{Illustration: Intuition behind the positive impact of the delay on the hit probability for periodic requests }
\begin{figure}[H]
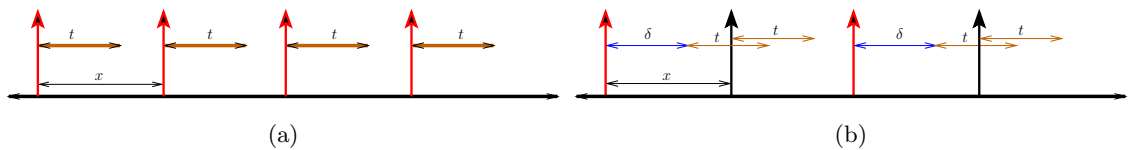

\centering
\begin{subfigure}[b]{0.45\textwidth}
\resizebox{1.0\linewidth}{!}{\input{Figs/periodic_miss}}
\caption{}
\end{subfigure}
\begin{subfigure}[b]{0.45\textwidth}
\resizebox{1.0\linewidth}{!}{\input{Figs/periodic_hit}}
\caption{}
\end{subfigure}
\caption{Assuming constant TTL and delay. (a)~Zero delay case: Zero hit probability. (b)~Non-zero delay: Positive hit probability.}
\label{fig:good_delay_0}
\end{figure}

\balance
\bibliographystyle{IEEEtran.bst}
\bibliography{IEEEabrv,bibliocache.bib}

\end{document}